\documentclass[aip,apl,longbibliography,reprint]{revtex4-1}

\usepackage{amssymb,amsmath}

\usepackage{graphicx}

\usepackage{xcolor}

\newcommand\beq{\begin{equation}}
\newcommand\eeq{\end{equation}}
\newcommand\beqa{\begin{eqnarray}}
\newcommand\eeqa{\end{eqnarray}}
\newcommand{\dd}{\text{d}}
\newcommand{\nn}{\nonumber\\}

\newcommand{\al}{\alpha}

\begin{document}
\title{Enskog kinetic theory in a model of confined granular mixtures. Heat flux and some applications} 
\author{David Gonz\'alez M\'endez}
\email{dgonzalezm@unex.es}
\affiliation{Departamento de F\'{\i}sica, Universidad de Extremadura, E-06071 Badajoz, Spain}
\author{Vicente Garz\'o}
\email{vicenteg@unex.es} \homepage{https://fisteor.cms.unex.es/investigadores/vicente-garzo-puertos/}
\affiliation{Departamento de
F\'{\i}sica and Instituto de Computaci\'on Cient\'{\i}fica Avanzada (ICCAEx), Universidad de Extremadura, E-06071 Badajoz, Spain}

\begin{abstract}

This work considers the Enskog kinetic theory for moderately dense confined granular mixtures within the framework of the $\Delta$-model, an effective collisional model that accounts for the transfer of energy from the vertical to the horizontal degrees of freedom of grains through an additional velocity increment during collisions. The main objective is to complete the Navier--Stokes hydrodynamic description of dense binary granular mixtures by explicitly determining the heat-flux transport coefficients, namely the Dufour and thermal conductivity coefficients, together with the first-order contributions to the partial temperatures and the cooling rate. These quantities are determined in the steady state from the Enskog kinetic equation by means of the Chapman--Enskog method and are expressed in terms of the masses and diameters of the particles, mixture composition, density, and coefficients of restitution. The results show that finite-density effects have a significant influence on transport, leading to behaviors that differ markedly from those observed in the dilute regime. The complete set of transport coefficients is considered further to study two different problems. First, we perform a linear stability analysis of the homogeneous steady state (HSS). The analysis demonstrates that no unstable transverse or longitudinal hydrodynamic modes are found over the parameter space explored, indicating that the HSS is linearly stable even at moderate densities. As a second problem, the violation of the Onsager reciprocal relations for a confined dense granular mixture is quantified in terms of the parameter space of the problem.

\end{abstract}



\date{\today}
\maketitle

\section{Introduction}
\label{sec1}

The study of transport properties in confined granular systems is an interesting and challenging problem that has
attracted the attention of many researchers in recent years.\cite{OU98,LCG99,PMEU04,CMS12,CMS15,GS18,CMSSGS19} In the experimental setup designed to analyze this problem, particles are confined in a box where its vertical $z$-direction is slightly larger than the diameter of a particle. The bottom plate of the box is a vibrating wall that supplies kinetic energy to the vertical degrees of freedom of particles when they collide with it. This energy is then dissipated and redistributed to the horizontal degrees of freedom of the grains when they collide with each other. Figure 1 of Ref.\ \onlinecite{GMG26} provides a schematic illustration of the studied system.

However, as we discussed in previous papers,\cite{GMG26,BSG26} the theoretical description of the confined systems using kinetic theory tools is a rather intricate problem due essentially to the restrictions imposed by confinement on the Boltzmann/Enskog collision operator. Thus, in spite of the recent advances made using this approach,\cite{MBGM22, MGB22, MPGM23} for the sake of simplicity it is quite common in granular literature to adopt a coarse-grained description where the effect of confinement on the motion of grains is accounted for via a collisional model. In this context, the so-called $\Delta$-model proposed years ago by Brito \emph{et al.}\cite{BRS13} has been considered by many researchers to assess the influence of confinement in an effective way on the dynamic properties of grains. As in the conventional inelastic hard sphere (IHS) model,\cite{G03,BP04,G19} the inelasticity in collisions is characterized in the case of smooth spheres by a (constant) coefficient of normal restitution. Additionally, there is an additional velocity increment of fixed magnitude ($\Delta>0$) along the normal collision direction. The factor $\Delta$ attempts to mimic the transfer of kinetic energy from the vertical to the horizontal degrees of freedom of grains due to their interaction with the vibrating bottom plate. More details on the $\Delta$-model can be found in Ref.\ \onlinecite{BSG26}.

Within the context of the Boltzmann equation, the $\Delta$-model has been considered to analyze the dynamic properties of simple confined granular gases.\cite{BGMB13,BMGB14,SRB14,BBMG15,BBGM16} An extension to moderate densities by starting from the inelastic Enskog kinetic equation has also been carried out.\cite{GBS18,GBS20} Apart from results in kinetic theory, the $\Delta$-model has also been employed in recent years in the study of systems with long-range interactions,\cite{JMV16} absorbing phase transitions in driven granular systems,\cite{MPSTSF24,MPSF25a} the formation of quasi-long-range ordered phases,\cite{PMFBRSF24,MP24} the non-equilibrium coexistence between a fluid and a crystal of granular hard disks\cite{MPSF25} and the study of hyperuniformity.\cite{MCh25}

The works concerning confined granular mixtures within the context of the $\Delta$-model have been more scarce. The reason is perhaps twofold. First, there is a large number of relevant parameters (such as diameters, masses, concentrations, and coefficients of restitution of each one of the species of the mixture) involved in the description of granular mixtures. Second, there is a wide array of complexities (such as the influence of non-equipartition of granular energy on transport properties)  that arise during the derivation of kinetic theory models. Due to the above difficulties, most studies have been restricted to the \emph{dilute} regime where the collisional contributions to the transport coefficients are negligibly small. In the low-density limit, explicit expressions for the Navier--Stokes transport coefficients of the binary mixture have been obtained\cite{GBS21} by solving the inelastic Boltzmann equation from the Chapman--Enskog perturbative method\cite{CC70} around the time-dependent homogeneous state.\cite{BSG20} These results have also been applied to study the stability of the homogeneous steady state (HSS) as well as to analyze a thermal diffusion segregation problem.\cite{GBS24,GBS24a}

An extension to granular mixtures at moderate densities has been recently published.\cite{GMG26} This work extends to arbitrary concentrations a previous study carried out in the tracer limit where the corresponding forms of the diffusion transport coefficients were explicitly determined by considering the lowest Sonine polynomial approximation.\cite{GGBS24} It is important to remark that the theoretical results derived for the tracer diffusion coefficient were compared against computer simulations showing in general a reasonably good agreement.\cite{GGBS24} Although in the study performed in Ref.\ \onlinecite{GMG26} the corresponding linear integral equations defining the transport coefficients were obtained, the explicit forms of the complete set of Navier-Stokes transport coefficients were not completely achieved. This was mainly due to the fact that for a multicomponent granular fluid the number of linear integrals to be solved to determine the transport coefficients is quite large. Thus, for instance, in the case of a binary mixture one has eight relevant transport coefficients plus two first-order contributions to the partial temperatures $T_i$ and the cooling rate $\zeta$. For this reason, the study of confined granular mixtures at moderate densities was in part carried out in Ref.\ \onlinecite{GMG26} where a complete study of the diffusion coefficients (associated with the mass flux) and the shear and bulk viscosities (associated with the pressure tensor) was worked out in steady-state conditions. However, the transport coefficients associated with the heat flux were not obtained. Thus, one of the primary objectives of the present paper is to determine the above coefficients (Dufour and thermal conductivity coefficients) as well as the first-order contributions to $T_i$ and $\zeta$.

As in previous papers,\cite{GBS21a,GBS24} the knowledge of the Navier--Stokes transport coefficients of the mixture opens up the possibility of analyzing the stability of the HSS. This study is of interest in its own right and also because the time-dependent homogeneous state is in fact the reference state in the Chapman--Enskog perturbative solution. In this context, this time-dependent homogeneous state plays a similar role to the so-called homogeneous cooling state (HCS) in the conventional inelastic hard sphere (IHS) model.\cite{BP04,G19} In this latter case, it is well known that there exists a certain critical length $L_c$ where the HCS becomes unstable when the linear size $L$ of the system is larger than $L_c$.\cite{GZ93,M93} Approximate theoretical predictions\cite{BDKS98,G05,GMD06,G15} for $L_c$ based on the leading Sonine approximations to the transport coefficients  have been shown to agree very well with computer simulations.\cite{BRM98,MDCPH11,MGHEH12,BR13,MGH14} This good agreement reinforces the usefulness of kinetic theory for describing granular flows, even for strong inelasticity conditions.

A relevant question is whether the HSS for \emph{confined} dense binary granular mixtures is unstable with respect to long-enough wavelength perturbations, as the HCS is.\cite{MGH14} However, in contrast to the HCS, a careful linear stability analysis of the linearized Navier--Stokes hydrodynamic equations (including the dependence of the transport coefficients and the first-order contributions to $T_i$ and $\zeta$ on the parameter space of the system) shows that the HSS for confined mixtures is stable. This conclusion agrees with previous stability analysis performed for monocomponent granular gases\cite{BBGM16,GBS21a} and for binary granular mixtures at low density.\cite{GBS24} Nevertheless, as expected, the forms of the hydrodynamic modes obtained here differ from the ones reported in the dilute regime.\cite{GBS24}

As an additional, simple application of the results derived here, this study examines the violation of Onsager reciprocal relations. Studying Onsager’s relations for dense granular mixtures in the context of the $\Delta$-model complements a previous analysis carried out in the low-density regime.\cite{GBS21} Even though time reversal invariance does not hold in granular systems (and consequently, Onsager’s relations do not apply to finite inelasticity), it is interesting to examine how the above relations deviate as inelasticity increases.

The plan of the paper is as follows. In Sec.\ \ref{sec2},  we introduce the $\Delta$-model and the inelastic Enskog equation for a multicomponent granular mixture. The balance equations for the densities of mass, momentum, and energy are also displayed along with the corresponding constitutive equations for the fluxes in the Navier--Stokes hydrodynamic order. Section \ref{sec3} deals with the determination of the transport coefficients associated with the heat flux as well as with the first-order contributions to the partial temperatures and the cooling rate. Expressions of these coefficients are obtained in terms of the dimensionality of the system, $d$, the masses and diameters of the granular mixtures, the concentrations, the volume fraction (or density), and the coefficients of restitution. For the sake of illustration, the dependence of the above transport coefficients
(scaled with respect to their values for elastic collisions) on inelasticity is
studied for the case of binary mixtures with a (common) coefficient of restitution
$\alpha$ and for different values of the solid volume fraction. The results show
that the thermal conductivity exhibits a relatively smooth dependence on density,
whereas the Dufour coefficients display a much more intricate behavior and may
even change sign for certain choices of the mechanical parameters. As expected, the influence of density on transport is in general important since their dependence on $\al$ differs from the one observed in the dilute regime.\cite{GBS21} The linear stability analysis of the HSS is addressed in Sec.\ \ref{sec4}. While the stability of the $d-1$ transverse shear modes is easily proved, the study of the time evolution of the four longitudinal hydrodynamic modes is much more intricate. To gain some insight, the limiting case $\mathbf{k}=\mathbf{0}$ is previously analyzed: the study shows that there are no unstable longitudinal modes in this limiting case. For $\mathbf{k} \neq \mathbf{0}$, one has to consider the terms coming from the spatial gradients in the constitutive equations and hence, one has to resort to a numerical analysis. A systematic study of the dependence of the longitudinal modes on the control parameters shows that these modes also decay in time and consequently, the HSS is linearly stable for finite densities as in the low-density limit.\cite{GBS24,GBS24a} The expected violation of the usual Onsager reciprocal relations is quantified in Sec.\ \ref{sec5} while a brief discussion on the results obtained in the paper is provided in Sec.\ \ref{sec6}.

\section{Hydrodynamics from Enskog kinetic theory in a model of confined granular mixtures}
\label{sec2}

\subsection{Model of confined quasi-two-dimensional granular mixtures}

We consider an $s$-component granular mixture of inelastic, smooth hard disks ($d=2$) or spheres ($d=3$) of masses $m_i$ and diameters $\sigma_i$. For the sake of simplicity, we assume that the spheres are completely \emph{smooth} and hence, the inelasticity in collisions is fully characterized by the constant (positive) coefficients of normal restitution $\al_{ij}\leq 1$ for collisions between particles of species (or components) $i$ and $j$. For moderate densities, the velocity distribution function $f_i(\mathbf{r}, \mathbf{v};t)$ of the species $i$ ($i=1,\cdots,s$) in the $\Delta$-model obeys the Enskog kinetic equation 
\beq
\label{2.1}
\frac{\partial}{\partial t}f_i+\mathbf{v}\cdot \nabla f_i
=\sum_{j=1}^s\; J_{ij}[\mathbf{r},\mathbf{v}|f_i,f_j],
\eeq
where the Enskog collision operators $J_{ij}$ for collisions between particles of species $i$ and $j$ in the $\Delta$-model read \cite{BSG20}
\beqa
\label{2.2}
& & J_{ij}[f_i,f_j]=\sigma_{ij}^{d-1} \int \mathrm{d}{\bf v}_{2}\int \mathrm{d} \widehat{\boldsymbol{\sigma}}\;
\Theta (-\widehat{{\boldsymbol {\sigma }}}\cdot {\bf g}_{12}-2\Delta_{ij})\nonumber\\
& &  \times (-\widehat{\boldsymbol {\sigma }}\cdot {\bf g}_{12}-2\Delta_{ij})
\al_{ij}^{-2}f_{ij}(\mathbf{r},\mathbf{r}+{\boldsymbol {\sigma}}_{ij},\mathbf{v}_1'',\mathbf{v}_2'';t)
\nonumber\\
& & 
-\sigma_{ij}^{d-1}\int \mathrm{d} {\bf v}_{2}\int \mathrm{d}\widehat{\boldsymbol{\sigma}}\;
\Theta (\widehat{{\boldsymbol {\sigma }}}\cdot {\bf g}_{12})
(\widehat{\boldsymbol {\sigma }}\cdot {\bf g}_{12})
\nonumber\\
& & \times  f_{ij}(\mathbf{r},\mathbf{r}+{\boldsymbol {\sigma}}_{ij},\mathbf{v}_1,\mathbf{v}_2;t).
\eeqa
Here, $\Theta(x)$ is the Heaviside step function, $\mathbf{g}_{12}=\mathbf{v}_1-\mathbf{v}_2$ is the relative velocity, $\widehat{\boldsymbol{\sigma}}$ is the unit collision vector joining the centers of the two colliding spheres, $\boldsymbol{\sigma}_{ij}=\sigma_{ij}\widehat{\boldsymbol{\sigma}}$ and $\sigma_{ij}=(\sigma_i+\sigma_j)/2$. The unit vector $\widehat{\boldsymbol{\sigma}}$ points from particle $1$ to particle $2$ and in Eq.\ \eqref{2.2} the particles are approaching if $\widehat{{\boldsymbol {\sigma }}}\cdot {\bf g}_{12}>0$. Moreover, the two-particle velocity distribution function $f_{ij}(\mathbf{r}_1, \mathbf{v}_1, \mathbf{r}_2, \mathbf{v}_2)$ is defined as
\beq
\label{2.14.3}
f_{ij}(\mathbf{r}_1, \mathbf{v}_1, \mathbf{r}_2, \mathbf{v}_2)=\chi_{ij}(\mathbf{r}_1, \mathbf{r}_2)f_i(\mathbf{r}_1, \mathbf{v}_1;t)f_j(\mathbf{r}_2, \mathbf{v}_2; t),
\eeq 
where $\chi_{ij}(\mathbf{r},\mathbf{r}+{\boldsymbol {\sigma}}_{ij})$ is the pair correlation function of two hard spheres, one of the species $i$ and the other of species $j$, at contact, i.e., when the distance between their centers is $\sigma_{ij}$. The quantities $\chi_{ij}$ account for volume excluded effects and spatial correlations not present in the Boltzmann equation. 

In Eq.\ \eqref{2.2}, the relationship between the velocities $\left(\mathbf{v}_1'',\mathbf{v}_2''\right)$ and $\left(\mathbf{v}_1,\mathbf{v}_2\right)$ is
\beq
\label{2.3a}
\mathbf{v}_1''=\mathbf{v}_1-\mu_{ji}\left(1+\alpha_{ij}^{-1}\right)(\widehat{{\boldsymbol {\sigma }}}\cdot \mathbf{g}_{12})\widehat{{\boldsymbol {\sigma }}}-2\mu_{ji}\Delta_{ij}\al_{ij}^{-1} \widehat{{\boldsymbol {\sigma }}},
\eeq
\beq
\label{2.3b}
\mathbf{v}_2''=\mathbf{v}_2+\mu_{ij}\left(1+\alpha_{ij}^{-1}\right)(\widehat{{\boldsymbol {\sigma }}}\cdot \mathbf{g}_{12})\widehat{{\boldsymbol {\sigma}}}+2\mu_{ij}\Delta_{ij}\al_{ij}^{-1} \widehat{{\boldsymbol {\sigma }}},
\eeq
where $\mu_{ij}=m_i/(m_i+m_j)$. From Eqs.\ \eqref{2.3a}--\eqref{2.3b}, one gets the identity
\beq
\label{2.4}
(\widehat{{\boldsymbol {\sigma}}}\cdot \mathbf{g}_{12}'')=-\al_{ij}^{-1}(\widehat{{\boldsymbol {\sigma}}}\cdot \mathbf{g}_{12})-2\Delta_{ij}\al_{ij}^{-1},
\eeq
where $\mathbf{g}_{12}''=\mathbf{v}_1''-\mathbf{v}_2''$. The quantity $\Delta_{ij}$ (which points outward in the normal direction $\widehat{{\boldsymbol {\sigma }}}$ as required by the conservation of angular momentum \cite{L04bis}) is an extra velocity added to the relative motion. It attempts to mimic the transfer of kinetic energy from the vertical degrees of freedom of grains (which has been gained by the collisions of particles of species $i$ and $j$ with the vibrating walls) to the horizontal ones. Additionally, as already noted in previous papers, \cite{GBS18,BSG20,GBS21,GGBS24} although the $\Delta$-model was mainly proposed to describe quasi-two dimensional systems, the calculations worked out in this paper will be performed for an arbitrary number of dimensions $d$.

For an $s$-component granular mixture, the relevant hydrodynamic fields are the number densities
\beq
\label{2.5}
n_{i}(\mathbf{r};t)=\int \dd{\bf v}\; f_{i}(\mathbf{r}, \mathbf{v};t),
\eeq
the flow velocity $\mathbf{U}$
\beq
\label{2.6}
\mathbf{U}(\mathbf{r};t)=\rho(\mathbf{r};t)^{-1}\sum_{i=1}^s\; m_{i}
\int \dd {\bf v}\; {\bf v}\; f_{i}(\mathbf{r}, \mathbf{v};t),
\eeq
and the granular temperature $T$
\beq
\label{2.7}
T(\mathbf{r};t)=\frac{1}{n(\mathbf{r};t)}\sum_{i=1}^s\frac{m_{i}}{d}\int \dd{\bf
v}\; V^{2}\; f_{i}(\mathbf{r}, \mathbf{v};t).
\eeq
Here, $\rho=\sum_i m_{i}n_{i}$ is the total mass density, ${\bf V}={\bf v}-{\bf U}$ is the peculiar velocity, and $n=\sum_i n_{i}$ is the total number density. Apart from the hydrodynamic fields, the partial temperatures $T_i$ $(i=1,\cdots, s)$ are defined as 
\beq
\label{2.8}
T_i(\mathbf{r};t)=\frac{m_i}{d n_i(\mathbf{r};t)}\int \dd{\bf v}\; V^{2}\; f_{i}(\mathbf{r}, \mathbf{v};t).
\eeq
The quantity $T_i$ measures the mean kinetic energy of species $i$.  

According to Eqs.\ \eqref{2.7} and \eqref{2.8}, the temperature of the mixture $T$ can be written in terms of the partial temperatures $T_i$ as
\beq
\label{2.8.1}
n T=\sum_{i=1}^s\; n_i T_i,
\eeq
where $n=\sum_i n_i$ is the total number density.

\subsection{Hydrodynamic equations}

The (macroscopic) hydrodynamic equations for the partial densities $n_i(\mathbf{r};t)$, the flow velocity $\mathbf{U}(\mathbf{r};t)$, and the granular temperature $T(\mathbf{r};t)$ were derived in Ref.\ \onlinecite{GMG26}. They are given by
\begin{equation}
\mathrm{D}_tn_{i}+n_{i}\nabla \cdot {\bf U}+\frac{\nabla \cdot {\bf j}_{i}}{m_{i}}
=0,  \label{2.9}
\end{equation}
\begin{equation}
\mathrm{D}_t{\bf U}+\rho ^{-1}\nabla \cdot \mathsf{P}=\mathbf{0},  \label{2.10}
\end{equation}
\begin{equation}
\mathrm{D}_tT-\frac{T}{n}\sum_{i=1}^s\frac{\nabla \cdot {\bf j}_{i}}{m_{i}}+\frac{2}{dn}
\left( \nabla \cdot {\bf q}+\mathsf{P}:\nabla {\bf U}\right) =-\zeta \,T.
\label{2.11}
\end{equation}
In the balance equations \eqref{2.9}--\eqref{2.11}, $\mathrm{D}_t=\partial_t+\mathbf{U}\cdot \nabla$ is the material derivative,
\begin{equation}
{\bf j}_{i}=m_{i}\int \mathrm{d}{\bf v}\,{\bf V}\,f_{i}({\bf v}),
\label{2.12}
\end{equation}
is the mass flux for the species $i$ relative to the local flow, $\mathsf{P}$ is the pressure tensor, and $\mathbf{q}$ is the heat flux. While the mass flux $\mathbf{j}_i$ has only \emph{kinetic} contributions, the pressure tensor and the heat flux have both \emph{kinetic} and \emph{collisional} transfer contributions, i.e., $\mathsf{P}=\mathsf{P}^\text
{k}+\mathsf{P}^\text{c}$ and $\mathbf{q}=\mathbf{q}^\text{k}+\mathbf{q}^\text{c}$. The kinetic contributions are given as usual by
\begin{equation}
\mathsf{P}^\text{k}=\sum_{i=1}^s\,\int \mathrm{d}{\bf v}\,m_{i}{\bf V}{\bf V}\,f_{i}({\bf v}),
\label{2.13}
\end{equation}
\begin{equation}
{\bf q}^\text{k}=\sum_{i=1}^s\,\int \mathrm{d}{\bf v}\,\frac{1}{2}m_{i}V^{2}{\bf V}\,f_{i}({\bf v}).
\label{2.14}
\end{equation}
The form of the collisional contribution $\mathsf{P}^\text{c}$ to the pressure tensor is given by Eq.\ (21) of Ref.\ \onlinecite{GMG26} while the collision contribution $\mathbf{q}^\text{c}$ to the heat flux  and the cooling rate $\zeta$ are given, respectively, by
\beqa
\label{2.14.1}
{\bf q}^\text{c}&=&\sum_{i,j}\frac{1+\alpha_{ij}}{8}m_{ij} \sigma_{ij}^{d}
\int \mathrm{d}\mathbf{v}_{1}\int \mathrm{d}\mathbf{v}_{2}\int
\dd\widehat{\boldsymbol {\sigma}}\,\Theta (\widehat{\boldsymbol{\sigma}}\cdot
\mathbf{g}_{12})\nonumber\\
& & \times
(\widehat{\boldsymbol {\sigma}}\cdot \mathbf{g}_{12})^{2}\widehat{\boldsymbol {\sigma}}\Big[4
(\widehat{\boldsymbol {\sigma}}\cdot {\bf G}_{ij})+(\mu_{ji}-\mu_{ij})(1-\al_{ij})\nonumber\\
& &\times
(\widehat{\boldsymbol{\sigma}}\cdot
\mathbf{g}_{12})\Big]\int_{0}^{1}\dd\lambda \nonumber\\
& & \times
f_{ij}\left[\mathbf{r}-
\lambda{\boldsymbol{\sigma}}_{ij},\mathbf{v}_{1},\mathbf{r}+(1-\lambda)
{\boldsymbol {\sigma}}_{ij},\mathbf{v}_{2},t\right]
\nonumber\\
& &-\sum_{i,j} \frac{m_i}{4} \sigma_{ij}^d \Delta_{ij}
\int \mathrm{d}\mathbf{v}_{1}\int \mathrm{d}\mathbf{v}_{2}\int
d\widehat{\boldsymbol {\sigma}}\Theta (\widehat{\boldsymbol{\sigma}}\cdot
\mathbf{g}_{12})\nonumber\\
& & \times (\widehat{\boldsymbol {\sigma}}\cdot \mathbf{g}_{12})
\widehat{\boldsymbol {\sigma}}
\Big[4\mu_{ji}^2\Delta_{ij} +4\mu_{ji}^2\al_{ij} (\widehat{\boldsymbol {\sigma}}\cdot \mathbf{g}_{12})\nonumber\\
& & -
4\mu_{ji} (\widehat{\boldsymbol {\sigma}}\cdot \mathbf{G}_{ij})\Big]
\int_{0}^{1}\dd\lambda \nonumber\\
& & \times
f_{ij}\left[\mathbf{r}-\lambda {\boldsymbol{\sigma}}_{ij},\mathbf{v}_{1},\mathbf{r}+(1-\lambda)
{\boldsymbol {\sigma}}_{ij},\mathbf{v}_{2},t\right].
\eeqa
The cooling rate $\zeta$ is
\beqa
\label{2.14.2}
\zeta&=&-\frac{2}{d n T}\sum_{i,j}\sigma_{ij}^{d-1}m_{ij}\int \dd \mathbf{v}_1\int \dd \mathbf{v}_2
\int \mathrm{d}\widehat{\boldsymbol {\sigma }}\,\Theta (\widehat{{\boldsymbol {\sigma}}}
\cdot \mathbf{g}_{12})\nonumber\\
& & \times (\widehat{\boldsymbol {\sigma}}\cdot {\bf g}_{12})\Big[\Delta_{ij}^2+\al_{ij}\Delta_{ij}(\widehat{{\boldsymbol {\sigma}}}\cdot \mathbf{g}_{12})-\frac{1-\al_{ij}^2}{4}(\widehat{{\boldsymbol {\sigma}}}\cdot \mathbf{g}_{12})^2\Big]
\nonumber\\
& & \times
f_{ij}(\mathbf{r},\mathbf{v}_1,\mathbf{r}+\boldsymbol{\sigma}_{ij},\mathbf{v}_2,t).
\eeqa
In Eqs.\ \eqref{2.14.1}--\eqref{2.14.2}, $m_{ij}=m_im_j/(m_i+m_j)$ is the reduced mass and $\mathbf{G}_{ij}=\mu_{ij}\mathbf{V}_1+\mu_{ji}\mathbf{V}_2$ is the center-of-mass velocity.

As in Ref.\ \onlinecite{GMG26}, in what follows we will use Latin indices to label the particle species and Greek indices to label the spatial dimensions ($d=2$ for hard disks and $d=3$ for hard spheres). In addition, Einstein summation convention over repeated Greek indices is assumed throughout this paper.

\subsection{Navier--Stokes hydrodynamic equations}

For a mixture consisting of $s$ species, there are $s+d+1$ independent hydrodynamic fields: the $s$ partial densities $n_i$ ($i=1,\cdots, s$), the granular temperature $T$, and the $d$ components of the flow velocity $\mathbf{U}$. On the other hand, as expected, the set of hydrodynamic equations \eqref{2.9}--\eqref{2.11} does not constitute a closed set of nonlinear differential equations for the fields. To close them, one needs to express $\mathbf{j}_i$, $\mathsf{P}$, $\mathbf{q}$, and $\zeta$ in terms of the above fields. Such expressions are called ``constitutive equations''. Up to the Navier--Stokes hydrodynamic order (first order in spatial gradients), the constitutive equations are
\beq
\label{2.15}
\mathbf{j}_i=-\sum_{j=1}^s \frac{m_im_j n_j}{\rho}D_{ij}\nabla \ln n_j-\rho D_i^T \nabla \ln T,
\eeq
\beq
\label{2.16}
P_{\lambda\beta}=p\delta_{\lambda\beta}-\eta\left(\frac{\partial U_\beta}{\partial r_\lambda} +\frac{\partial U_\lambda}{\partial r_\beta}-\frac{2}{d}\delta_{\lambda\beta}\nabla \cdot \mathbf{U}\right) -
\delta_{\lambda\beta} \eta_b  \nabla \cdot \mathbf{U},
\eeq
\begin{equation}
\mathbf{q}=-\sum_{i=1}^{s}\sum_{j=1}^{s}T^{2}D_{q,ij}\nabla \ln n_{j}
-T\kappa \nabla \ln T,  \label{2.17}
\end{equation}
\beq
\label{2.18}
\zeta=\zeta^{(0)}+\zeta_U \nabla \cdot \mathbf{U}.
\eeq
In Eq.\ \eqref{2.15}, $D_{ij}$ are the mutual diffusion coefficients and $D_i^T$ are the thermal diffusion coefficients. In Eq.\ \eqref{2.16}, $p$ is the hydrostatic pressure while $\eta$ and $\eta_b$ are the shear and bulk viscosity coefficients, respectively. In Eq.\ \eqref{2.17}, the coefficients
\beq
\label{2.18.1}
D_{q,i}=\sum_{\ell=1}^s\; D_{q,\ell i}
\eeq
are the Dufour coefficients while $\kappa$ is the thermal conductivity coefficient. Finally, in Eq.\ \eqref{2.18}, $\zeta^{(0)}$ and $\zeta_U$ are the zeroth- and first-order contributions to the cooling rate, respectively. Moreover, the partial temperatures are expanded as
\beq
\label{2.18.2}
T_i=T_i^{(0)}+T_i^{(1)},
\eeq
where the first-order contribution $T_i^{(1)}$ is
\beq
\label{2.19}
T_i^{(1)}=\varpi_i \nabla \cdot \mathbf{U}.
\eeq
Relation \eqref{2.8.1} imposes the constraints
\beq
\label{2.20}
T=\sum_{i=1}^s\; x_i T_i^{(0)}, \quad \sum_{i=1}^s\; x_i\varpi_i=0,
\eeq
where $x_i=n_i/n$ is the concentration or mole fraction of species $i$.

In a previous paper, \cite{GMG26} explicit expressions for the set of Navier--Stokes transport coefficients $(D_{ij}, D_i^T,\eta,\eta_b)$ were derived by solving the integral equations verifying these coefficients. These forms were approximately obtained in the steady state by considering the leading terms in a Sonine polynomial expansion. In addition, for the sake of simplicity, the case $\Delta_{ij}\equiv\Delta$ ($i,j=1,\cdots,s)$ was considered. The determination of the remaining heat-flux transport coefficients $D_{q,ij}$ and $\kappa$ together with the first-order contributions to the partial temperatures $\varpi_i$ and the cooling rate $\zeta_U$ is accomplished in Sec.\ \ref{sec3} of the present paper. As in the results reported in Ref.\ \onlinecite{GMG26}, in reduced forms, the coefficients $(D_{q,ij},\kappa,\varpi_i)$ are given in terms of the masses $m_i/\overline{m}$ and diameter $\sigma_i/\overline{\sigma}$ ratios, the concentrations $x_i$, the coefficients of restitution $\al_{ij}$, and the volume fraction $\phi$. Here, $\overline{m}=\sum_i m_i/s$ and $\overline{\sigma}=\sum_i \sigma_i/s$.

The nonlinear Navier--Stokes hydrodynamic equations for the confined granular mixture can be obtained by substituting the constitutive equations \eqref{2.15}--\eqref{2.18} into the exact balance equations \eqref{2.9}--\eqref{2.11}. The explicit form of the Navier--Stokes equations will be displayed in Sec.\ \ref{sec4} when we analyze the stability of the HSS.

\section{Heat flux transport coefficients and cooling rate}
\label{sec3}

\subsection{Heat flux}

The constitutive equation for the heat flux to first-order in spatial gradients is given by Eq.\ \eqref{2.17}. As said before, the heat flux transport coefficients $D_{q,ij}$ and $\kappa $ have kinetic and
collisional contributions:
\begin{equation}
D_{q,ij}=D_{q,ij}^{\text{k}}+D_{q,ij}^{\text{c}},\quad \kappa =\kappa^{\text{k}}+\kappa^{\text{c}}.  \label{3.1}
\end{equation}
As discussed in the low-density regime,\cite{GBS21} the evaluation of the kinetic contributions to those coefficients would require in principle to consider the second Sonine approximation. However, this is a rather intricate and lengthy problem in the context of the $\Delta$-model for mixtures. For this reason, to provide simple forms for the kinetic coefficients, we consider here the first Sonine approximation for estimating these coefficients. In this approximation, one gets the expressions
\begin{equation}
\kappa^{\text{k}}=\frac{d+2}{2T}\sum_{i=1}^{s}\frac{\rho T_{i}^{(0)}}{m_{i}}D_{i}^{T},  \label{3.2}
\end{equation}
\begin{equation}
D_{q,ij}^{\text{k}}=\frac{d+2}{2T^{2}}\frac{m_{j}n_{j}T_{i}^{(0)}}{\rho }D_{ij}.
\label{3.3}
\end{equation}
The diffusion transport coefficients $D_i^T$ and $D_{ij}$ were obtained in Ref.\ \onlinecite{GMG26} in the first Sonine approximation. Note that for mechanically equivalent particles, $D_i^T=0$ and so, the approximation \eqref{3.2} yields a zero thermal conductivity for monocomponent granular gases. In other words, the first Sonine approximation \eqref{3.2} to the heat flux transport coefficients is not able to reproduce the correct result in this limiting case.\cite{GBS18,BSG26} Nevertheless, the expressions \eqref{3.2} and \eqref{3.3} are consistent in the order of approximation used to obtain the mass flux transport coefficients and, consequently, can be employed to analyze for instance the stability of the HSS or to quantify the violation of  Onsager's relations in granular mixtures.

We now focus our attention on the collisional contributions to the heat flux transport coefficients. These contributions can be even more significant than the kinetic ones for relatively high densities. Since the determination of these contributions involves several lengthy manipulations, we only offer here their final expressions. Some technical details on this derivation are provided in the Appendix \ref{appA}.

The collisional coefficient $D_{q,ij}^{\text{c}}$ can be written as
\beq
\label{3.4}
D_{q,ij}^{\text{c}}=D_{q,ij}^{\text{c}(\Delta=0)}+D_{q,ij}^{\text{c}(\Delta\neq 0)},
\eeq
where $D_{q,ij}^{\text{c}(\Delta=0)}$ refers to the expression of this coefficient obtained in the conventional IHS model (namely, when $\Delta_{ij}=0$) while $D_{q,ij}^{\text{c}(\Delta\neq 0)}$ corresponds to the contributions to $D_{q,ij}^{\text{c}}$ proportional to $\Delta_{ij}$. The coefficient $D_{q,ij}^{\text{c}(\Delta=0)}$ can be written as\cite{GDH07,GHD07,G19} 
\begin{widetext}
\beqa
\label{3.5}
D_{q,ij}^{\text{c}(\Delta=0)}&=&\frac{3\pi^{\frac{d}{2}}}{d\Gamma\left(\frac{d}{2}\right)}
\sum_{\ell=1}^{s}m_{i\ell}\sigma_{i\ell}^d \chi_{i \ell}(1+\al_{i \ell})\Bigg\{\frac{1}{4}\frac{n_i \rho_j}{\rho T}D_{\ell j}\Bigg[\frac{\gamma_\ell}{m_\ell}\Big((1-\al_{i\ell})\mu_{i\ell}
+(3+\al_{i\ell})\mu_{\ell i}\Big)+\frac{\gamma_i}{m_i}\Bigg(\left(\frac{1}{3}+\al_{i\ell}\right)\mu_{\ell i}\nonumber\\
& &
-\left(\frac{5}{3}+\al_{i\ell}\right)\mu_{i\ell}\Bigg)\Bigg]
-\frac{1}{3\sqrt{\pi}}\sigma_{i\ell}\frac{n_i n_\ell \upsilon_{\text{th}}}{\overline{m}}T
C_{i \ell j}^*
\Bigg\},
\eeqa
where $\gamma_i=T_i^{(0)}/T$ is the temperature ratio of species $i$ and $\upsilon_{\text{th}}=\sqrt{2T/\overline{m}}$ is the thermal velocity. In Eq.\ \eqref{3.5} we have introduced the quantity
\beqa
\label{3.7}
C_{i \ell j}^*&=&
(\theta_i+\theta_\ell)^{-1/2}(\theta_i\theta_\ell)^{-3/2}\left\{
\left(\delta_{j\ell}+\frac{1}{2}I_{i \ell j}\right)
\beta_{i\ell}(\theta_i+\theta_\ell) -\frac{1}{2}\theta_i\theta_\ell \left[1+\frac{\mu_{\ell\;i}(\theta_i+\theta_\ell)-2 \beta_{i\ell}}{\theta_\ell}\right]
\frac{\partial \ln \gamma_\ell}{\partial \ln
n_j}\right\}\nonumber\\
& & +\frac{1}{4}(1-\alpha_{i\;\ell})(\mu_{\ell\; i}-\mu_{i\ell})
\left(\frac{\theta_i+\theta_\ell}{\theta_i\theta_\ell}\right)^{3/2} \left( \delta_{j\ell}+\frac{1}{2}I_{i \ell j}
+\frac{3}{2}\frac{\theta_i}{\theta_i+\theta_\ell}\frac{\partial \ln \gamma_\ell}{\partial \ln
n_j}\right),\nonumber\\
\eeqa
where $\theta_i=m_i/(\overline{m}\gamma_i)$, $\beta_{ij}=\mu_{ij}\theta_j-\mu_{ji}\theta_i$ and the expressions of the quantities  $I_{i\ell j}$ for a binary mixture ($s=2$) of hard disks are displayed in the Appendix \ref{appC}. The coefficient $D_{q,ij}^{\text{c}(\Delta\neq 0)}$ is given by
\beqa
\label{3.8}
D_{q,ij}^{\text{c}(\Delta\neq 0)}&=&\frac{\pi^{\frac{d-1}{2}}}{8d\Gamma\left(\frac{d}{2}\right)}T^{-2}m_i \upsilon_\text{th}^3\sum_{\ell=1}^s
\sigma_{i\ell}^{d+1}\chi_{i\ell}\Delta_{i\ell}^*n_in_\ell \Bigg\{
\left[4\mu_{\ell i}(\mu_{\ell i}-\mu_{i\ell})\delta_{j\ell}+4 \mu_{\ell i}^2 I_{i\ell j}\right]\Bigg[\left(\frac{\theta_i+\theta_\ell}{\theta_i\theta_\ell}\right)^{1/2}\Delta_{i\ell}^*\nonumber\\
& & +\frac{\sqrt{\pi}}{2}
\al_{i\ell}\left(\frac{\theta_i+\theta_\ell}{\theta_i\theta_\ell}\right)\Bigg]-4\sqrt{\pi} \frac{\overline{m}}{m_i+m_\ell}\mu_{\ell i}(\gamma_i-\gamma_\ell)
\Big(\delta_{j\ell}+\frac{1}{2}I_{i\ell j}\Big)
+\Bigg[2\mu_{\ell i}(\mu_{\ell i}-\mu_{i\ell})\Delta_{i\ell}^*
\nonumber\\
& & \times (\theta_i+\theta_\ell)^{-1/2}\theta_i^{1/2}\theta_\ell^{-1/2}+
2\sqrt{\pi}\frac{\overline{m}\gamma_\ell}
{m_i}\mu_{i\ell}\left[\al_{i\ell}(\mu_{\ell i}-\mu_{i\ell})+2\mu_{\ell i}\right]\Bigg]
n_j
\frac{\partial\ln\gamma_\ell}{\partial n_j}
\Bigg\}
\nonumber\\
& &+\sum_{\ell=1}^s\frac{n_i m_{i\ell}\rho_j}{\rho T^2}\sigma_{i\ell}^d\chi_{i\ell}\upsilon_{\text{th}}\Delta_{i\ell}^*
D_{\ell j}\mathcal{Q}_{i\ell},
\eeqa
where $\Delta_{ij}^*=\Delta_{ij}/\upsilon_{\text{th}}$ and the quantity $\mathcal{Q}_{ij}$ is defined as
\beq
\label{3.8.1}
\mathcal{Q}_{ij}=\frac{\pi^{\frac{d-1}{2}}}{d\Gamma\left(\frac{d}{2}\right)}\upsilon_{\text{th}}\Bigg\{\sqrt{\pi}(\mu_{ji}-\mu_{ij})\Delta_{ij}^* +
2\left(\frac{\theta_i+\theta_j}{\theta_i\theta_j}\right)^{1/2}
\Bigg[\alpha_{ij}(\mu_{ji}-\mu_{ij})+\frac{(\theta_j-2\beta_{ij})}{\theta_i+\theta_j}\Bigg]\Bigg\}.
\eeq

We consider now the collisional coefficient $\kappa_c$. It can be written as
\beq
\label{3.9}
\kappa_\text{c}=\kappa_\text{c}^{(\Delta=0)}+\kappa_\text{c}^{(\Delta\neq 0)},
\eeq
where
\beqa
\label{3.10}
\kappa_\text{c}^{(\Delta=0)}&=&\frac{3\pi^{\frac{d}{2}}}{d\Gamma\left(\frac{d}{2}\right)}\sum_{i=1}^{s}\sum_{j=1}^{s}
\mu_{ij}\sigma
_{ij}^{d}\chi_{ij}\left(1+\alpha_{ij}\right)\Bigg\{\frac{1}{4}n_i\rho D_j^T\Bigg[\frac{\gamma_j}{m_j}\Big((1-\al_{ij})\mu_{ij}
+(3+\al_{ij})\mu_{ji}\Big)\nonumber\\
& & +\frac{\gamma_i}{m_i}\Bigg(\left(\frac{1}{3}+\al_{ij}\right)\mu_{ji}
-\left(\frac{5}{3}+\al_{ij}\right)\mu_{ij}\Bigg)\Bigg]+\frac{1}{6\sqrt{\pi}}\sigma_{ij}\frac{n_i\rho_j}
{\overline{m}}\upsilon_{\text{th}}C_{ij}^*\Bigg\},
\eeqa
\begin{align}
\label{3.11}
\kappa_\text{c}^{(\Delta\neq 0)}=\frac{\pi^{\frac{d-1}{2}}}{8d\Gamma\left(\frac{d}{2}\right)}\sum_{i=1}^s\sum_{j=1}^s T^{-1}m_in_in_j
\sigma_{ij}^{d+1} \chi_{ij} \Delta_{ij}^*\upsilon_\text{th}^3\Bigg\{2\mu_{ji}
(\mu_{ji}-\mu_{ij})\Delta_{ij}^*
(\theta_i+\theta_j)^{-1/2}\theta_i^{1/2}\theta_j^{-1/2}\nonumber\\ +\,2\sqrt{\pi}\frac{\overline{m}\gamma_j}{m_i}\mu_{ij}\left[\al_{ij}(\mu_{ji}-\mu_{ij})+
2\mu_{ji}\right]\Bigg\}\Bigg(1-\frac{1}{2}\Delta^*\frac{\partial \ln \gamma_j}{\partial \Delta^*}\Bigg)  +\sum_{i=1}^s\sum_{j=1}^s \frac{n_i\mu_{ij}}{T}\sigma_{ij}^d\chi_{ij}\upsilon_{\text{th}}\Delta_{ij}^*\rho D_j^T\mathcal{Q}_{ij}.
\end{align}
In Eqs. \eqref{3.10} and \eqref{3.11}, we have introduced the quantity
\beqa
\label{3.13}
C_{ij}^*&=&\Bigg\{(\theta_{i}+\theta_{j})^{-1/2}(\theta _{i}\theta_{j})^{-3/2}\Big( 2\beta_{ij}^{2}+\theta_{i}\theta_{j}+(\theta_{i}+
\theta_{j})\left[ (\theta_{i}+\theta_{j})\mu_{ij}\mu_{ji}+\beta_{ij}
(1+\mu_{ji})\right] \Big)  \notag \\
&&+\frac{3}{4}(1-\alpha_{ij})(\mu_{ji}-\mu
_{ij})\left( \frac{\theta_{i}+\theta_{j}}{\theta _{i}\theta _{j}}\right) ^{3/2}\left[ \mu_{ji}+\beta_{ij}(\theta_{i}+\theta_{j})^{-1}\right]\Bigg\}\Bigg(1-\frac{\Delta^*}{2}\frac{\partial \ln \gamma_j}{\partial \Delta^*}\Bigg),
\eeqa
\end{widetext}
and have used the shorthand notation
\beq
\label{3.13.1}
\Delta^* \frac{\partial \ln \gamma_i}{\partial \Delta^*}\equiv \sum_{j=1}^s\sum_{\ell=1}^s \Delta_{\ell j}^* \frac{\partial \ln \gamma_i}{\partial \Delta_{\ell j}^*}.
\eeq
During the present calculations, we realized that there is a nonzero contribution to $\mathbf{q}_c^{(\Delta=0)}$ that was not considered in previous papers.\cite{GDH07,GHD07} This contribution is proportional to the terms $I_{ij\ell}$ and has been included in the expression \eqref{3.5} of $D_{q,ij}^{\text{c}(\Delta=0)}$. Additionally, some typos present in the previous forms of  $D_{q,ij}^{\text{c}(\Delta=0)}$ and $\kappa_\text{c}^{(\Delta=0)}$ have also been fixed.

For elastic collisions ($\al_{ij}=1$), $\gamma_i=1$, $\beta_{ij}=0$, and $\Delta_{ij}^*=0$. In this limiting case, Eqs.\ \eqref{3.5} and \eqref{3.10} lead to the following expressions for the collisional coefficients $D_{q,ij}^{\text{c}}$ and $\kappa_\text{c}$:
\beq
\label{3.14}
D_{q,ij}^c=\frac{2\pi^{\frac{d}{2}}}{d\Gamma\left(\frac{d}{2}\right)}\frac{\rho_j}{\rho T}
\sum_{\ell=1}^{s}n_i\sigma_{i\ell}^d \mu_{\ell i}\chi_{i\ell}D_{\ell j},
\eeq
\beqa
\label{3.15}
\kappa_\text{c}&=&\frac{2\pi^{\frac{d}{2}}}{d\Gamma\left(\frac{d}{2}\right)}\sum_{i=1}^s\sum_{j=1}^s\Bigg[
\frac{\rho n_i\sigma_{ij}^d \chi_{ij}}{m_i+m_j}D_j^T+ \frac{n_in_j\sigma_{ij}^{d+1}\chi_{ij}}{m_i+m_j}
\nonumber\\
& &\times \left(\frac{2T m_i m_j}{\pi (m_i+m_j)}\right)^{1/2}\Bigg].
\eeqa
Expressions \eqref{3.14} and \eqref{3.15} are consistent with the results derived years ago for a molecular mixture of hard spheres.\cite{LCK83}

\subsection{First order contributions to the partial temperatures and the cooling rate}

The first-order contribution to the partial temperature $T_i$ is defined as
\beq
\label{3.16}
T_i^{(1)}=\frac{m_i}{d n_i}\int \dd\mathbf{v}\; V^2 f_i^{(1)}(\mathbf{V}),
\eeq
where the first-order velocity distribution function $f_i^{(1)}(\mathbf{r}, \mathbf{v};t)$ is given by
\beqa
\label{3.16.1}
f_i^{(1)}&=&\boldsymbol{\mathcal{A}}_i\left(\mathbf{V}\right)\cdot  \nabla \ln
T+\sum_{j=1}^s\boldsymbol{\mathcal{B}}_{ij}\left(\mathbf{V}\right) \cdot \nabla \ln n_j\nonumber\\
& & +\mathcal{C}_{i,\lambda \beta}(\mathbf{V})\frac{1}{2}\left(\partial_\beta U_\lambda+\partial_\lambda U_\beta-\frac{2}{d}\delta_{\lambda\beta}\nabla \cdot \mathbf{U}\right)\nonumber\\
& & +\mathcal{D}_i
\left(\mathbf{V}\right) \nabla \cdot \mathbf{U}.
\eeqa
The unknowns $\boldsymbol{\mathcal{A}}_i$, $\boldsymbol{\mathcal{B}}_{ij}$, $\mathcal{C}_{i,\lambda \beta}$, and $\mathcal{D}_i$ are the solutions of the set of coupled linear integral equations (55)-(58) of Ref.\ \onlinecite{GMG26}, respectively.

Since $T_i^{(1)}$ is a scalar, it can only be coupled to the divergence of the flow velocity $\nabla \cdot \mathbf{U}$ since $\nabla n_i$ and $\nabla T$ are vectors and the tensor $\partial_\lambda U_\beta+\partial_\beta U_\lambda-
(2/d)\delta_{\lambda\beta}\nabla\cdot\mathbf{U}$ is a traceless tensor. Thus, $T_i^{(1)}$ is defined by Eq.\ \eqref{2.19} where
\beq
\label{3.17}
\varpi_i=\frac{m_i}{d n_i}\int\; \dd \mathbf{v}\;  V^2 \mathcal{D}_i(\mathbf{V}).
\eeq

In steady state conditions, the zeroth-order contribution $\zeta^{(0)}$ to the cooling rate vanishes and hence, the integral equations for the unknowns $\mathcal{D}_i$ become\cite{GMG26}
\begin{equation}
\label{3.18}
-\zeta^{(1,1)}T \frac{\partial f_i^{(0)}}{\partial T}
-\sum_{j=1}^s\left(J_{ij}^{(0)}[\mathcal{D}_{i},f_j^{(0)}]+
J_{ij}^{(0)}[f_i^{(0)},\mathcal{D}_{j}]\right)=D_{i}',
\end{equation}
where
\beqa
\label{3.19}
D_i'\left(\mathbf{V}\right)&=&\frac{1}{d}\mathbf{V}\cdot \frac{\partial f_i^{(0)}}{\partial \mathbf{V}}
+\left(\zeta^{(1,0)}+\frac{2}{d}\frac{p}{nT}\right)T\frac{\partial f_i^{(0)}}{\partial T}+\nonumber\\
& & +\sum_{j=1}^s \left\{n_j \frac{\partial f_i^{(0)}}{\partial n_j}+
\frac{1}{d}\mathcal{K}_{ij,\beta}\left[\frac{\partial f_j^{(0)}}{\partial V_\beta}\right]\right\}.
\eeqa
The operator $\boldsymbol{\mathcal{K}}[X]$  is defined in Eq.\ (A4) of Ref.\ \onlinecite{GMG26} while the coefficients $\zeta^{(1,0)}$ and $\zeta^{(1,1)}$ are defined as\cite{GMG26}
\beqa
\label{3.19.1}
\zeta^{(1,0)}&=&\frac{4\pi^{d/2}}{d^2\Gamma\left(\frac{d}{2}\right)}\sum_{i,j}\chi_{ij}\frac{m_{ij}}{\overline{m}}x_ix_j n\sigma_{ij}^d\Delta_{ij}^{*2}+\frac{4\pi^{(d-1)/2}}{d^2\Gamma\left(\frac{d+1}{2}\right)}\frac{v_{\text{th}}}{nT}\nonumber\\
& & \times \sum_{i,j}\chi_{ij}
m_{ij}\sigma_{ij}^d \al_{ij}\Delta_{ij}^*\int \mathrm{d}\mathbf{v}_1 \int \mathrm{d}\mathbf{v}_2\; g_{12} f_i^{(0)}(\mathbf{V}_1)\nonumber\\
& & \times f_j^{(0)}(\mathbf{V}_2)-\frac{3\pi^{d/2}}{d^2\Gamma\left(\frac{d}{2}\right)}\sum_{i,j}\chi_{ij}\mu_{ji}x_i x_j n\sigma_{ij}^d\gamma_i (1-\al_{ij}^2),
\nonumber\\
\eeqa
\beqa
\label{3.19.2}
\zeta^{(1,1)}&=&-\frac{4\pi^{(d-1)/2}}{d n T}\sum_{i,j}\chi_{ij}m_{ij}\sigma_{ij}^{d-1}\int \mathrm{d}\mathbf{v}_1 \int \mathrm{d}\mathbf{v}_2\;
f_i^{(0)}(\mathbf{V}_1)\nonumber\\
& & \times \mathcal{D}_j(\mathbf{V}_2)\Bigg[\frac{\Delta_{ij}^2}{\Gamma\left(\frac{d+1}{2}\right)}
g_{12}+\frac{\sqrt{\pi}}
{d\Gamma\left(\frac{d}{2}\right)}\al_{ij}\Delta_{ij}g_{12}^2\nonumber\\
& & -
\frac{1-\al_{ij}^2}{4\Gamma\left(\frac{d+3}{2}\right)}g_{12}^3\Bigg].
\eeqa

According to Eq.\ \eqref{3.19.1}, the coefficient $\zeta^{(1,0)}$ is given in terms of the distribution $f_i^{(0)}$, whose exact form is not known to date. A good estimate of $\zeta^{(1,0)}$ can be obtained by replacing the true $f_i^{(0)}(\mathbf{V})$ by its Maxwellian form 
\beq
\label{3.20}
f_{i,\text{M}}(\mathbf{V})=n_i\left(\frac{m_i}{2\pi T_i^{(0)}}\right)^{d/2} \exp\left(-\frac{m_iV^2}{2T_i^{(0)}}\right).
\eeq
In the Maxwellian approximation, $\zeta^{(1,0)}$ is
\beqa
\label{3.21}
\zeta^{(1,0)}&=&\frac{4 \pi^{d/2}}{d^2\Gamma\left(\frac{d}{2}\right)}\sum_{i=1}^s\sum_{j=1}^s\chi_{ij}x_i x_j n \sigma_{ij}^{d}\Bigg\{\frac{m_{ij}}{\overline{m}}\Delta_{ij}^*\Bigg[\Delta_{ij}^*\nonumber\\
& & +\frac{2}{\sqrt{\pi}}\al_{ij}
\left(\frac{\theta_i+\theta_j}{\theta_i\theta_j}\right)^{1/2}\Bigg]-\frac{3}{4}\mu_{ji}\gamma_i(1-\al_{ij}^2)\Bigg\}.
\nonumber\\
\eeqa

To determine $\zeta^{(1,1)}$, one takes 
the leading Sonine approximation to $\mathcal{D}_{i}(\mathbf{V})$ given by
\beq
\label{3.22}
\mathcal{D}_i(\mathbf{V})\rightarrow f_{i\text{M}}(\mathbf{V})W_i(\mathbf{V})\frac{\varpi_i}{T_i^{(0)}},
\eeq
where
\beq
\label{3.23}
W_i(\mathbf{V})=\frac{m_iV^2}{2T_i^{(0)}}-\frac{d}{2}.
\eeq 
By using approximation \eqref{3.22} in Eq.\ \eqref{3.19.2}, one gets the result 
\beq
\label{3.24}
\zeta^{(1,1)}=\sum_{i=1}^s\;\xi_i \varpi_i,
\eeq
with
\beqa
\label{3.25}
\xi_i&=&\frac{\pi^{(d-1)/2}}{d\Gamma\left(\frac{d}{2}\right)}\frac{\upsilon_{\text{th}}}{n T}\sum_{j=1}^s
n_i n_j \sigma_{ij}^{d-1}\chi_{ij}\mu_{ji}\Bigg\{3\left(\frac{\theta_i+\theta_j}{\theta_i\theta_j}\right)^{1/2}\nonumber\\
& & \times
(1-\al_{ij}^2)
-4\Delta_{ij}^*\Bigg[\sqrt{\pi}\al_{ij}+\Delta_{ij}^*
\left(
\frac{\theta_i\theta_j}{\theta_i+\theta_j}
\right)^{1/2}\Bigg]\Bigg\}.
\nonumber\\
\eeqa
The first-order contribution $\zeta_U$ to the cooling rate is given by\cite{GMG26}
\beq
\label{3.25.1}
\zeta_U=\zeta^{(1,0)}+\zeta^{(1,1)}.
\eeq

It still remains to get the coefficients $\varpi_i$. To obtain them, we substitute first $\mathcal{D}_i$ by its Sonine approximation \eqref{3.22} into the integral equation \eqref{3.18}. Then, we multiply both sides of Eq.\ \eqref{3.18} by $m_i V^2$ and integrate over $\mathbf{v}$. The result is
\beq
\label{3.26}
\sum_{j=1}^s \Bigg[\omega_{ij}+T\xi_j\Big(\gamma_i-\frac{1}{2}\Delta^*\frac{\partial \gamma_i}{\partial \Delta^*}\Big)\Bigg]\varpi_j=\Lambda_i,
\eeq
where
\beqa
\label{3.27}
\omega_{ii}&=&\frac{1}{dn_iT_i^{(0)}}\Bigg(\sum_{j=1}^s\int \dd\mathbf{v}m_iV^2J_{ij}^{(0)}\left[f_{i,\text{M}}W_i,f_j^{(0)}\right]\nonumber\\
& &
+\int \dd\mathbf{v}m_iV^2J_{ii}^{(0)}\left[f_i^{(0)},f_{i,\text{M}}W_i\right]\Bigg),
\eeqa
\beq
\label{3.28}
\omega_{ij}=\frac{1}{dn_iT_j^{(0)}}\int \dd\mathbf{v}m_iV^2J_{ij}^{(0)}\left[f_i^{(0)},f_{j,\text{M}}W_j\right], \quad (i\neq j),
\eeq
\beqa
\label{3.29}
& & \Lambda_i=\frac{2}{d}T\gamma_i-T\Bigg(\gamma_i-\frac{1}{2}\Delta^*\frac{\partial \gamma_i}{\partial \Delta^*}\Bigg)
\Bigg(\zeta^{(1,0)}+\frac{2}{d}p^*\Bigg)\nonumber\\
& & -\sum_{j=1}^s T n_j\frac{\partial \gamma_i}{\partial n_j}-\frac{1}{d^2 n_i}\sum_{j=1}^s\int \dd\mathbf{v} \, m_i V^2 \mathcal{K}_{ij,\beta}\left[\frac{\partial f_j^{(0)}}{\partial V_\beta}\right].
\nonumber\\
\eeqa
In Eqs.\ \eqref{3.27} and \eqref{3.28},
the Boltzmann--Enskog collision operator $J_{ij}^{(0)}[X,Y]$ is defined in Eq.\ (36) of Ref.\ \onlinecite{GMG26}. In Eq.\ \eqref{3.29}, an approximate expression for the (reduced) hydrostatic pressure $p^*=p/(nT)$ is achieved by replacing $f_i^{(0)}$ by $f_{i,\text{M}}$. It is given by
\cite{GMG26}
\beqa
\label{3.30}
p^*&=&1+\frac{\pi^{d/2}}{d\Gamma\left(\frac{d}{2}\right)}\sum_{i,j}\mu_{ji}n\sigma_{ij}^d\chi_{ij}x_ix_j\Bigg[(1+\al_{ij})
\gamma_i
\nonumber\\
& &+\frac{2}{\sqrt{\pi}}\frac{m_i}{\overline{m}}\Delta_{ij}^*\left(\frac{\theta_i+\theta_j}{\theta_i\theta_j}\right)^{1/2}
\Bigg].
\eeqa
For mechanically equivalent particles ($m_i=m$, $\sigma_i=\sigma$, and $\al_{ij}=\al$), $\Lambda_i=0$ and so, Eq.\ \eqref{3.26} yields $T_i^{(1)}=0$ as expected.

Approximate forms for $\omega_{ii}$, $\omega_{ij}$, and $\Lambda_i$ are provided in Appendix \ref{appB}. Once these forms are known, the coefficients $\varpi_i$ can be easily obtained by solving Eq.\ \eqref{3.26}. The knowledge of coefficients $\varpi_i$ allows us to get the first-order contribution $\zeta_U$ from Eqs.\ \eqref{3.21}, \eqref{3.24}, and \eqref{3.25}.
In the case of a binary mixture ($s=2$), $\varpi_1$ can be written as
\beq
\label{3.31}
\varpi_1=\frac{\Lambda_1}{\omega_{11}-\frac{x_1}{x_2}\omega_{12}+T
\left(\xi_1-\frac{x_1}{x_2}\xi_2\right)\Big(\gamma_1-\frac{1}{2}\Delta^*\frac{\partial \gamma_1}{\partial \Delta^*}\Big)},
\eeq
where the relation $\varpi_2=-(x_1/x_2) \varpi_1$ has been accounted for. The expression for $\varpi_2$ can be easily obtained from Eq.\ \eqref{3.31} by making the changes $1\leftrightarrow 2$. Equation \eqref{3.31} along with its counterpart for $\varpi_2$ are indeed consistent with the requirement $x_1 \varpi_1+x_2 \varpi_2=0$. For later use, it is convenient to introduce the reduced coefficients
\begin{equation}
\varpi_i^*=\frac{\nu}{T}\varpi_i,
\label{varpi_reduced}
\end{equation}
where $\nu=n\overline{\sigma}^{d-1}v_{\rm th}$ is an effective collision frequency.

\subsection{Some illustrative mixtures}

The results derived in the preceding subsections apply to a mixture with an arbitrary number of species. To illustrate the dependence of the heat-flux transport coefficients and the first-order contributions to the partial temperatures and the cooling rate on the parameters of the mixture, we consider here binary mixtures of hard disks ($d=2$). For simplicity, we assume an equimolar composition ($x_1=x_2=1/2$), a common
coefficient of restitution 
($\alpha_{11}=\alpha_{22}=\alpha_{12}\equiv\alpha$), and a common scaled injection parameter ($\Delta_{11}^*=\Delta_{22}^*=\Delta_{12}^*\equiv\Delta^*$).
Thus, the effective mechanism transferring the energy injected by
the vertical vibration to the horizontal degrees of freedom is taken to be the same for all collision pairs. For each value of $\alpha$ and the parameters of the mixture, the temperature ratios $\gamma_i$ and the parameter $\Delta^*$ are evaluated in the HSS. We consider three representative solid volume fractions, namely, the dilute limit $\phi\to0$ and the moderately dense
values $\phi=0.1$ and $0.2$. Moreover, to assess the impact of inelasticity on heat transport, we reduce the transport coefficients with respect to their elastic values. 

\begin{figure}[t]
\centering
\includegraphics[width=0.40\textwidth]{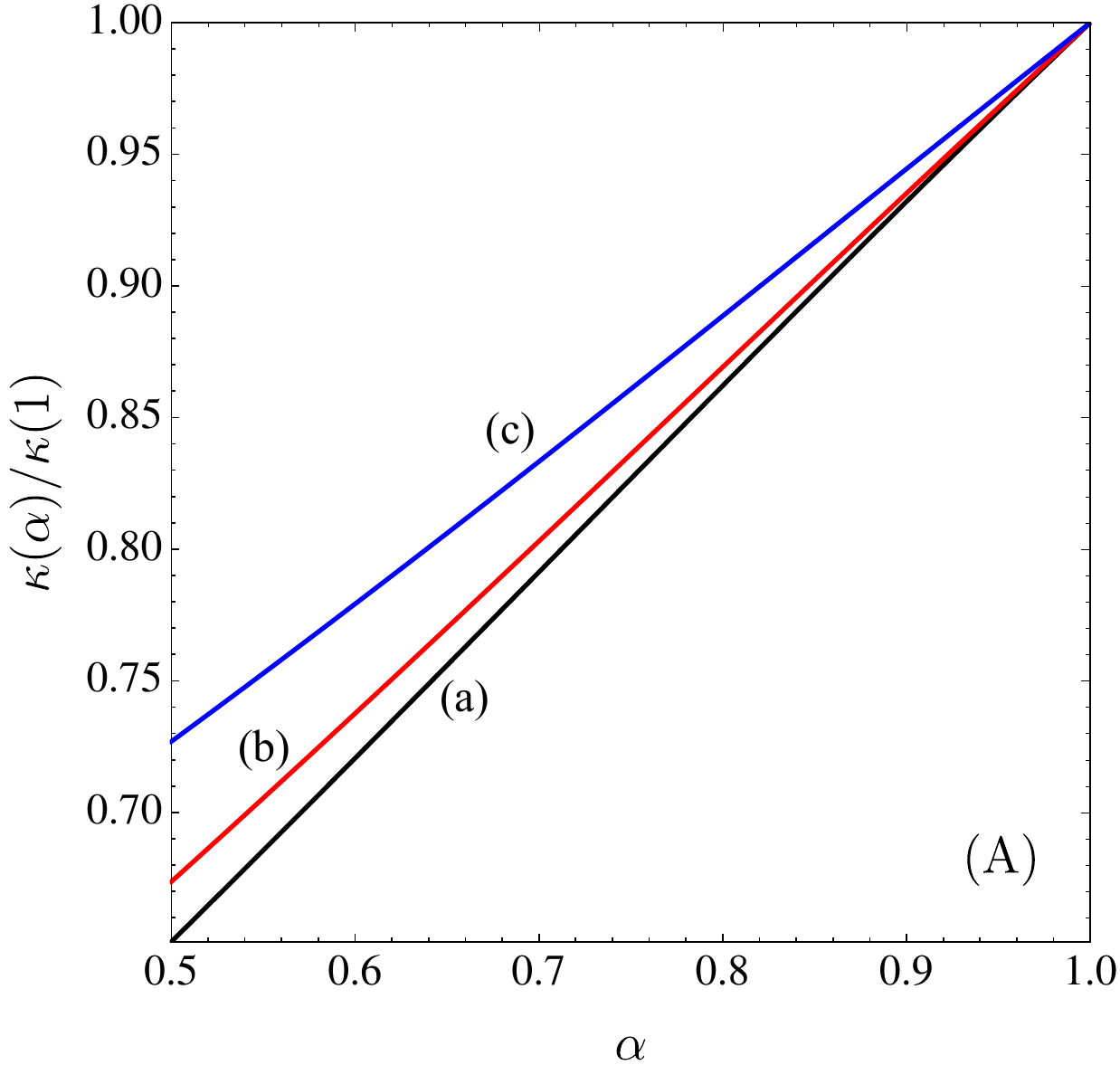}
\includegraphics[width=0.40\textwidth]{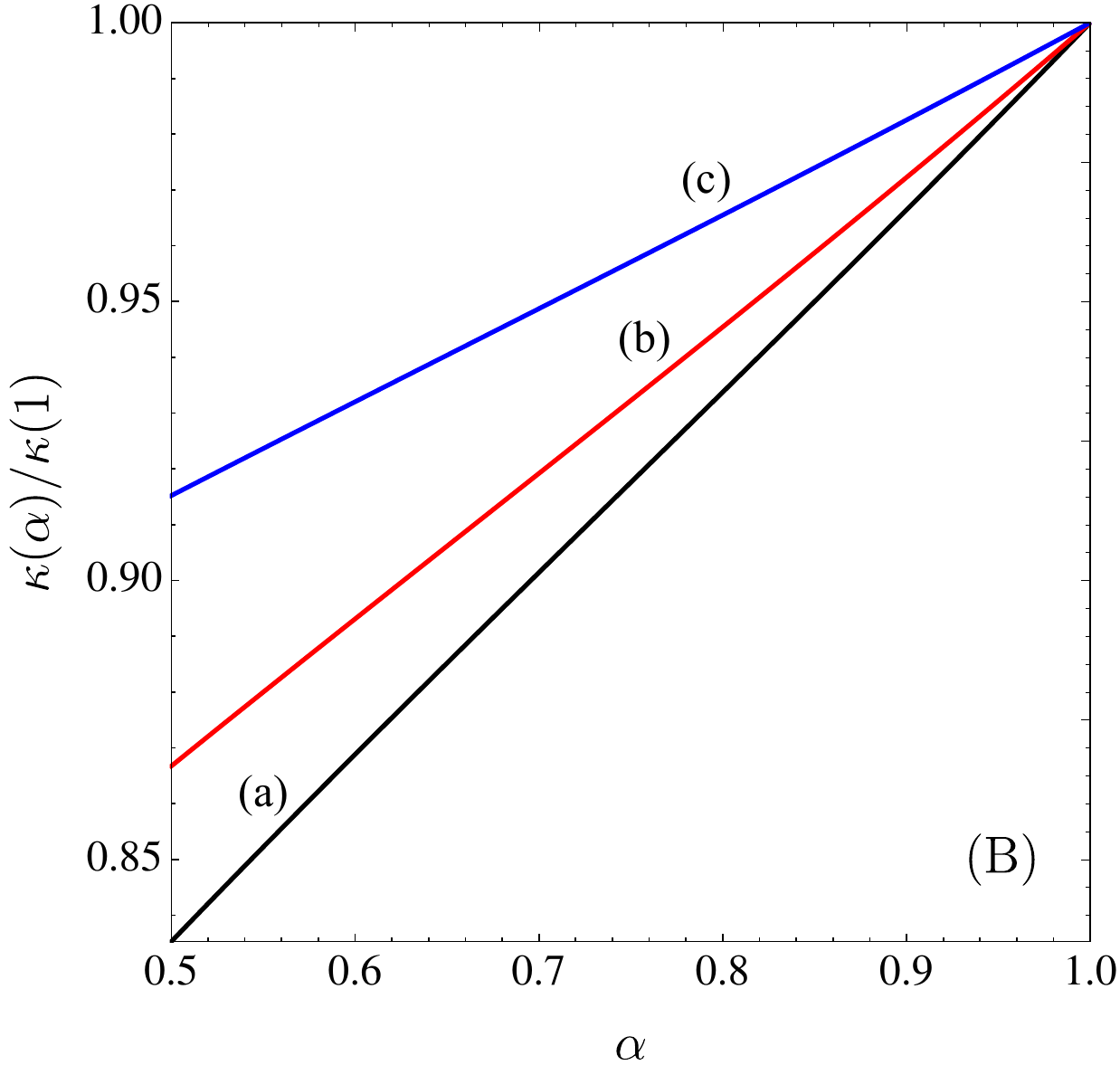}
\caption{Plot of the scaled thermal conductivity coefficient, $\kappa(\alpha)/\kappa(1)$, as a function of the common coefficient of restitution, $\alpha\equiv\alpha_{ij}$, for a binary equimolar mixture with $x_1=0.5$ and $\sigma_1/\sigma_2=1$. (A): $m_1/m_2=4$. (B): $m_1/m_2=0.5$. In both panels, the curves correspond to three different solid volume fractions: $\phi\to0$ (a), $\phi=0.1$ (b), and $\phi=0.2$ (c).
\label{fig_kappa}}
\end{figure}

Figure~\ref{fig_kappa} shows the scaled thermal conductivity
$\kappa(\alpha)/\kappa(1)$ for two equimolar mixtures with equal
diameters, $\sigma_1/\sigma_2=1$, and mass ratios $m_1/m_2=4$
[panel (A)] and $m_1/m_2=0.5$ [panel (B)]. Here, $\kappa(1)$ refers to the value of $\kappa$ for elastic collisions. In both mixtures, we observe that inelasticity reduces the thermal conductivity
relative to its elastic value, and the departure from unity becomes
larger as $\alpha$ decreases. On the other hand, increasing the
density shifts the scaled conductivity upward. Thus, although
collisional transfer contributions become increasingly important at
finite density, their net effect in these mixtures is to weaken the
relative sensitivity of $\kappa$ to inelasticity. This influence is
particularly more pronounced for the mixture with $m_1/m_2=0.5$.

\begin{figure}[b]
\centering
\includegraphics[width=0.40\textwidth]{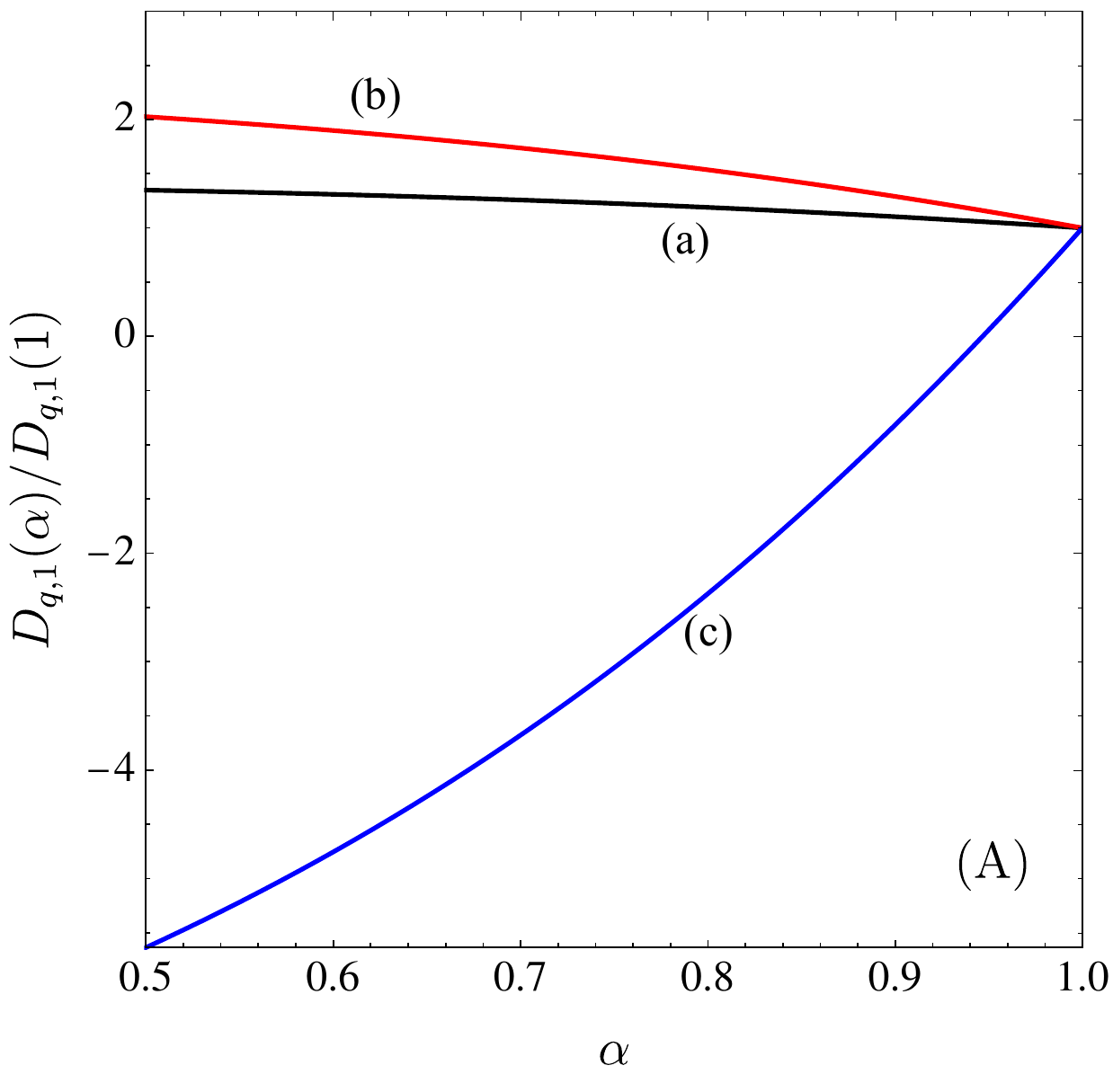}
\includegraphics[width=0.40\textwidth]{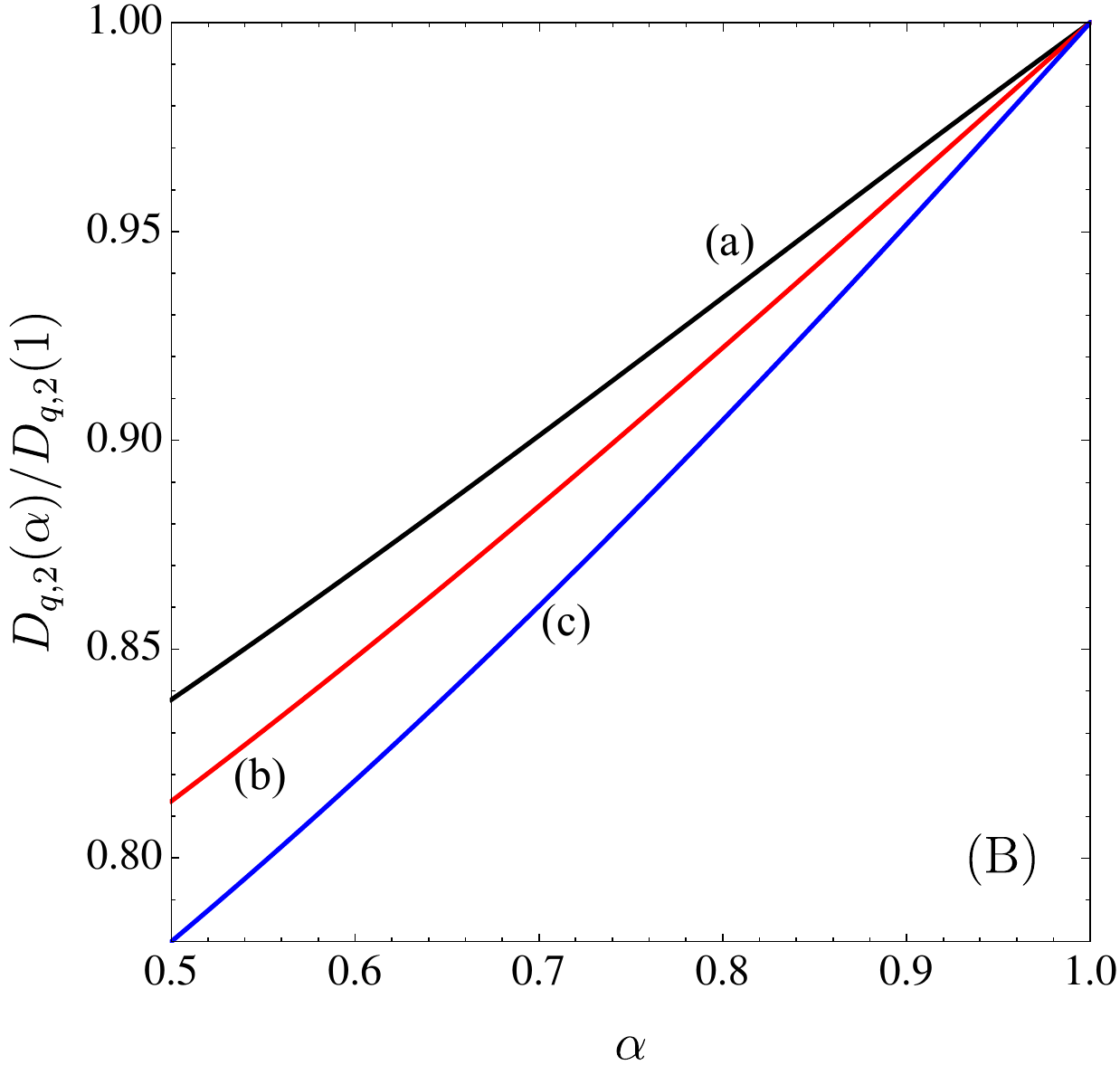}
\caption{Plot of the scaled Dufour coefficients as functions of the common coefficient of restitution, $\alpha\equiv\alpha_{ij}$, for a binary equimolar mixture with $x_1=0.5$, $m_1/m_2=4$, and $\sigma_1/\sigma_2=2$. (A): $D_{q,1}(\alpha)/D_{q,1}(1)$. (B): $D_{q,2}(\alpha)/D_{q,2}(1)$. In both panels, the curves correspond to three different solid volume fractions: $\phi\to0$ (a), $\phi=0.1$ (b), and $\phi=0.2$ (c).
\label{fig1_dufour}}
\end{figure}

The behavior of the Dufour coefficients exhibits a considerably
stronger dependence on the mechanical parameters of the particles. Figure~\ref{fig1_dufour} shows $D_{q,1}(\alpha)/D_{q,1}(1)$ and $D_{q,2}(\alpha)/D_{q,2}(1)$ for the mixture $m_1/m_2=4$ and $\sigma_1/\sigma_2=2$. As seen in panel (A), the density dependence of $D_{q,1}$ is highly nontrivial. In the dilute limit and at $\phi=0.1$, its magnitude is enhanced by inelasticity, since the corresponding scaled values are larger than unity.
However, at $\phi=0.2$, the coefficient changes sign for sufficiently inelastic collisions. The negative values of
$D_{q,1}(\alpha)/D_{q,1}(1)$ imply that $D_{q,1}(\alpha)$ has a sign
opposite to that of its elastic counterpart. This sign reversal shows
that collisional transfer can modify not only the magnitude of a Dufour coefficient but also qualitatively reverse the associated
contribution to the heat flux. In contrast, panel (B) shows that
$D_{q,2}(\alpha)/D_{q,2}(1)$ remains positive and below unity over
the whole range of $\alpha$ considered. For this coefficient,
increasing density enhances the reduction produced by inelasticity.

\begin{figure}[t]
\centering
\includegraphics[width=0.40\textwidth]{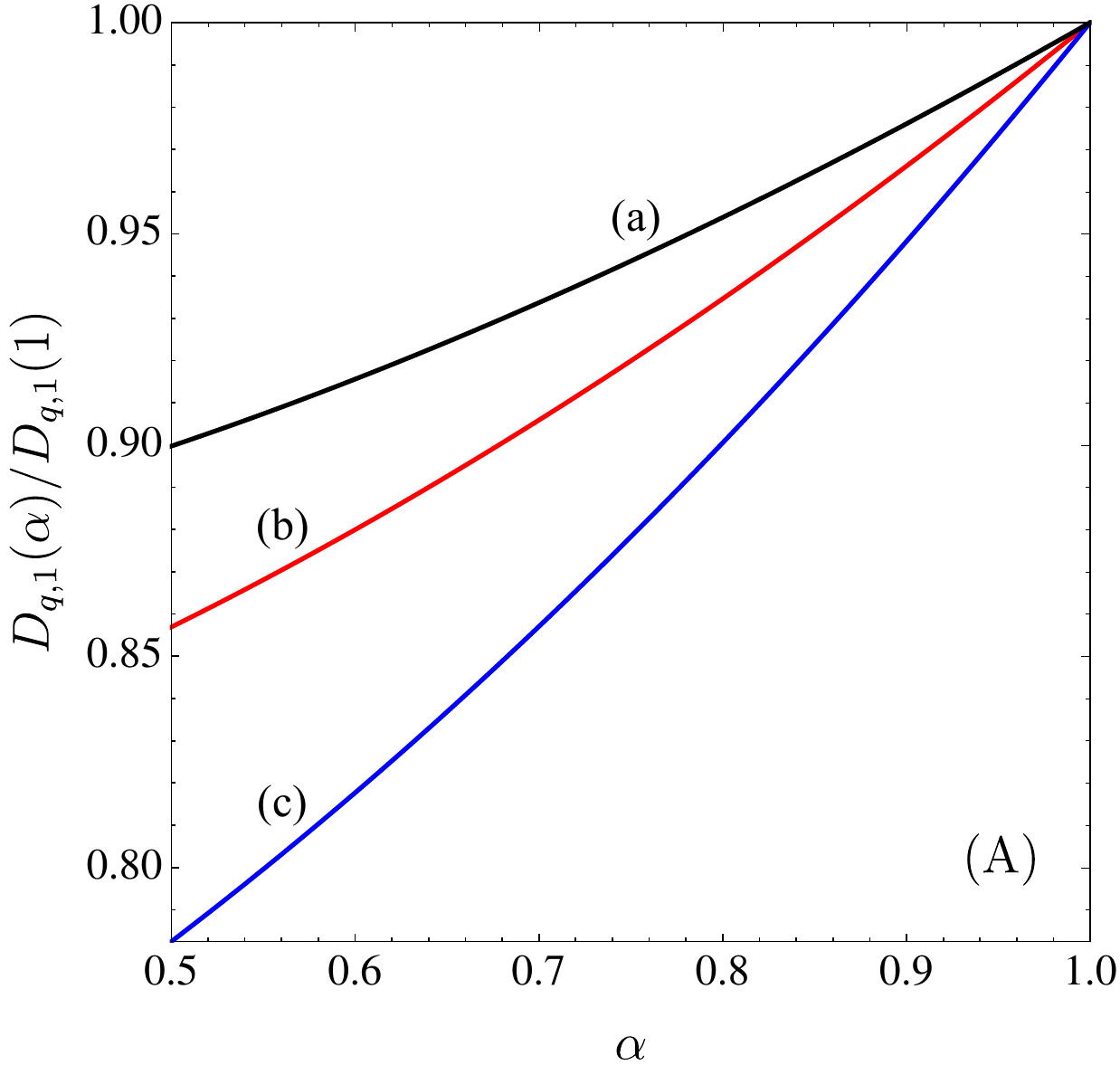}
\includegraphics[width=0.40\textwidth]{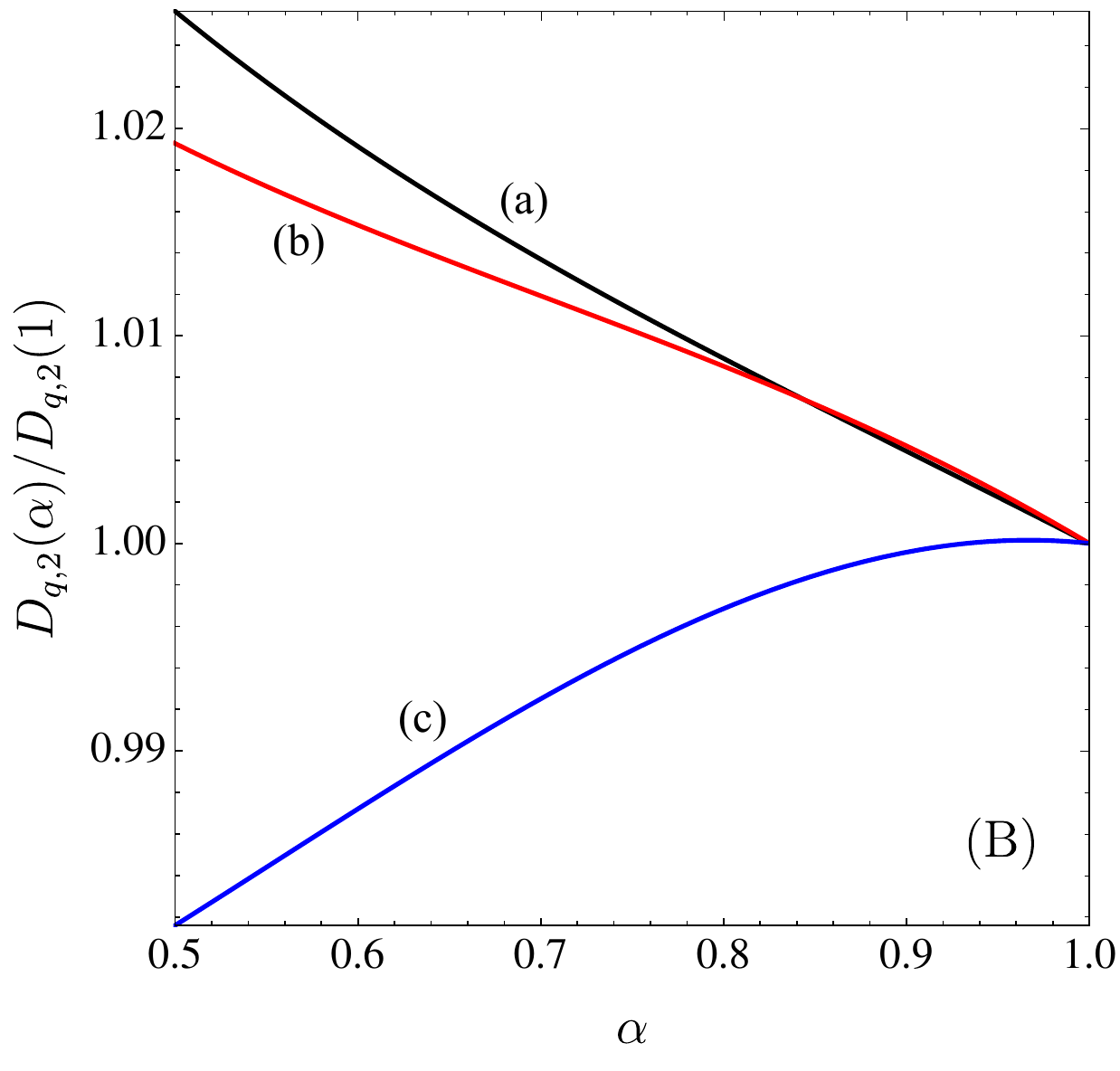}
\caption{Plot of the scaled Dufour coefficients as functions of the common coefficient of restitution, $\alpha\equiv\alpha_{ij}$, for a binary equimolar mixture with $x_1=0.5$, $m_1/m_2=0.5$, and $\sigma_1/\sigma_2=0.4$. (A): $D_{q,1}(\alpha)/D_{q,1}(1)$. (B): $D_{q,2}(\alpha)/D_{q,2}(1)$. In both panels, the curves correspond to three different solid volume fractions: $\phi\to0$ (a), $\phi=0.1$ (b), and $\phi=0.2$ (c).
\label{fig2_dufour}}
\end{figure}

A different behavior is found for the mixture defined by $m_1/m_2=0.5$ and $\sigma_1/\sigma_2=0.4$, whose results are displayed in Fig.~\ref{fig2_dufour}. In this case, $D_{q,1}(\alpha)/D_{q,1}(1)$ remains positive and smaller than unity, and its departure from the elastic value increases with density.
The second Dufour coefficient is much less sensitive to inelasticity.
For $\phi\to0$ and $0.1$, $D_{q,2}(\alpha)$ is slightly larger than its
elastic value, whereas for $\phi=0.2$ it becomes slightly smaller.
Thus, finite-density effects reverse the sign of the deviation of the ratio 
$D_{q,2}(\alpha)/D_{q,2}(1)$ from 1, although the coefficient itself does
not change sign. Comparison between Figs.~\ref{fig1_dufour} and
\ref{fig2_dufour} therefore illustrates that no universal density
correction can be assigned to the Dufour coefficients: both their
magnitude and sign depend sensitively on the mass and diameter
ratios and on which species-density gradient is considered. Thus, due to the complex nonlinear dependence of the Dufour coefficients on the parameters of the mixture, it is quite difficult to provide a simple physical picture of the trends observed for those transport coefficients.

\begin{figure}[t]
\begin{center}
\begin{tabular}{lr}
\resizebox{7.7cm}{!}{\includegraphics{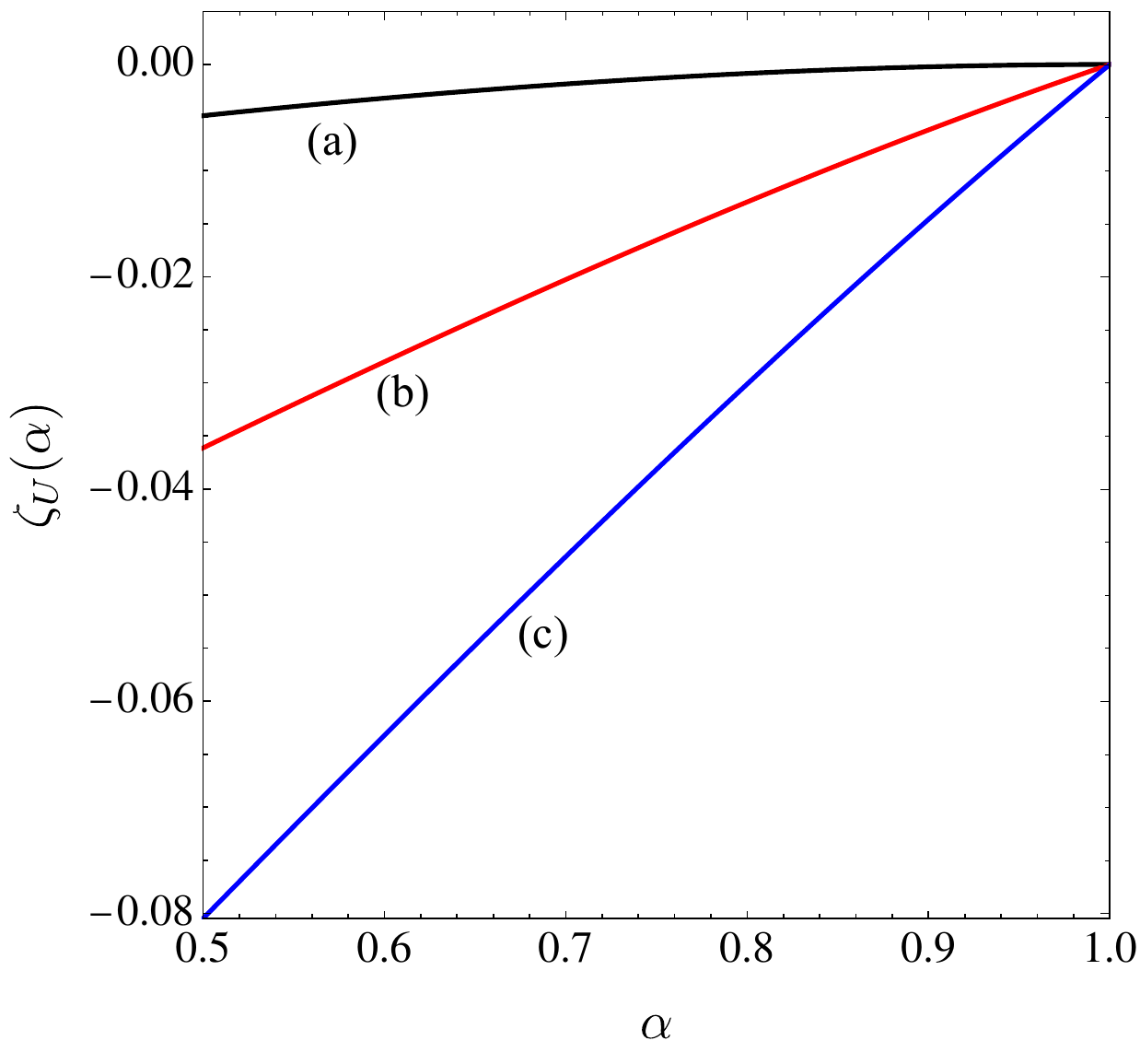}}
\end{tabular}
\end{center}
\caption{First-order contribution $\zeta_U$ of the cooling rate as a
function of the common coefficient of restitution $\alpha\equiv\alpha_{ij}$ for a binary equimolar mixture
($x_1=0.5$) of hard disks with $m_1/m_2=0.5$ and
$\sigma_1/\sigma_2=1$. The curves correspond to three different solid
volume fractions: the dilute limit $\phi
\to0$ (a), $\phi=0.1$ (b),
and $\phi=0.2$ (c).
\label{fig_zetaU}}
\end{figure}

Figure ~\ref{fig_zetaU} displays the first-order contribution
$\zeta_U$ to the cooling rate for an equimolar mixture with $m_1/m_2=0.5$ and $\sigma_1/\sigma_2=1$. The coefficient vanishes in the elastic limit and is negative for all inelastic systems considered. At a fixed value of $\alpha<1$, the magnitude of $\zeta_U$ increases markedly with density. For instance, the relatively weak dependence found in the dilute limit becomes a sizable contribution at $\phi=0.2$.
According to Eq.\ \eqref{2.18}, the negative sign of $\zeta_U$
implies that a local compression, $\nabla\cdot\mathbf{U}<0$, increases
the cooling rate, whereas a local expansion reduces it. The strong
density dependence observed in Fig.~\ref{fig_zetaU} also anticipates
that this contribution may have a non-negligible influence on the
longitudinal hydrodynamic modes analyzed in the next section.

\begin{figure}[h]
\centering
\includegraphics[width=0.42\textwidth]{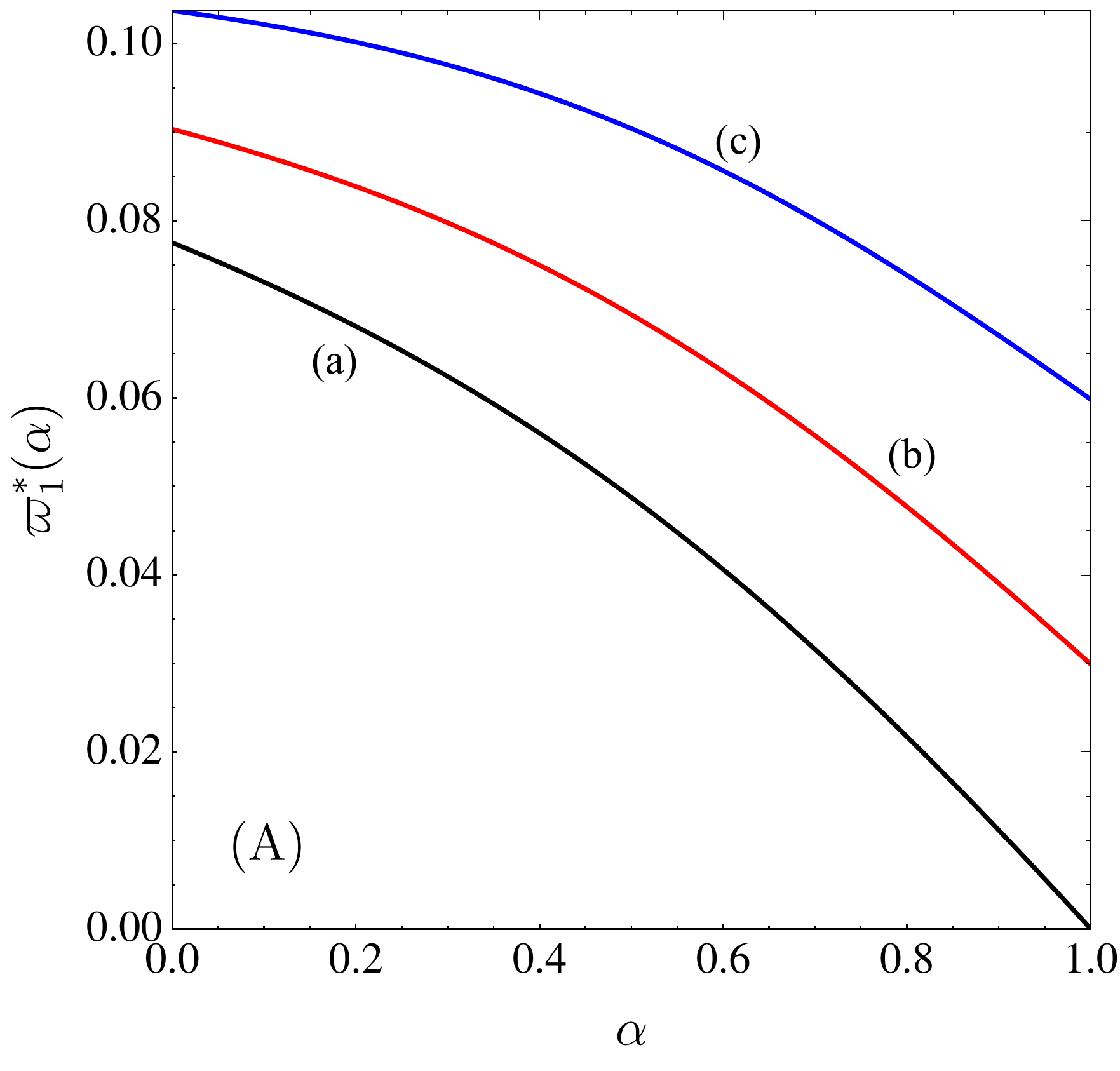}
\includegraphics[width=0.42\textwidth]{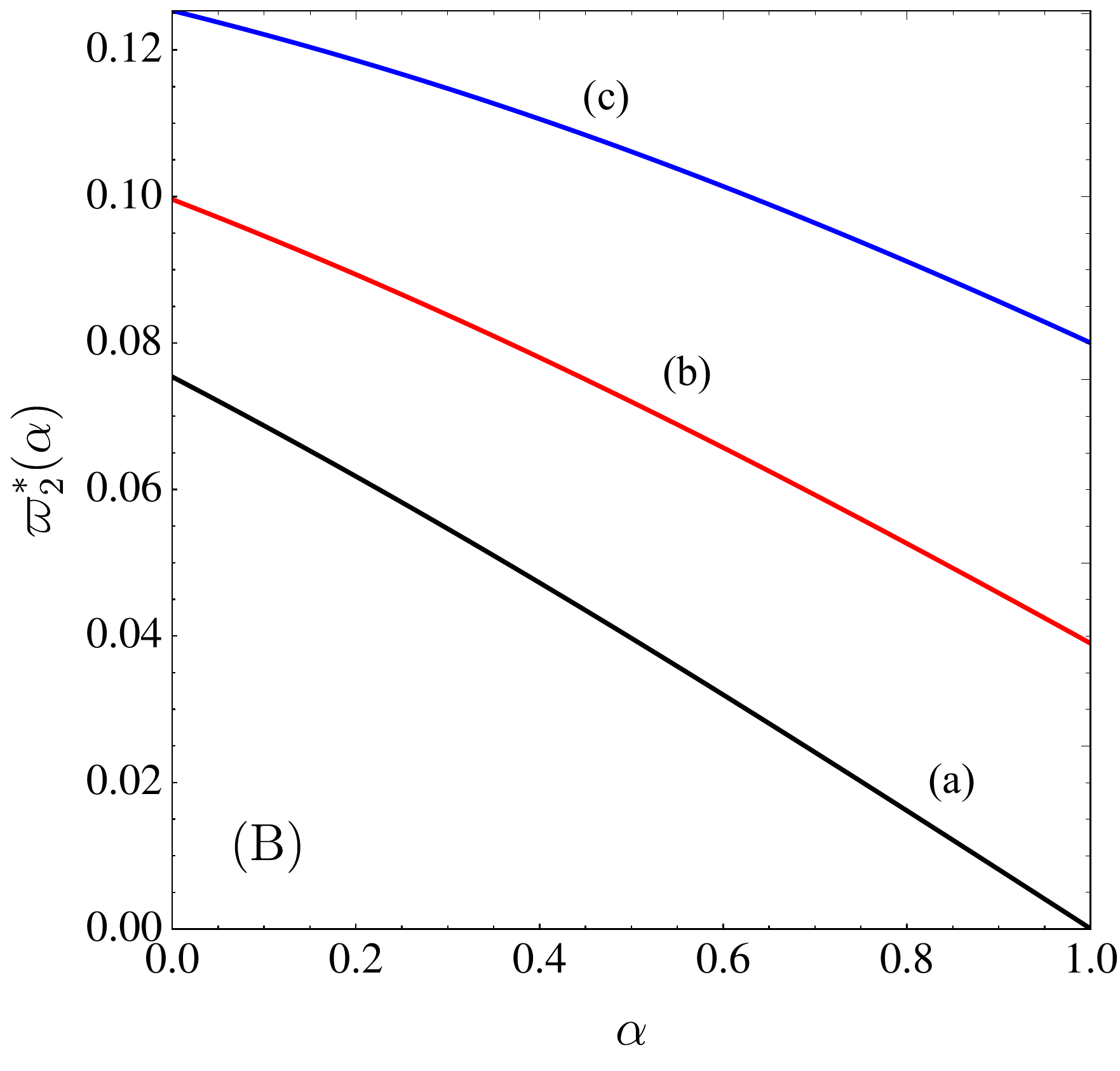}
\caption{Reduced first-order contributions $\varpi_i^*$ to the partial temperatures as functions of the common coefficient of restitution $\alpha\equiv\alpha_{ij}$ for binary equimolar mixtures ($x_1=0.5$) of hard disks ($d=2$). Panel (A) shows
$\varpi_1^*$ for $m_1/m_2=4$ and $\sigma_1/\sigma_2=1$, whereas panel (B) shows $\varpi_2^*$ for $m_1/m_2=0.5$ and $\sigma_1/\sigma_2=2$. In both panels,
the labels (a), (b), and (c) refer to the dilute limit $\phi\to0$, $\phi=0.1$,
and $\phi=0.2$, respectively.} \label{fig:varpi}
\end{figure}

Finally, Fig.~\ref{fig:varpi} illustrates the behavior of the (reduced)
first-order contributions $\varpi_i^*$ to the partial temperatures for
two representative binary mixtures. Panel (A) shows $\varpi_1^*$ for
$m_1/m_2=4$ and $\sigma_1/\sigma_2=1$, whereas panel (B) displays
$\varpi_2^*$ for $m_1/m_2=0.5$ and $\sigma_1/\sigma_2=2$. A clear density effect is observed: at a given value of $\alpha$, increasing $\phi$ systematically
enhances the magnitude of $\varpi_i^*$. As for molecular mixtures of hard spheres,\cite{KS79b} while the dilute
contribution tends to zero in the elastic limit, finite-density effects
lead to nonzero values of $\varpi_i^*$ at $\alpha=1$.  Thus, collisional
transfer has a significant influence not only on the heat-flux
coefficients and the cooling rate, but also on the first-order corrections
to the partial temperatures.

\section{Linear stability analysis of the HSS}
\label{sec4}

It is quite apparent that the knowledge of the complete set of Navier--Stokes transport coefficients of the confined granular mixtures allows us to perform a linear stability analysis of the HSS. Within the context of the $\Delta$-model, previous theoretical studies for monocomponent dense granular gases \cite{BBGM16,GBS21a} and dilute granular mixtures\cite{GBS24,GBS24a} have shown that the HSS is linearly stable. A natural question arises then as to whether, and if so to what extent, the conclusion achieved in the low-density regime for a confined granular mixture\cite{GBS24} can be changed at moderate densities. This is the main goal of the present Section.

\subsection{Homogeneous steady state (HSS)}

It is well known that a particular solution to the balance equations \eqref{2.9}--\eqref{2.11} corresponds to the HSS solution. In this situation, $n_{i\text{H}}\equiv \text{constant}$, $T_{\text{H}}\equiv \text{constant}$, and $\mathbf{U}_\text{H}=\mathbf{0}$ (without loss of generality). Here, the subscript $\text{H}$ means that the hydrodynamic fields are evaluated in a homogeneous steady state. Moreover, in the absence of spatial gradients, the mass and heat fluxes vanish and the pressure tensor $P_{k\ell,\text{H}}=n_\text{H} T_\text{H} p_\text{H}^*\delta_{k \ell}$, where the (reduced) hydrostatic pressure $p_\text{H}^*$ is given by Eq.\ \eqref{3.30}.

In the steady state, $\partial_t T_{\text{H}}=0$ and hence, according to Eq.\ \eqref{2.11} $\zeta_{\text{H}}=0$. Additionally, in the homogeneous state,\cite{BSG20} the partial temperatures $T_i$ obey the equations $\partial_t \ln T_{i\text{H}}=-\zeta_{i\text{H}}$, where $\zeta_{i\text{H}}$ refer to the partial cooling rates (i.e., those associated with the time evolution of the partial temperatures). In the steady state, $\zeta_{i\text{H}}=0$ ($i=1,\cdots,s$) and so the temperature ratios $\gamma_i$ are determined from the constraints:
\beq
\label{4.1}
\zeta_{\text{H}}=\zeta_{1\text{H}}=\cdots =\zeta_{s\text{H}}=0.
\eeq
As in the case of the hydrostatic pressure, the determination of the partial cooling rates $\zeta_{i\text{H}}$ requires the knowledge of the HSS distribution function. A good estimate of $\zeta_{i\text{H}}$ can be obtained by considering the Maxwellian approximation \eqref{3.20} to the HSS distribution. In this approximation,
the cooling rates $\zeta_{i\text{H}}$ are given by \cite{BSG20}
\begin{widetext}
\beqa
\label{4.2}
\zeta_{i\text{H}}&=&\frac{4\pi^{(d-1)/2}}{d\Gamma\left(\frac{d}{2}\right)}\nu_\text{H}\sum_{j=1}^s
x_{j\text{H}}\chi_{ij}
\left(\frac{\sigma_{ij}}{\overline{\sigma}}\right)^{d-1}\mu_{ji}(1+\al_{ij})\theta_i^{-1/2}
\left(1+\theta_{ij}\right)^{1/2}
\left[1-\frac{1}{2}\mu_{ji}(1+\alpha_{ij})(1+\theta_{ij}) \right]\nonumber\\
& &-\frac{4\pi^{d/2}}{d\Gamma\left(\frac{d}{2}\right)}\nu_\text{H}\sum_{j=1}^s x_{j\text{H}}\chi_{ij}
\left(\frac{\sigma_{ij}}{\overline{\sigma}}\right)^{d-1}\mu_{ji}\Delta_{ij}^*\left[
\frac{2\mu_{ji}\Delta_{ij}^*}{\sqrt{\pi}}\theta_i^{1/2}\left(1+\theta_{ij}\right)^{1/2}
-1+\mu_{ji}(1+\al_{ij})\left(1+\theta_{ij}\right)\right],
\eeqa
\end{widetext}
where $\nu_\text{H}=n_\text{H}\overline{\sigma}_{12}^{d-1}\upsilon_{\text{th},\text{H}}$ and $v_{\text{th,H}}=
\sqrt{2T_\text{H}/\overline{m}}$. Moreover, 
$\theta_{ij}=m_iT_{j\text{H}}/m_jT_{i\text{H}}$ gives the ratio between the mean-square
velocity of the particles of the species $j$ relative to that of the particles of the species $i$. Note that $\zeta_{i\text{H}}=\zeta_i^{(0)}$ and $T_{i\text{H}}=T_i^{(0)}$.

\subsection{Linearized hydrodynamic equations}

We now investigate whether the HSS is linearly stable or not with respect to sufficiently long-wavelength perturbations. To give an answer to this question we carry out a linear stability analysis of the Navier--Stokes hydrodynamic equations. On the other hand, it is well known that the expressions \eqref{2.15} and \eqref{2.17} for mass and heat fluxes, respectively, can be defined in a variety of equivalent ways depending on the choice of the spatial gradients of the hydrodynamic fields. Here, to perform an analysis parallel to the one carried out before in the conventional IHS model for dense granular mixtures,\cite{G15} the hydrodynamic fields $x_i=n_i/n$ and $n=\sum_{i=1}^s n_i$ are chosen instead of the partial densities $n_i$. Additionally, for the sake of simplicity, henceforth a binary system ($s=2$) is considered. In this particular case, the balance equations \eqref{2.9} for the concentration $x_1$ and the total number density $n$ are  
\begin{equation}
\label{4.3}
\mathrm{D}_t x_1+\frac{\rho}{n^2m_1m_2}\nabla \cdot \mathbf{j}_1=0,
\end{equation}
\begin{equation}
\label{4.4}
\mathrm{D}_t n+n\nabla \cdot \mathbf{U}+\frac{m_2-m_1}{m_1m_2}\nabla \cdot \mathbf{j}_1=0.
\end{equation}
Moreover, in terms of $\nabla x_1$ and $\nabla n$, the mass and heat fluxes can be rewritten as
\begin{equation}
\mathbf{j}_{1}=-\frac{m_1 m_{2}n}{\rho} D\nabla x_{1}-
\frac{m_1 m_2}{\rho} D_{n}\nabla n
-\frac{\rho}{T} D_1^{T}\nabla  T, \quad \mathbf{j}_2=-\mathbf{j}_1,
\label{4.5}
\end{equation}
\begin{equation}
\label{4.6}
\mathbf{q}=-T^2D''\nabla  x_1-\frac{T^2}{n} D_{qn}\nabla  n-\kappa
\nabla T,
\end{equation}
where the transport coefficients $D$, $D_n$, $D^{\prime \prime}$, and $D_{qn}$ are given in terms of $D_{ij}$ and $D_{q,i}$ as
\begin{equation}
\label{4.7}
D=\mu D_{11}-D_{12}, \quad D_n=x_1 \mu D_{11}+x_2 D_{12},
\end{equation}
\begin{equation}
\label{4.8}
D''=\frac{D_{q,1}}{x_1}-\frac{D_{q,2}}{x_2}, \quad D_{qn}=D_{q,1}+ D_{q,2}.
\end{equation}
Here, $x_2=1-x_1$ and $\mu \equiv m_1/m_2$ is the mass ratio. We recall that the Dufour coefficients $D_{q,i}$ are defined by Eq.\ \eqref{2.18.1}.

When the expressions \eqref{2.16}, \eqref{2.18}, \eqref{4.5}, and \eqref{4.6} for the pressure tensor, the cooling rate, the mass flux and the heat flux, respectively, are substituted into the \emph{exact} balance equations \eqref{2.10}, \eqref{2.11}, \eqref{4.3}, and \eqref{4.4} one gets the corresponding Navier-Stokes hydrodynamic equations for the hydrodynamic fields $x_1$, $n$, $\mathbf{U}$, and $T$. They are given by
\beqa
\mathrm{D}_tx_{1}&=&\frac{\rho}{n^{2}m_{1}m_{2}}\nabla \cdot \left( \frac{
m_{1}m_{2}n}{\rho}D\nabla x_{1}+\frac{
m_{1}m_{2}}{\rho}D_{n}\nabla n\right. \nn
& &\left. +\frac{\rho }{T}D_1^T\nabla T\right),  \label{4.9}
\eeqa
\beqa
\mathrm{D}_tn+n\nabla \cdot \mathbf{U}&=&\frac{m_2-m_1}{m_{1}m_{2}}\nabla \cdot
\left( \frac{m_{1}m_{2}n}{\rho}D\nabla x_{1}\right.\nn
& & \left.+\frac{m_{1}m_{2}}{\rho}D_{n}\nabla n+\frac{\rho }{T}D_1^T\nabla T\right),  \nn
\label{4.10}
\eeqa
\beqa
\rho \mathrm{D}_tU_{\lambda}+\nabla_{\lambda}p &=&\nabla_{\beta}\Big[\eta \left(
\nabla_{\lambda}U_{\beta}+\nabla_{\beta}U_{\lambda}-\frac{2}{d}\delta_{\lambda\beta}\nabla
\cdot \mathbf{U}\right)\nn
& &
+\eta_b \delta_{\lambda\beta}\nabla \cdot \mathbf{U}\Big],  \label{4.11}
\eeqa
\begin{eqnarray}
& &n\left(\mathrm{D}_t+\zeta^{(0)}\right) T+\frac{2}{d}p\nabla \cdot \mathbf{U} =-\frac{T(m_2-m_1)}{m_1m_2} \nn
& & \times \nabla \cdot \left( \frac{
m_{1}m_{2}n}{\rho}D\nabla x_{1}+\frac{m_{1}m_{2}}{\rho}D_{n}\nabla n+
\frac{\rho }{T}D_1^T\nabla T\right)  \nonumber \\
&&+\frac{2}{d}\nabla \cdot \left( T^{2}D^{\prime \prime}\nabla
x_{1}+\frac{T^2}{n}D_{qn}\nabla n+\kappa \nabla T\right)  \nonumber \\
&&+\frac{2}{d} \left[ \eta\left(\nabla_{\ell }U_{k}+\nabla_{k}U_{\ell}-\frac{2
}{d}\delta_{k\ell}\nabla \cdot \mathbf{U}\right)+\delta_{k\ell}\eta_b\nabla \cdot \mathbf{U}\right] \nn
& & \nabla_{\ell}U_{k}-nT\zeta_U\nabla \cdot \mathbf{U}.
\label{4.12}
\end{eqnarray}
In Eq.\ \eqref{4.11}, $\nabla_\lambda\equiv \partial/\partial r_\lambda$.

As discussed in previous works,\cite{G05} consistency would require to consider up to second order in the spatial gradients in the expression \eqref{2.18} for the cooling rate, since this is the order of the terms appearing in the balance equation \eqref{4.12} for the granular temperature coming from the mass flux, the pressure tensor and the heat flux. Thus, since the cooling rate $\zeta$ is a scalar, its most general form at this order for a granular binary dense mixture is
\beqa
\label{4.13}
\zeta&=&\zeta^{(0)}+\zeta _{U}\nabla \cdot \mathbf{U}+\zeta_{n_1}\nabla^2 n_1+
\zeta_{n_2}\nabla^2 n_2+\zeta_T \nabla^2 T\nonumber\\
 & & +\zeta_{TT}(\nabla T)^2+\zeta_{n_1 n_1}(\nabla n_1)^2+\zeta_{n_2 n_2}(\nabla n_2)^2 \nonumber\\
& & +\zeta_{T n_1}(\nabla T)\cdot (\nabla n_1)+\zeta_{T n_2}(\nabla T)\cdot (\nabla n_2)\nonumber\\
& & +\zeta_{n_1 n_2}(\nabla n_1)\cdot (\nabla n_2)
+\zeta_{1,UU} (\nabla_i U_j)(\nabla_i U_j)\nonumber\\
& & +\zeta_{2,UU} (\nabla_i U_j)(\nabla_j U_i).
\eeqa
The first (linear) second-order terms ($\zeta_{n_i}$ and $\zeta_T$) have been determined for a one-component \emph{dilute} gas of inelastic hard spheres\cite{BDKS98} while the \emph{complete} set of coefficients (linear and nonlinear terms) have been computed for
granular monodisperse gases of viscoelastic particles.\cite{BP03} The evaluation of the above set of coefficients for granular mixtures is a quite intricate problem. To the best of our knowledge, no explicit results for these coefficients have been reported for granular binary mixtures, even in the simplest case of a low-density mixture ($\phi=0$). On the other hand, it has
been shown in the context of the conventional IHS model for dilute gases\cite{BDKS98} that the contributions of the second-order terms to the cooling rate $\zeta$ are negligible, as compared with the corresponding zeroth-order contribution $\zeta^{(0)}$ (the first-order contribution $\zeta_U$ vanishes for IHS in the low-density regime). It is assumed here that the same holds in the dense case
and so, for practical applications these second-order contributions can be in principle neglected in the Navier-Stokes hydrodynamic equations. In fact, the good agreement found between the theoretical results for mixtures of inelastic hard spheres (where the nonlinear contributions to $\zeta$ are not accounted for) and computer simulations\cite{BR13,MGH14} for the onset of velocity vortices for strong inelasticity, finite density, and particle dissimilarity supports the above expectation.

We assume that the deviations
\beq
\label{4.13.1}
\delta y_{\beta}({\bf r},t)=y_{\beta}({\bf r},t)-y_{\text{H} \beta}
\eeq
are small, where $\delta y_{\beta}({\bf r},t)$ denotes the deviation of $\{x_1, n, T, \mathbf{U}\}$ from their values in the HSS. If the initial spatial perturbation is sufficiently small, then for some initial time interval these deviations will remain small and the
hydrodynamic equations \eqref{4.9}--\eqref{4.12} can be linearized with respect to $\delta y_{\beta}(\mathbf{r},t)$. As in the monocomponent case, for the sake of convenience,\cite{GBS21a} we introduce the time and space dimensionless variables:
\beq
\label{4.14}
\tau=\frac{1}{2}\nu_\text{H}t, \quad \mathbf{r}'=\frac{\mathbf{r}}{2\ell_\text{H}}.
\eeq
The dimensionless time scale $\tau$ is a measure of the average number of collisions per particle
in the time interval between $0$ and $t$. The unit length $\ell_\text{H}=\upsilon_{\text{th,H}}/\nu_{\text{H}}=1/(n_\text{H}\sigma_{12}^{d-1})$ is proportional to the time-independent mean free path of gas particles.

The linearized Navier--Stokes hydrodynamic equations for the perturbations $(\delta x_1, \delta n, \delta T, \delta \mathbf{U})$
can be obtained by substituting the relation \eqref{4.13.1} into Eqs.\ \eqref{4.9}--\eqref{4.12} and neglecting second-
and higher-order terms in these perturbations. After some algebra, one achieves the result
\beqa
\label{4.15}
& & \frac{\partial}{\partial \tau}\frac{\delta x_1}{x_1}=\frac{1}{8}\frac{x_2+\mu x_1}{\mu_{12}}\Bigg(D^*
\nabla^{'2}\frac{\delta x_1}{x_1}+\frac{D_n^*}{x_1}
\nabla^{'2}\frac{\delta n}{n}\nonumber\\
& & +\frac{D_1^{T*}}{x_1}\nabla^{'2}\frac{\delta T}{T}\Bigg),
\eeqa
\beqa
\label{4.16}
& & \frac{\partial}{\partial \tau}\frac{\delta n}{n}+\nabla'\cdot \frac{\delta \mathbf{U}}{\upsilon_{\text{th}}}=
\frac{1}{8}\frac{(1-\mu)}{\mu_{12}}\Bigg(x_1 D^*
\nabla^{'2}\frac{\delta x_1}{x_1}\nonumber\\
& & +D_n^*
\nabla^{'2}\frac{\delta n}{n}+D_1^{T*}\nabla^{'2}\frac{\delta T}{T}\Bigg),
\eeqa
\beqa
\label{4.17}
& & \frac{\partial}{\partial \tau}\frac{\delta U_\beta}{\upsilon_\text{th}}+\frac{1}{4}\frac{1+\mu}{x_2+x_1\mu}\Bigg[x_1 \left(\frac{\partial p^*}{\partial x_1}\right)_{\phi,\Delta^*} \nabla_\beta'\frac{\delta x_1}{x_1}+ \nonumber\\
& & p^*\Bigg(1+\phi \left(\frac{\partial \ln p^*}{\partial \phi}\right)_{x_1,\Delta^*}\Bigg) \nabla_\beta'\frac{\delta n}{n}
+p^*\Bigg(1-\frac{\Delta^*}{2} \nonumber\\
& & \times \left(\frac{\partial \ln p^*}{\partial \Delta^*}\right)_{x_1,\phi}\Bigg)\nabla_\beta'\frac{\delta T}{T}\Bigg]=\frac{1}{8}\frac{1+\mu}{x_2+x_1\mu}\Bigg[\eta^*\nabla^{'2}\frac{\delta U_\beta}{\upsilon_\text{th}}\nonumber\\
& & +
\Bigg(\frac{d-2}{d}\eta^*+\eta_b^*\Bigg) \nabla_\beta'\nabla'\cdot \frac{\delta \mathbf{U}}{\upsilon_\text{th}}\Bigg],
\eeqa
\beqa
\label{4.18}
& & \frac{\partial}{\partial \tau}\frac{\delta T}{T}+\frac{2}{d}p^*\nabla'\cdot \frac{\delta \mathbf{U}}{v_\text{th}}
+2 x_1 \left(\frac{\partial \zeta_0^*}{\partial x_1}\right)_{\phi,\Delta^*} \frac{\delta x_1}{x_1}+2\phi \nonumber\\
& &\times \left(\frac{\partial \zeta_0^*}{\partial \phi}\right)_{x_1,\Delta^*}\frac{\delta n}{n}-\Delta^* \left(\frac{\partial \zeta_0^*}{\partial \Delta^*}\right)_{x_1,\phi}\frac{\delta T}{T}
=-\frac{x_1}{8}\nonumber\\
& & \times \Big[\frac{1-\mu}{\mu_{12}}D^*-(d+2)D^{"*}\Big]\nabla^{'2} \frac{\delta x_1}{x_1}-\frac{1}{8}\Big[\frac{1-\mu}{\mu_{12}}D_n^*\nonumber\\
& & -(d+2)D_{qn}^*\Big]\nabla^{'2} \frac{\delta n}{n}-\frac{1}{8}\Big[\frac{1-\mu}{\mu_{12}}D_1^{T*}-(d+2)\kappa^*\Big]\nonumber\\
& & \times \nabla^{'2} \frac{\delta T}{T}-\zeta_U \nabla'\cdot \frac{\delta \mathbf{U}}{v_\text{th}}.
\eeqa
The subscript H has been omitted in Eqs.\ \eqref{4.15}--\eqref{4.18} for the sake of brevity. In these equations, $\nabla_\beta'=\partial/\partial r_\beta'$, $\nabla^{'2}=\partial^2/\partial r_\beta'\partial r_\beta'$, and $\phi=\phi_1+\phi_2$ is the total solid volume fraction, where  
\beq
\label{4.19}
\phi_i=\frac{\pi^{d/2}}{2^{d-1}d \Gamma(d/2)} n_i\sigma_i^d.
\eeq
Moreover, $\zeta_0^*=x_1\gamma_1 \zeta_1^*+x_2\gamma_2 \zeta_2^*$, $\zeta_i^*=\zeta_i/\nu$, and we have introduced the dimensionless transport coefficients
\begin{equation}
\label{4.20}
D^*\equiv \frac{m_1m_2\nu}{\rho T}D, \quad D_n^*\equiv \frac{m_1m_2\nu}{\rho T}D_{n},
\end{equation}
\beq
\label{4.21}
D_1^{T*}\equiv \frac{\rho \nu}{n T}D_1^T, \quad
D^{''*}\equiv \frac{4}{d(d+2)}\frac{\overline{m}\nu}{n}D'',
\eeq
\begin{equation}
\label{4.22}
D_{qn}^*\equiv \frac{4}{d(d+2)}\frac{\overline{m}\nu}{n}D_{qn},
\quad \kappa^{*}\equiv \frac{4}{d(d+2)}\frac{\overline{m}\nu}{n T}\kappa,
\end{equation}
\begin{equation}
\eta^*\equiv \frac{\nu \eta}{n T}, \quad
\eta_b^*\equiv \frac{\nu \eta_b}{n T}. \label{4.23}
\end{equation}

Then, a set of Fourier transformed dimensionless variables  is introduced as
\begin{equation}
\rho_{1,\mathbf{k}}(\tau)=\frac{\delta x_{1\mathbf{k}}(\tau)}{x_{1\text{H}}},\quad
\rho_{\mathbf{k}}(\tau)=\frac{\delta n_{\mathbf{k}}(\tau)}{n_{\text{H}}},  \label{4.24}
\end{equation}
\begin{equation}
\mathbf{w}_{\mathbf{k}}(\tau)=\frac{\delta \mathbf{U}_{\mathbf{k}}(\tau)}{
\upsilon_{\text{th,H}}}, \quad
\theta_{\mathbf{k}}(\tau)=\frac{\delta T_{\mathbf{k}}(\tau)}{T_{\text{H}}},
\label{4.25}
\end{equation}
where $\delta y_{\mathbf{k}\beta}(\tau)\equiv \left\{\rho_{1,\mathbf{k}}(\tau), \rho_{\mathbf{k}}(\tau), \bf{w}_{\mathbf{k}}(\tau), \theta_{\mathbf{k}}(\tau)\right\}$ is defined as
\begin{equation}
\delta y_{\mathbf{k}\beta}(\tau)=\int \dd\mathbf{r}'\;e^{-\imath \mathbf{k} \cdot \mathbf{r}'}\delta y_{\beta}(\bf{r}',\tau).
\label{4.26}
\end{equation}
Note that in Eq.\ \eqref{4.26} the wave vector $\mathbf{k}$ is dimensionless.

\subsection{Transverse shear modes}

As expected, in terms of the above dimensionless variables, the $d-1$ transverse velocity components
${\bf w}_{{\bf k}\perp}={\bf w}_{{\bf k}}-({\bf w}_{{\bf k}}\cdot
\widehat{{\bf k}})\widehat{{\bf k}}$ (orthogonal to the wave vector ${\bf k}$)
decouple from the other four modes. Their evolution equation can be easily obtained when one takes the Fourier transform of Eq.\ \eqref{4.17}. The result is
\begin{equation}
\label{4.27}
\frac{\partial {\bf w}_{{\bf k}\perp}}{\partial \tau}=\lambda_\perp {\bf w}_{{\bf k}\perp},
\eeq
where the eigenvalue $\lambda_\perp$ is
\beq
\label{4.27.1}
\lambda_\perp=
-\frac{1}{8}\frac{1+\mu}{x_2+x_1 \mu}\eta^* k^2 .
\end{equation}
The solution to Eq.\ (\ref{4.27}) is
\begin{equation}
\label{4.28}
{\bf w}_{{\bf k}\perp}({\bf k}, \tau)={\bf w}_{{\bf k}\perp}(0)\exp[\lambda_{\perp}(k)\tau].
\end{equation}
This identifies $d-1$ shear (transverse) modes analogous to the elastic ones.\cite{RL77} Since $\eta^*>0$, the eigenvalue $\lambda_{\perp}(k)<0$ and so, the transverse shear modes are always (linearly) stable (provided that $k>0$). The mode with $k=0$ is neutral, as expected. This conclusion agrees with previous results for monocomponent granular gases\cite{BBGM16,GBS21a} and for dilute granular binary mixtures.\cite{GBS24}

\subsection{Four longitudinal modes}

The remaining (longitudinal) modes correspond to the composition field $\rho_{1,\mathbf{k}}$, the density field $\rho_{\mathbf{k}}$, the longitudinal component of the velocity field $\mathbf{w}_{\mathbf{k}||}=({\bf w}_{{\bf k}}\cdot
\widehat{{\bf k}})\widehat{\mathbf{k}}$ (i.e., parallel to $\mathbf{k}$), and the temperature field $\theta_{\mathbf{k}}$. These modes are coupled and obey the time-dependent equation
\begin{equation}
\frac{\partial \delta z_{\mathbf{k}\beta}(\tau)}{\partial \tau}=\left(
M_{\beta\gamma}^{(0)}+i kM_{\beta\gamma}^{(1)}+k^{2}M_{\beta\gamma
}^{(2)}\right) \delta z_{\mathbf{k}\gamma}(\tau),  \label{4.29}
\end{equation}
where now $\delta z_{\mathbf{k}\beta}(\tau)$ denotes the four variables $
\left( \rho_{1,\mathbf{k}},\rho_{\mathbf{k}}, \theta_{\mathbf{k}}, w_{\mathbf{
k}||}\right)$. The matrices in Eq.\ \eqref{4.29} are defined as
\begin{widetext}
\begin{equation}
\mathsf{M}^{(0)}=\left(
\begin{array}{cccc}
0 & 0 & 0 & 0 \\
0 &0&0&0\\
-2x_{1}\left(\frac{\partial \zeta_0^*}{\partial x_1}\right)_{\phi,\Delta^*}&
-2\phi \left(\frac{\partial \zeta_0^*}{\partial \phi}\right)_{x_1,\Delta^*}& \Delta^* \left(\frac{\partial \zeta_0^*}{\partial \Delta^*}\right)_{x_1,\phi}&0\\
0 & 0 & 0 &0
\end{array}
\right),  \label{4.30}
\end{equation}
\begin{equation}
\mathsf{M}^{(1)}=\left(
\begin{array}{cccc}
0 & 0 & 0 & 0 \\
0 & 0 & 0 & -1\\
0 & 0 & 0 & -\left(\frac{2}{d}p^*+\zeta_U \right)\\
-\frac{x_1}{4}\frac{(1+\mu)}{x_2+\mu x_1}\left(\frac{\partial p^*}{\partial x_{1}}\right)_{\phi,\Delta^*}
&-\frac{p^*}{4}\frac{(1+\mu)}{x_2+\mu x_1}\Bigg(1+\phi\left(\frac{\partial \ln p^*}{\partial \phi}\right)_{x_1,\Delta^*}\Bigg)&-\frac{p^*}{4}\frac{(1+\mu)}{x_2+\mu x_1}\Bigg(1-\frac{\Delta^*}{2}\left(\frac{\partial \ln p^*}{\partial \Delta^*}\right)_{x_1,\phi}\Bigg) & 0
\end{array}
\right) ,  \label{4.31}
\end{equation}
\begin{equation}
\mathsf{M}^{(2)}=\left(
\begin{array}{cccc}
-\frac{x_2+\mu x_1}{8\mu_{12}}D^* &-\frac{x_2+\mu x_1}{8x_1\mu_{12}}D_n^*
 &-\frac{x_2+\mu x_1}{8x_1\mu_{12}}D_1^{T*}& 0 \\
-\frac{x_1(1-\mu)}{8\mu_{12}}D^*
&-\frac{(1-\mu)}{8\mu_{12}}D_n^*&-\frac{(1-\mu)}{8\mu_{12}}D_1^{T*}& 0\\
\frac{x_1}{8}\left[\frac{(1-\mu)}{\mu_{12}}D^*-(d+2)D^{''*}\right] & \frac{1}{8}\left[\frac{(1-\mu)}{\mu_{12}}D_n^*-(d+2)D_{qn}^{*}\right]& \frac{1}{8}\left[\frac{(1-\mu)}{\mu_{12}}D_1^{T*}-(d+2)\kappa^{*}\right]& 0 \\
0 & 0 & 0 & -\frac{1}{4}\frac{(1+\mu)}{x_2+\mu x_1}\Big(\frac{d-1}{d}\eta^* + \frac{1}{2}\eta_b^*\Big)
\end{array}
\right).  \label{4.32}
\end{equation}
\end{widetext}

In the particular case of mechanically equivalent particles ($m_i=m$, $\sigma_i=\sigma$, $\al_{ij}=\al$, and $\Delta_{ij}=\Delta$), Eqs.\ \eqref{4.30}--\eqref{4.32} are consistent with the results obtained in the $\Delta$-model for monocomponent granular gases.\cite{GBS21a} In addition, when $\Delta_{ij}=0$, Eqs.\ \eqref{4.30}--\eqref{4.32} agree with the results obtained in the IHS model for mixtures.\cite{G15}

The time evolution of the four longitudinal modes has the form $e^{\lambda_n(k)\tau}$ for $n=$1, 2, 3, and 4. The quantities $\lambda_n(k)$ are the eigenvalues of the matrix
\beq
\label{4.33}
M_{\beta\gamma}=M_{\beta\gamma}^{(0)}+ik M_{\beta\gamma}^{(1)}+k^2 M_{\beta\gamma}^{(2)}.
\eeq
Thus, the eigenvalues $\lambda_n(k)$ are the solutions of the quartic equation
\beq
\label{4.34}
\det \left(\lambda \openone - \mathsf{M}\right)=0,
\eeq
where $\openone$ is the matrix identity.

It is quite apparent that the determination of the dependence of the eigenvalues $\lambda_n(k)$ on the (dimensionless) wave vector $k$ and the parameters of the mixture is not really a simple problem. Therefore, to gain some insight into the general problem, it is convenient to study first the solution to the quartic equation \eqref{4.34}  in the extreme long wavelength limit, $k=0$.

\begin{figure}[t]
\centering
\includegraphics[width=0.40\textwidth]{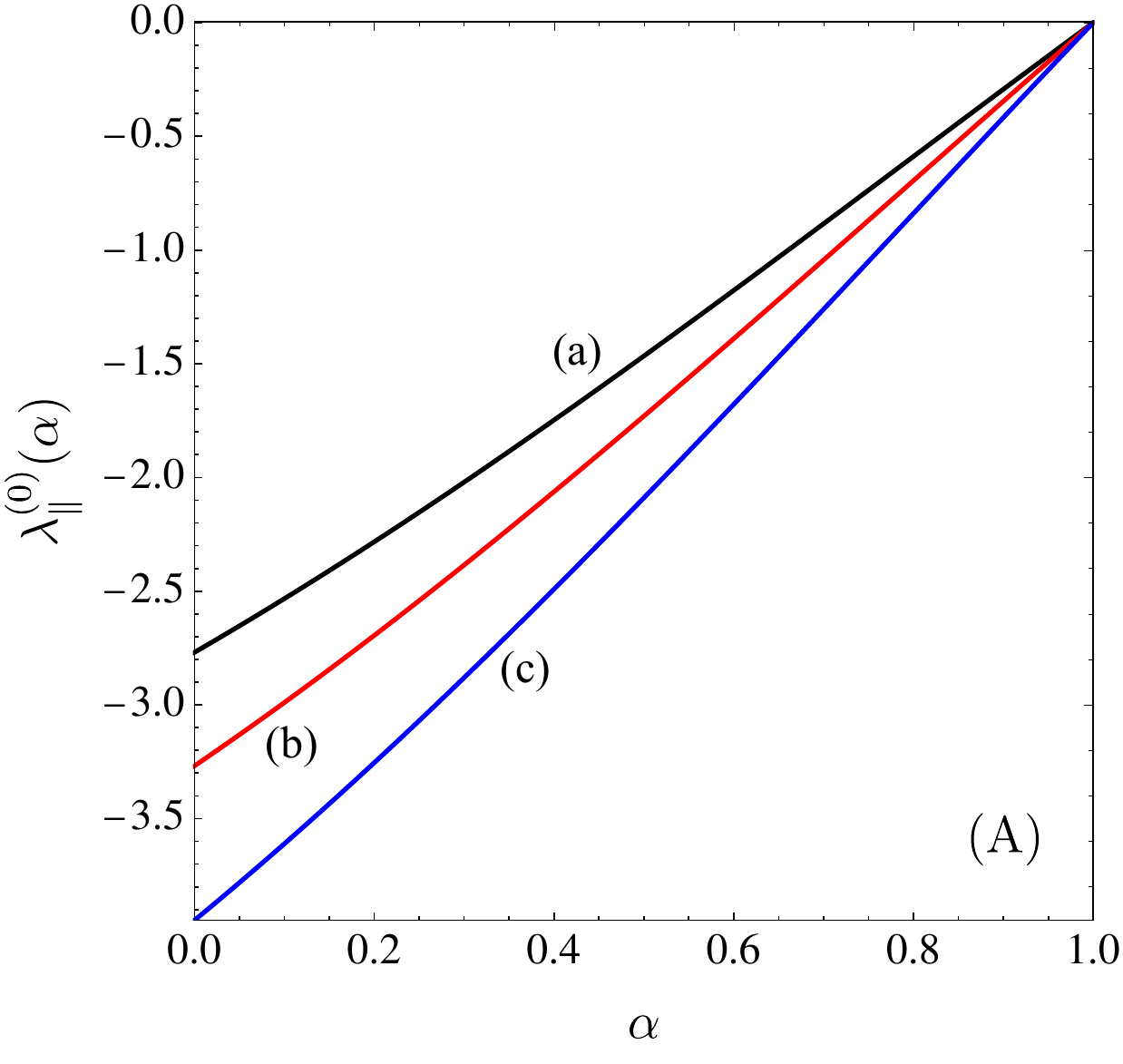}
\includegraphics[width=0.40\textwidth]{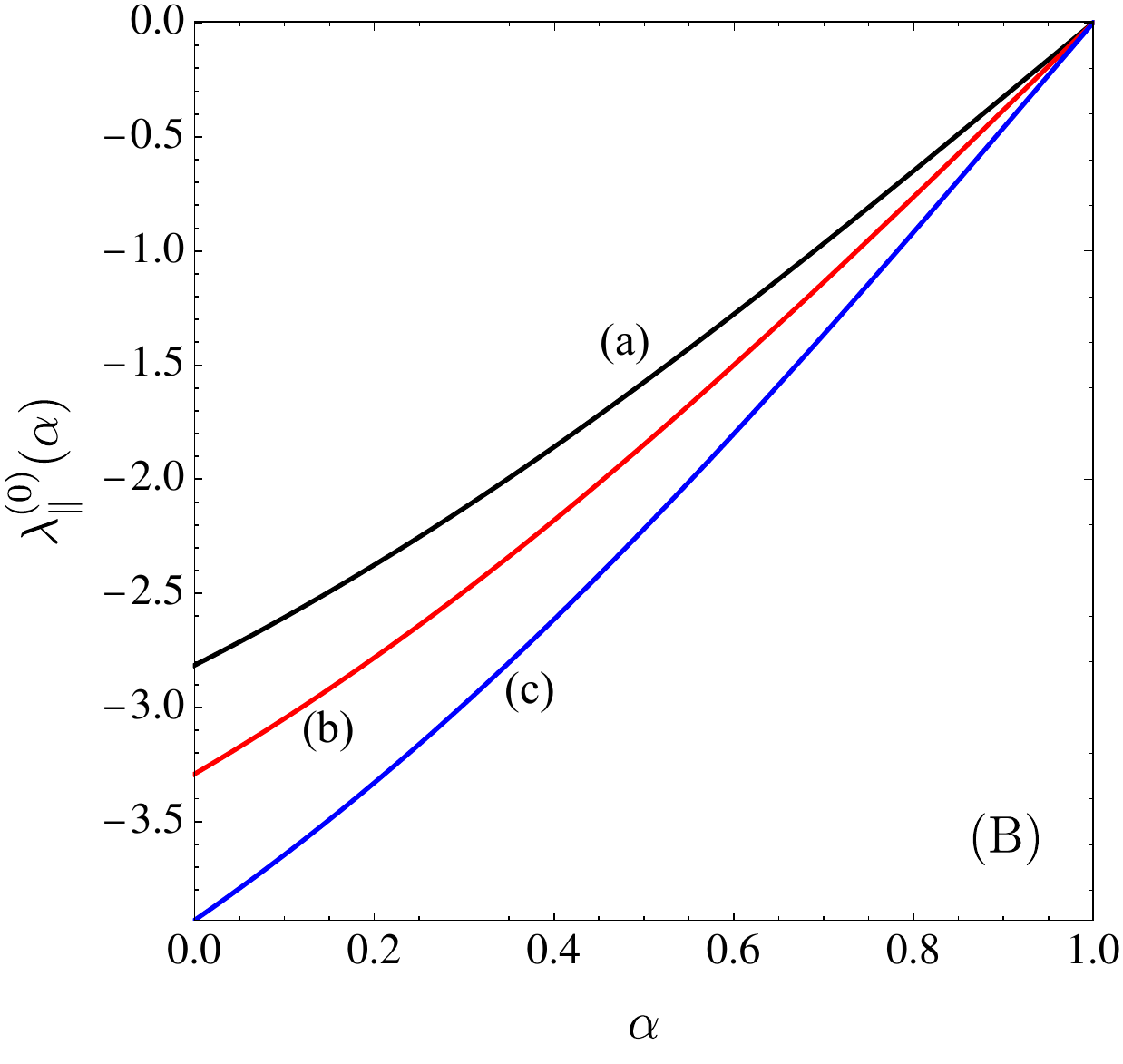}
\caption{Homogeneous temperature eigenvalue
    $\lambda_{\parallel}^{(0)}
    $
    as a function of the common coefficient of restitution
    $\alpha\equiv\alpha_{11}=\alpha_{22}=\alpha_{12}$ for binary
    mixtures of hard disks ($d=2$).
    Panel (A) corresponds to $x_1=0.8$, $m_1/m_2=0.5$, and
    $\sigma_1/\sigma_2=1$, while panel (B) corresponds to
    $x_1=0.3$, $m_1/m_2=2$, and $\sigma_1/\sigma_2=0.5$.
    In both panels, the curves refer to three different solid volume
    fractions: the dilute limit $\phi \to 0$ (a), $\phi=0.1$ (b),
    and $\phi=0.2$ (c).
\label{fig_eigen0}}
\end{figure}

\subsection{Extreme long wavelength limit ($k=0$)}

In the case of a homogeneous fluid (namely, when $k=0$ in Eq.\ \eqref{4.29}), the square matrix $\mathsf{M}$ reduces to the lower triangular 
matrix $\mathsf{M}^{(0)}$, whose eigenvalues are
\beq
\label{4.35.1}
\lambda_{\parallel}=\left\{0,0,0,\lambda_{\parallel}^{(0)} \right\},
\eeq
where
\beq
\label{4.35}
\lambda_{\parallel}^{(0)}=\Delta^* \left(\frac{\partial \zeta_0^*}{\partial \Delta^*}\right)_{x_1,\phi}.
\eeq
The physical meaning of the four modes in the limit $k=0$ is transparent. The first two null scalar modes are associated with the conservation of the concentration and with the conservation of the global number density. Since the perturbations are spatially uniform, there are no gradients and, consequently, neither diffusion nor heat conduction can relax them. The third null mode corresponds to a uniform velocity perturbation. Such a perturbation represents a uniform translation of the whole system and is not damped by viscosity; this reflects total momentum conservation. The only nontrivial eigenvalue is therefore the homogeneous temperature mode, whose sign is controlled by the slope of the homogeneous energy balance around the steady state. This means that the homogeneous temperature mode is linearly stable provided that
\beq
\lambda_\parallel^{(0)} < 0 \Longleftrightarrow \left(\frac{\partial\zeta_0^*}{\partial T}\right)_{x_1,\phi} > 0.
\eeq

The dependence of the eigenvalue $\lambda_{\parallel}^{(0)}$ on the parameters of the mixture is still complex.

A simpler situation corresponds to the limiting case of mechanically equivalent particles where
\beq
\label{4.36}
\zeta_1^*=\zeta_2^*=\zeta_0^*=\frac{\sqrt{2}\pi^{(d-1)/2}}{d\Gamma\left(\frac{d}{2}\right)}\chi\left(1-\al^2-2 \Delta^{*2}-\sqrt{2\pi}
\al \Delta^* \right),
\eeq
and so,
\beq
\label{4.37}
\lambda_{\parallel}^{(0)}=-\frac{\sqrt{2}\pi^{(d-1)/2}}{d\Gamma\left(\frac{d}{2}\right)}\chi\Delta^*\left(4\Delta^*+\sqrt{2\pi}
\al\right)<0.
\eeq
In Eq.\ \eqref{4.37}, the $\al$-dependence of $\Delta^*$ in the steady state is\cite{GBS18}
\beq
\label{4.38}
\Delta^*(\al)=\frac{1}{2}\sqrt{\frac{\pi}{2}}\al\left[\sqrt{1+
\frac{4(1-\al^2)}{\pi \al^2}}-1\right].
\eeq
Thus, according to Eq.\ \eqref{4.37} the longitudinal mode $\lambda_{\parallel}^{(0)}$ is (linearly) stable in agreement with the results reported for simple granular gases.\cite{GBS21a}

In the case of a dense granular mixture, a detailed analysis of the dependence of $\lambda_\parallel^{(0)}$ on the parameter space of the system shows that this mode is always negative. Hence, there are no unstable longitudinal modes for vanishing wave number ($k=0$): the temperature mode is damped, whereas the remaining three modes are neutral as a consequence of the conservation laws. As an illustration, Fig.~\ref{fig_eigen0} shows
$\lambda_{\parallel}^{(0)}$ as a function of the common coefficient
of restitution $\alpha$ for two representative binary mixtures and
three different solid volume fractions. Panel (A) corresponds to
$x_1=0.8$, $m_1/m_2=0.5$, and $\sigma_1/\sigma_2=1$, whereas panel
(B) refers to $x_1=0.3$, $m_1/m_2=2$, and $\sigma_1/\sigma_2=0.5$. In both cases, $\lambda_{\parallel}^{(0)}$ increases monotonically with $\alpha$, remaining negative throughout the inelastic range and vanishing at $\alpha=1$. At a fixed value of $\alpha<1$, increasing the density makes the eigenvalue more negative and hence enhances the relaxation rate of the homogeneous temperature perturbation. Moreover, the very similar behavior observed for the two mixtures suggests that, for the representative systems considered, the qualitative dependence of this mode on inelasticity and density is only weakly affected by particle dissimilarity.

\subsection{General case}

It is quite apparent that the analysis at finite (but small) values of the wave number $k$ is quite intricate and requires the numerical solution of Eq.\ \eqref{4.34}. Nevertheless, a complete
stability criterion can be formulated without explicitly solving the quartic equation. To that end, we write the characteristic polynomial in the form
\begin{align}
\mathfrak{P}(\lambda;k) & \equiv
\det \left(\lambda \openone - \mathsf{M} \right) \nonumber \\
& =
\lambda^4+a_1(k)\lambda^3+a_2(k)\lambda^2
+a_3(k)\lambda+a_4(k),
\label{eq:quartic_RH}
\end{align}
where, according to Newton's identities,\cite{Prasolov04}
\begin{equation}
    a_1(k) = - \mathrm{Tr}(\mathsf{M}), \quad a_2(k) = \frac{1}{2}\left[\mathrm{Tr}^2(\mathsf{M}) -\mathrm{Tr}(\mathsf{M}^2) \right],
\end{equation}
\begin{equation}
    a_3(k) = -\frac{1}{6}\left[\mathrm{Tr}^3(\mathsf{M}) - 3\mathrm{Tr}(\mathsf{M})\mathrm{Tr}(\mathsf{M}^2) + 2\mathrm{Tr}(\mathsf{M}^3) \right],
\end{equation}
\begin{equation}
    a_4(k) = \det(\mathsf{M}).
\end{equation}
Although the matrix $\mathsf{M}(k)$ contains imaginary terms
proportional to $ik$, the coefficients $a_\ell(k)$ are real
functions of $k^2$. Since the longitudinal perturbations evolve as
$\exp[\lambda_n(k)\tau]$, linear stability requires
$\Re[\lambda_n(k)]<0$. According to the Routh--Hurwitz
criterion, \cite{N20} this condition is fulfilled for $k>0$ if and only if
\begin{equation}
a_1(k)>0,\quad a_2(k)>0,\quad
a_3(k)>0,\quad a_4(k)>0, \label{eq:RH1}
\end{equation}
\begin{equation}
g(k) \equiv a_1(k)a_2(k) - a_3(k)>0, \label{eq:RH2}
\end{equation}
\begin{equation}
h(k) \equiv a_1(k)a_2(k)a_3(k)-a_3^2(k)-a_1^2(k)a_4(k)>0
. \label{eq:RH3}
\end{equation}
The case $k=0$, for which the characteristic polynomial has three vanishing roots, has been considered separately in the
preceding subsection.

To assess the above conditions, we have carried out a systematic
numerical exploration over the parameter space of the mixture. Since the present stability analysis is based on the Navier--Stokes hydrodynamic equations, attention has been restricted to the long-wavelength regime (let's say, $0<k\lesssim 1$). The numerical analysis shows that the four coefficients
$a_\ell(k)$ ($\ell=1,\ldots,4$), as well as the two Hurwitz
combinations $g(k)$ and $h(k)$, remain positive for all the systems
and wave numbers explored. Therefore, the Routh--Hurwitz conditions
\eqref{eq:RH1}--\eqref{eq:RH3} are fulfilled throughout the
Navier--Stokes hydrodynamic range. It follows that all four roots of the
characteristic polynomial \eqref{eq:quartic_RH} lie in the left half of the complex plane,
\begin{equation}
\Re\left[\lambda_n(k)\right]<0,
\qquad n=1,\ldots,4.
\end{equation}
Consequently, no instability threshold is found for the longitudinal modes within the range of validity of the Navier--Stokes description. Together with the stability of the transverse shear modes established above, this result shows that the HSS of a confined dense binary granular mixture is linearly stable over the parameter space
investigated.

\begin{figure}
    \centering
    \includegraphics[width=0.46\textwidth]{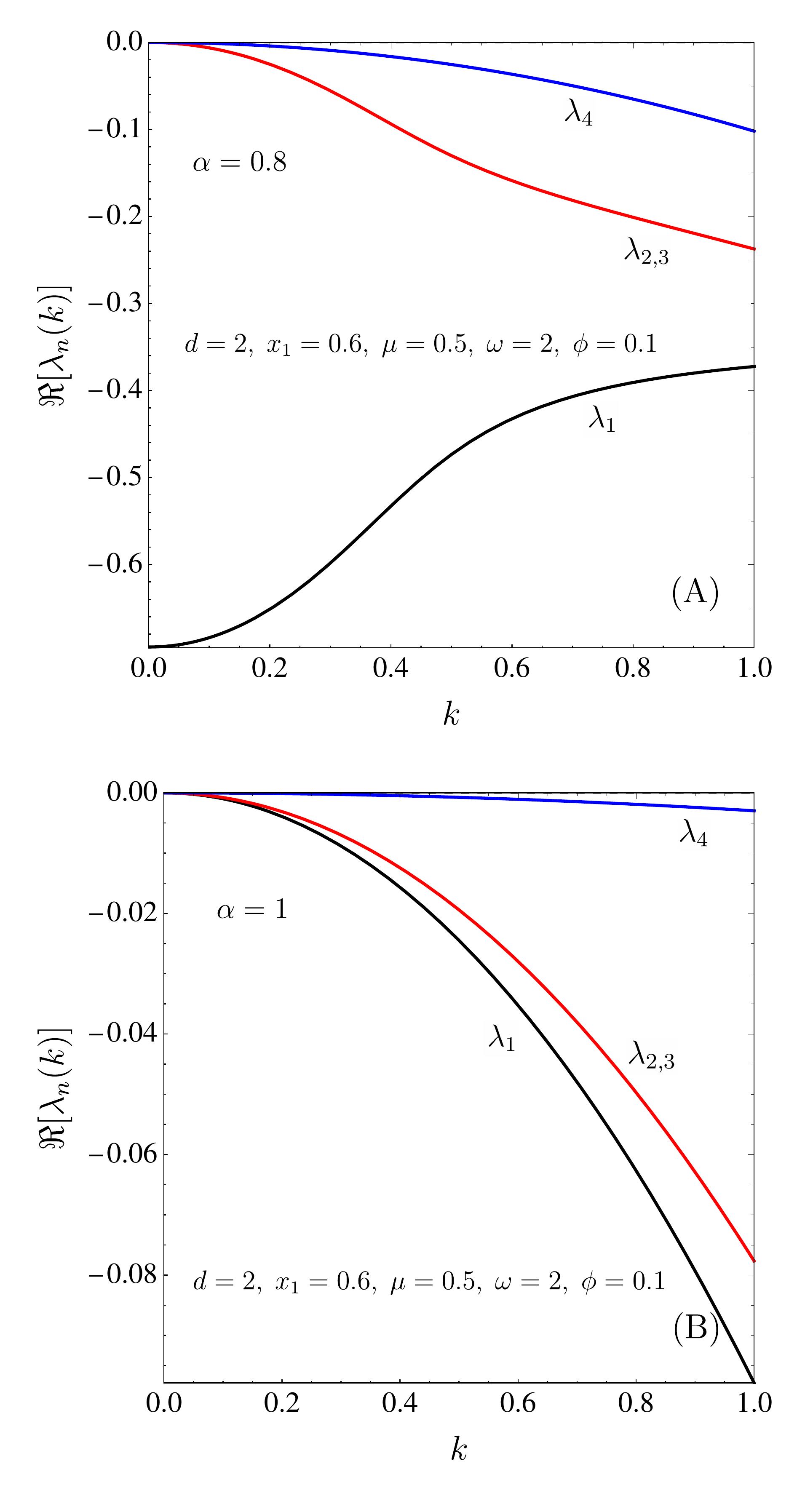}
    \caption{Real parts of the longitudinal eigenvalues $\lambda_n(k)$
    as functions of the dimensionless wave number $k$ for a binary
    mixture of hard disks ($d=2$) with $x_1=0.6$, $m_1/m_2=0.5$,
    $\sigma_1/\sigma_2=2$, and $\phi=0.1$. Panel (A) corresponds to an inelastic
    mixture with $\alpha_{ij}=\alpha=0.8$, whereas panel (B) refers
    to the elastic case $\alpha=1$. The modes
    $\lambda_2$ and $\lambda_3$ form a complex-conjugate pair and
    therefore have the same real part, denoted by
    $\Re(\lambda_{2,3})$. All the real parts remain negative
    for $k>0$ in the hydrodynamic range considered.}
    \label{fig:longitudinal_modes}
\end{figure}

For illustrative purposes, Fig.~\ref{fig:longitudinal_modes}
shows the real parts of the four longitudinal modes as functions of the
wave number $k$ for a representative binary mixture. Results for an
inelastic system ($\alpha=0.8$) are compared with those obtained in the elastic limit ($\alpha=1$). We observe first that the dependence of $\Re(\lambda_{n})$ on $k$ is quite different in both systems. Moreover, we see that
two of the eigenvalues form
a complex-conjugate pair and consequently have the same real part,
$\Re(\lambda_2)=\Re(\lambda_3)$. In agreement with the
Routh--Hurwitz analysis, all the modes remain damped for $k>0$.

\section{Onsager's reciprocal relations for dense mixtures}
\label{sec5}

As a second application, the quantification of the possible violation of the Onsager reciprocal relations for confined granular dense mixtures is studied in this Section. As discussed in previous works,\cite{GMD06,GBS21} since collisions in granular mixtures are inelastic, time reversal symmetry is broken in these systems. As a consequence, Onsager’s relations are expected to fail as the inelasticity increases. However, we think that the assessment of the expected violation and the influence of inelasticity on it in confined dense granular mixtures is still an interesting problem.

Let us consider a binary mixture for the sake of simplicity. In the usual language of the linear irreversible thermodynamics
for ordinary fluid mixtures,\cite{GM84, MDST13} to first order in spatial gradients, the constitutive equations for the mass and heat fluxes of a binary mixture are written as
\begin{equation}
\mathbf{j}_{i}=-\sum_{j=1}^2 L_{ij} \frac{\left(\nabla \mu_{j}\right)_{T}}{m_jT}
-L_{iq}\frac{\nabla T}{T^{2}},  \label{5.1}
\end{equation}
\beqa
\mathbf{J}_{q}&=&\mathbf{q}-\frac{d}{2}T \sum_{i=1}^2 \frac{\mathbf{j}_i}{m_i}
\nonumber\\
&=&-L_{qq}\nabla T-\sum_{i=1}^2 L_{qi}
\frac{\left(\nabla \mu _{i}\right)_{T}}{m_i T}.  \label{5.3}
\eeqa
Here, $\mu_{i}$ is the chemical potential of the species $i$. The coefficients $L_{ij}$, $L_{iq}$, and $L_{qi}$ are the so-called Onsager phenomenological coefficients.

It is worth clarifying the factor $d/2$ appearing in the
first line of the definition of $\mathbf{J}_q$. In the conventional representation in terms
of gradients of composition, pressure, and temperature, 
the corresponding modified heat flux $\mathbf{J}_{q}$ involves the partial enthalpies
$h_i$.\cite{GM84,MDST13} In the present representation, based on
the gradients of $n_1$, $n_2$, and $T$, the analogous decomposition of the entropy
production involves the partial internal energies
$u_i=\mu_i-T(\partial\mu_i/\partial T)_{n_1,n_2}$ instead of the partial enthalpies. In the case of hard disks or
spheres, the internal energy is purely kinetic, so that $u_i=dT/2$,
which leads to the definition of $\mathbf{J}_q$ given in Eq.~(\ref{5.3}).
The two representations are thermodynamically equivalent provided that
the conjugate fluxes and forces are transformed consistently.

For elastic collisions ($\al_{ij}=1$), Onsager showed that time reversal invariance of the underlying microscopic equations of motion leads to the following relations:
\beq
\label{5.4}
L_{ij}=L_{ji},\quad L_{iq}=L_{qi}.
\eeq 
These relations are called reciprocal relations as they relate transport coefficients for different processes. Thus, the coefficients $L_{iq}$ link the mass flux to the thermal gradient, while the coefficients $L_{qi}$ link the heat flux to the gradient of the chemical potentials.

In order to analyze the violation of Onsager's relations, one has first to express the spatial gradients $\nabla \ln n_1$ and $\nabla \ln n_2$ in terms of the chemical potential gradients $(\nabla \mu_1)_T$ and $(\nabla \mu_2)_T$. Since    
the chemical potentials $\mu_{i}$ $ (i=1,2)$ depend on space through their dependence on the concentration $x_1$ and the volume fraction $\phi$ at a given temperature, then
\beq
\label{5.5}
\left(\nabla \mu_{i}\right)_{T}=\left(\frac{\partial \mu _{i}}{\partial x_1}\right)
_{\phi,T}\nabla x_1+\left(\frac{\partial \mu_{i}}{\partial \phi}\right)
_{x_1,T}\nabla \phi.
\eeq
In addition,
\beq
\label{5.6}
\nabla x_1=x_1 x_2\left(\nabla \ln n_1-\nabla \ln n_2\right),
\eeq
and 
\beq
\label{5.6.1}
\nabla \phi=\phi_1 \nabla \ln n_1+\phi_2 \nabla \ln n_2.
\eeq
Thus, from Eqs.\ \eqref{5.5}--\eqref{5.6.1}, one achieves the identities
\beqa
\label{5.7}
(\nabla \mu_i)_T&=&\Bigg[x_1 x_2  \left(\frac{\partial \mu _{i}}{\partial x_1}\right)_{\phi,T}+\phi_1 \left(\frac{\partial \mu_{i}}{\partial \phi}\right)_{x_1,T}\Bigg]\nabla \ln n_1
\nonumber\\
&-& \Bigg[x_1 x_2  \left(\frac{\partial \mu _{i}}{\partial x_1}\right)_{\phi,T}-\phi_2 \left(\frac{\partial \mu_{i}}{\partial \phi}\right)_{x_1,T}\Bigg]\nabla \ln n_2.
\nonumber\\
\eeqa
According to Eq.\eqref{5.7}, the relationship between the gradients of the chemical potentials and the gradients of the partial densities can be written as
\beq
\label{5.8}
\nabla \ln n_1=\widetilde{A}_1\left(\nabla \mu_{1}\right)_{T}+\widetilde{B}_1\left(\nabla \mu_{2}\right)_{T}, 
\eeq
\beq
\label{5.8.1}
\nabla \ln n_2=\widetilde{A}_2\left(\nabla \mu_{1}\right)_{T}+\widetilde{B}_2\left(\nabla \mu_{2}\right)_{T},
\eeq
where 
\beq
\label{5.9}
\widetilde{A}_1=\frac{B_2}{A_1B_2-A_2B_1},
\quad \widetilde{B}_1=-\frac{A_2}{A_1B_2-A_2B_1},
\eeq
\beq
\label{5.10}
\widetilde{A}_2=-\frac{B_1}{A_1B_2-A_2B_1}, \quad \widetilde{B}_2=\frac{A_1}{A_1B_2-A_2B_1}.
\eeq
Here, we have introduced the quantities 
\beq
\label{5.11}
A_1=x_1 x_2 \left(\frac{\partial \mu _{1}}{\partial x_1}\right)_{\phi,T}+\phi_1 \left(\frac{\partial \mu_{1}}{\partial \phi}\right)_{x_1,T}, 
\eeq
\beq
\label{5.11.1}
A_2=-x_1 x_2 \left(\frac{\partial \mu _{1}}{\partial x_1}\right)_{\phi,T}+\phi_2 \left(\frac{\partial \mu_{1}}{\partial \phi}\right)_{x_1,T},
\eeq
\beq
\label{5.12}
B_1=x_1 x_2 \left(\frac{\partial \mu _{2}}{\partial x_1}\right)_{\phi,T}+\phi_1 \left(\frac{\partial \mu_{2}}{\partial \phi}\right)_{x_1,T},
\eeq
\beq
\label{5.12.1}
B_2=-x_1 x_2 \left(\frac{\partial \mu _{2}}{\partial x_1}\right)_{\phi,T}+\phi_2 \left(\frac{\partial \mu_{2}}{\partial \phi}\right)_{x_1,T}.
\eeq

In the case of a binary mixture ($s=2$) and in terms of the gradients of $n_1$, $n_2$, and $T$, the mass and heat fluxes are given by 
\beq
\label{5.13}
\mathbf{j}_1=-\frac{m_1 \rho_1}{\rho}D_{11}\nabla \ln n_1-\frac{m_1 \rho_2}{\rho}D_{12}\nabla \ln n_2-\rho D_1^T \nabla \ln T, 
\eeq
\beq
\label{5.14}
\mathbf{j}_2=-\frac{m_2 \rho_1}{\rho}D_{21}\nabla \ln n_1-\frac{m_2 \rho_2}{\rho}D_{22}\nabla \ln n_2-\rho D_2^T \nabla \ln T,
\eeq
\beq
\label{5.15}
\mathbf{q}=-T^2 D_{q,1}\nabla \ln n_1-T^2 D_{q,2}\nabla \ln n_2-T\kappa \nabla \ln T.
\eeq
The expressions of the coefficients $D_{ij}$ and $D_i^T$ are provided in the Appendix D of Ref.\ \onlinecite{GMG26}.

\begin{figure}[h!]
    \centering
    \includegraphics[width=0.405\textwidth]{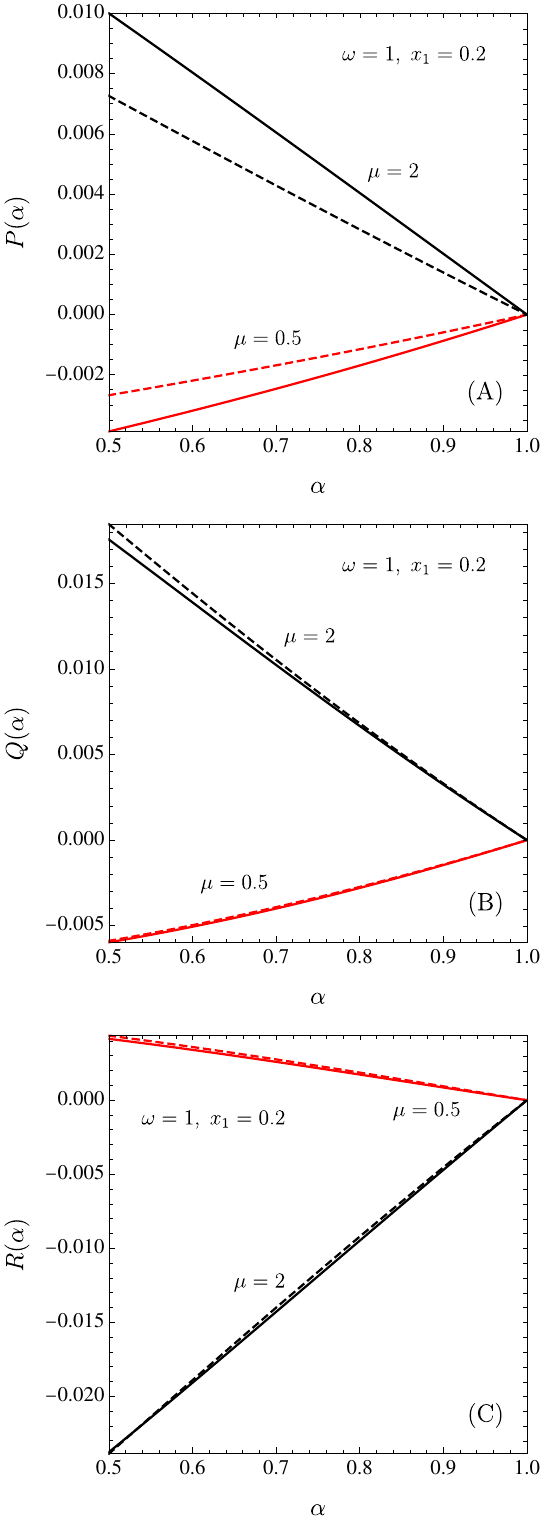}
    \caption{Plot of the dimensionless coefficients $P(\alpha)$ (A), 
    $Q(\alpha)$ (B), and $R(\alpha)$ (C) versus the (common) coefficient 
    of restitution $\alpha_{ij}=\alpha$ for $d=2$, $\sigma_1/\sigma_2=1$, and $x_1=0.2$. 
    The red and black curves correspond to the mass ratios $\mu=0.5$ and 
    $\mu=2$, respectively. Solid lines refer to the dilute limit 
    ($\phi\to0$), whereas dashed lines correspond to a moderately 
    dense mixture with $\phi=0.15$.}
    \label{fig:Onsager}
\end{figure}

The Onsager coefficients ($L_{ij}$, $L_{iq}$, $L_{qi}$) can be written in terms 
of both the diffusion ($D_{ij}$, $D_i^T$) and heat flux ($D_{q,i}$, $\kappa$) transport coefficients when one takes into account the relations \eqref{5.8} and Eqs.\ \eqref{5.1}-- \eqref{5.3} and \eqref{5.13}--\eqref{5.15}. After some algebra, the Onsager coefficients can be identified as 
\beq
\label{5.16}
L_{11}=m_1 T\Bigg(\frac{m_1\rho_1}{\rho}\widetilde{A}_1 D_{11}+\frac{m_1\rho_2}{\rho}\widetilde{A}_2 D_{12}\Bigg),
\eeq
\beq
\label{5.16.1}
L_{12}=m_2 T\Bigg(\frac{m_1\rho_1}{\rho}\widetilde{B}_1 D_{11}+\frac{m_1\rho_2}{\rho}\widetilde{B}_2 D_{12}\Bigg), 
\eeq
\beq
\label{5.17}
L_{22}=m_2 T\Bigg(\frac{m_2\rho_2}{\rho}\widetilde{B}_2 D_{22}+\frac{m_2\rho_1}{\rho}\widetilde{B}_1 D_{21}\Bigg), 
\eeq
\beq
\label{5.18}
L_{21}=m_1 T\Bigg(\frac{m_2\rho_2}{\rho}\widetilde{A}_2 D_{22}+\frac{m_2\rho_1}{\rho}\widetilde{A}_1 D_{21}\Bigg), 
\eeq
\beq
\label{5.19}
L_{1q}=\rho T D_1^T, \quad L_{2q}=\rho T D_2^T,
\eeq
\beqa
\label{5.20}
L_{q1}&=&m_1 T\Bigg[\Bigg(T^2 D_{q,1}-\frac{d}{2}\frac{m_2-m_1}{m_2}T\frac{\rho_1}{\rho}D_{11}\Bigg)\widetilde{A}_1\nonumber\\
&+ & \Bigg(T^2 D_{q,2}-\frac{d}{2}\frac{m_2-m_1}{m_2}T\frac{\rho_2}{\rho}D_{12}\Bigg)\widetilde{A}_2\Bigg],
\nonumber\\
\eeqa
\beqa
\label{5.21}
L_{q2}&=&m_2 T\Bigg[\Bigg(T^2 D_{q,1}-\frac{d}{2}\frac{m_2-m_1}{m_2}T\frac{\rho_1}{\rho}D_{11}\Bigg)\widetilde{B}_1\nonumber\\
&+&\Bigg(T^2 D_{q,2}-\frac{d}{2}\frac{m_2-m_1}{m_2}T\frac{\rho_2}{\rho}D_{12}\Bigg)\widetilde{B}_2\Bigg],
\nonumber\\
\eeqa
\beq
\label{5.22}
L_{qq}=\kappa-\frac{d}{2}\frac{m_2-m_1}{m_1 m_2}\rho D_1^T.
\eeq

To quantify the possible violation of the Onsager reciprocal relations $L_{12}=L_{21}$, $L_{1q}=L_{q1}$, and $L_{2q}=L_{q2}$, we introduce the three dimensionless functions
\begin{equation}
P(\alpha_{ij})\equiv
\frac{\nu}{\rho T}\left(L_{12}-L_{21}\right),
\label{5.23a}
\end{equation}
\begin{equation}
Q(\alpha_{ij})\equiv
\frac{\overline{m}\nu}{\rho T^2}
\left(L_{q1}-L_{1q}\right),
\label{5.23b}
\end{equation}
and
\begin{equation}
R(\alpha_{ij})\equiv
\frac{\overline{m}\nu}{\rho T^2}
\left(L_{q2}-L_{2q}\right).
\label{5.23c}
\end{equation}
Using the expressions of the phenomenological coefficients given
above, these functions can be written as
\begin{align}
P(\alpha_{ij})={}&
\frac{m_2}{m_1}\frac{\rho_1}{\rho}D_{11}^*
\left(\widetilde{B}_1^*+\mu\widetilde{A}_1^*\right)+\frac{\rho_2}{\rho}D_{12}^*
\left(\widetilde{B}_2^*+\mu\widetilde{A}_2^*\right),
\label{5.24}
\end{align}
\begin{align}
Q(\alpha_{ij})={}&
\Bigg(
\frac{\overline{m}}{m_1}D_{q,1}^*
-\frac{d}{2}
\frac{\overline{m}}{m_2}
\frac{m_2-m_1}{m_1}
\frac{\rho_1}{\rho}D_{11}^*
\Bigg)\widetilde{A}_1^*
\nonumber\\
&+
\Bigg(
\frac{\overline{m}}{m_2}\mu D_{q,2}^*
-\frac{d}{2}
\frac{\overline{m}}{m_2}
\frac{m_2-m_1}{m_2}
\frac{\rho_2}{\rho}D_{12}^*
\Bigg)\widetilde{A}_2^*
\nonumber\\
&-\frac{n\overline{m}}{\rho}D_1^{T*},
\label{5.25}
\end{align}
\begin{align}
R(\alpha_{ij})={}&
\Bigg(
\frac{\overline{m}}{m_1}\frac{m_2}{m_1}D_{q,1}^*
-\frac{d}{2}
\frac{\overline{m}}{m_1}
\frac{m_2-m_1}{m_1}
\frac{\rho_1}{\rho}D_{11}^*
\Bigg)\widetilde{B}_1^*
\nonumber\\
&+
\Bigg(
\frac{\overline{m}}{m_2}D_{q,2}^*
-\frac{d}{2}
\frac{\overline{m}}{m_2}
\frac{m_2-m_1}{m_1}
\frac{\rho_2}{\rho}D_{12}^*
\Bigg)\widetilde{B}_2^*
\nonumber\\
&-\frac{n\overline{m}}{\rho}D_2^{T*}.
\label{5.26}
\end{align}
Here, $\widetilde{A}_i^*=T\widetilde{A}_i$,
$\widetilde{B}_i^*=T\widetilde{B}_i$, and
\beq
\label{76}
D_{q,i}^*=\frac{m_i^2 \nu}{\rho}D_{q,i}, \quad D_{ij}^*=\frac{m_i m_j \nu}{\rho T}D_{ij}, \quad D_i^{T*}=\frac{\rho \nu}{n T}D_i^T.
\eeq
By construction, $Q$, $P$, and $R$ vanish when the corresponding Onsager reciprocal relations are fulfilled. In the elastic molecular limit ($\alpha_{ij}=1$ and $\Delta_{ij}^*=0$), microscopic time-reversal invariance requires all three functions to vanish; i.e., $Q(1)=P(1)=R(1)=0$. This provides a useful consistency check of the present expressions
against the transport coefficients of molecular hard-sphere mixtures.\cite{LCK83}
To illustrate the departure from Onsager symmetry away from the elastic limit, 
Fig.~\ref{fig:Onsager} shows the functions $P$, $Q$, and $R$ as functions of the 
(common) coefficient of restitution $\alpha_{ij}=\alpha$ for two different mass 
ratios. Results are shown both in the dilute limit ($\phi\to 0$) and at finite 
density ($\phi=0.15$). As expected, all three functions tend to zero as 
$\alpha\to 1$ for both densities, thus recovering the Onsager reciprocal 
relations in the elastic limit. For inelastic collisions, finite-density effects 
modify the magnitude of the deviations from Onsager symmetry. In addition, within the $\Delta$-model, we observe that in general the breakdown of Onsager's relations is much smaller than the one reported in a previous work\cite{GMD06} in the conventional IHS model. In this context, the use of the reciprocal relations of the Onsager coefficients for confined granular systems could be considered as a reliable approximation for moderately inelastic collisions.


\section{Discussion}
\label{sec6}

In this work, the Enskog kinetic description of confined granular mixtures within the $\Delta$-model has been completed at Navier--Stokes order.  The results derived here complement those obtained in our previous work\cite{GMG26} for the mass flux and the pressure tensor by providing explicit expressions for the transport coefficients associated with the heat flux, as well as for the first-order contributions to the partial temperatures and the cooling rate. Consequently, the complete set of Navier--Stokes transport coefficients required to describe a moderately dense confined granular mixture is now available within the same kinetic-theory framework of the $\Delta$-model.

One of the main results of the present study is the important role played by finite-density effects in heat transport. The collisional contributions to the thermal conductivity and the Dufour coefficients can be comparable to, or even more important than, their kinetic counterparts. The illustrative results for binary mixtures show that the influence of density cannot in general be regarded as a simple quantitative correction to the dilute behavior. While the scaled thermal conductivity exhibits a comparatively smooth dependence on density and inelasticity, the Dufour coefficients show a much richer behavior, strongly dependent on the mass and diameter ratios. In particular, for some mixtures the Dufour coefficient can change sign as the density and inelasticity increase. The first-order contribution $\zeta_U$ to the cooling rate is also found to be strongly enhanced by density. The first-order contributions to the partial temperatures also exhibit a marked density dependence; in particular, they remain finite at nonzero density in the elastic limit, consistently with the behavior of ordinary dense mixtures. In addition, the present calculation has revealed a nonzero contribution to the collisional heat flux proportional to the quantities $I_{ij\ell}$ that had not been included in previous calculations.\cite{GDH07,GHD07,G19} The recovery of the known expressions for molecular mixtures in the elastic limit provides a useful consistency check of these collisional contributions.

The availability of the complete set of transport coefficients has made it possible to revisit the linear stability of the HSS. The transverse shear modes are always damped for finite wave number $k$, while in the extreme long-wavelength limit ($k=0$) the only nontrivial longitudinal eigenvalue, associated with a homogeneous temperature perturbation, is found to be negative. For finite wave numbers, the Routh--Hurwitz conditions\cite{N20} have been evaluated numerically over a broad range of the parameter space relevant to the Navier--Stokes description. No violation of these conditions has been found for $0<k\lesssim 1$.
Thus, within the range explored, both the transverse and longitudinal modes remain linearly stable. This extends to moderately dense binary mixtures the stability previously found for confined monocomponent systems and dilute granular mixtures.

As a second application, the possible violation of Onsager's reciprocal relations has been quantified through the dimensionless functions $P$,
$Q$, and $R$. As expected from the lack of microscopic time reversal invariance, these quantities are in general different from zero for inelastic collisions. Their magnitude depends on both particle dissimilarity and density, showing that collisional transfer also affects the departure from Onsager symmetry. An important consistency test is that
all three quantities vanish when the elastic limit is approached, not only in the dilute regime but also at finite density. The recovery of the Onsager reciprocal relations in the dense elastic limit lends further support to the collisional contributions to the transport coefficients and
to the choice of the quantities $I_{ij\ell}$ entering the revised Enskog
description.

Some limitations of the present calculations should nevertheless be mentioned. The kinetic contributions to the heat-flux transport coefficients have been evaluated in the first Sonine approximation. As already noted, this approximation yields a vanishing kinetic thermal conductivity for mechanically equivalent particles and therefore a higher-order Sonine approximation would be required to recover the correct monocomponent limit. Maxwellian approximations have also been employed in the evaluation of several collision integrals. Moreover, the stability analysis at finite wave number is necessarily restricted to the long-wavelength regime where the Navier--Stokes description is expected to apply. Finally, the chemical potentials employed in the Onsager analysis are those of the corresponding elastic hard-disk mixture and hence their use for inelastic granular systems should be understood as an extension of the molecular description rather than as a genuine thermodynamic formulation.

A natural continuation of the present work would be to assess the theoretical predictions for the heat-flux transport coefficients against
computer simulations, particularly in the parameter regions where strong finite-density effects and sign changes of the Dufour coefficients are predicted. It would also be interesting to improve the kinetic contributions to the heat flux transport coefficients by considering the so-called second Sonine approximation and to extend the analysis to mixtures where the injection parameters $\Delta_{ij}$ are different for all collision pairs. These studies would provide further tests of the range of validity of the Enskog kinetic description for confined granular mixtures.

\acknowledgments

We acknowledge financial support from grant no. PID2024-156352NB-I00 funded by MCIU/ AEI/10.13039/501100011033/FEDER, UE and from grant no. GR24022 funded by Junta de Extremadura (Spain) and by European Regional Development Fund (ERDF) ``A way of making Europe''. The research of David Gonz\'alez M\'endez has been supported by the predoctoral fellowship FPU24/01056 from the Spanish Government.\\
 
\section*{AUTHOR DECLARATIONS}

\subsection*{Conflict of Interest}

The authors have no conflicts to disclose.

\subsection*{Author Contributions}

David Gonz\'alez M\'endez: Formal analysis (equal); Investigation (equal);
Methodology (equal); Writing -- original draft (equal); Writing -- review
\& editing (equal).

Vicente Garz\'o: Formal analysis (equal); Investigation (equal);
Methodology (equal); Writing -- original draft (equal); Writing -- review
\& editing (equal).

\section*{DATA AVAILABILITY}

The data that support the findings of this study are available from the
corresponding author upon reasonable request.

\appendix

\section{Collisional contributions to the heat flux}
\label{appA}

In this Appendix we provide some of the mathematical steps needed to get the collisional contributions to the heat flux. This allows us to identify the forms of the collisional transport coefficients $D_{q,ij}^{\text{c}}$ and $\kappa_\text{c}$. The exact expression of the collisional heat flux $\mathbf{q}^c$ is given by Eq.\ \eqref{2.14.1}. To get the first-order contributions to $\mathbf{q}^c$ one has to consider the following expansion:
\begin{widetext}
\begin{eqnarray}
&& \int_{0}^{1}\dd\lambda\; f_{ij}(\mathbf{r}-\lambda\boldsymbol {\sigma}_{ij},\mathbf{r}+\left(
1-\lambda\right) \boldsymbol {\sigma}_{ij},\mathbf{v}_{1},\mathbf{v}_{2},t)=\int_{0}^{1}\dd\lambda\;\chi_{ij}\left(
\mathbf{r}-\lambda\boldsymbol {\sigma }_{ij}, \mathbf{r}+\left( 1-\lambda\right) \boldsymbol {\sigma }_{ij}\mid
\left\{n_{i}\right\} \right)f_{i}(\mathbf{r}-\lambda\boldsymbol {\sigma
}_{ij},\mathbf{v}_{1};t)\nonumber\\
& &\times f_{j}( \mathbf{r}+\left( 1-\lambda\right)
\boldsymbol {\sigma}_{ij},\mathbf{v}_{2})\rightarrow\chi _{ij}\left(\sigma _{ij}\right)
f_{i}^{(0)}(\mathbf{v}_{1})f_{j}^{(0)}(\mathbf{v}_{2})
+\frac{1}{2}\chi_{ij}\left[f_{i}^{(0)}(\mathbf{v}_{1})\boldsymbol {\sigma
}_{ij}\cdot \nabla f_{j}^{(0)}(\mathbf{v}_{2})-f_{j}^{(0)}(\mathbf{v}_{2})\boldsymbol {\sigma}_{ij}\cdot \nabla f_i^{(0)}( \mathbf{v}_{1})\right]
\nonumber\\
& & +\chi _{ij}\left[f_{i}^{(0)}(\mathbf{v}_{1})f_{j}^{(1)}(\mathbf{v}_{2})+
f_{i}^{(1)}(\mathbf{v}_{1})f_{j}^{(0)}(\mathbf{v}
_{2})\right]+f_{i}^{(0)}(\mathbf{v}_{1})f_{j}^{(0)}(\mathbf{v}
_{2})\delta \chi _{ij}\nonumber\\
& &\equiv \chi _{ij}\left(\sigma _{ij}\right)
f_{i}^{(0)}(\mathbf{v}_{1})f_{j}^{(0)}(\mathbf{v}_{2})+\Xi_{ij},  \label{a1}
\end{eqnarray}
\end{widetext}
where the explicit form of $\Xi_{ij}$ can be easily identified. The quantity
$\delta \chi _{ij}$ is defined by
\beq
\label{a2}
\delta \chi_{ij}=\frac{1}{2}
\chi_{ij}\sum_{\ell=1}^s\; I_{ij\ell}\boldsymbol {\sigma}_{ij}\cdot \nabla \ln n_\ell.
\eeq
In Eq.\ \eqref{a1}, the dependence of the distributions $f_i^{(0)}$, $f_j^{(0)}$, $f_i^{(1)}$, and $f_j^{(1)}$ on $\mathbf{r}$ and $t$ has been omitted for the sake of brevity. Thus, according to Eq.\ \eqref{a1}, the collisional contribution to the heat flux to first order in spatial gradients can be written as
\begin{widetext}
\beqa
\label{a3}
{\bf q}^\text{c}&=&\sum_{i,j}\frac{1+\alpha_{ij}}{8}m_{ij} \sigma_{ij}^{d}
\int \dd\mathbf{v}_{1}\int \mathrm{d}\mathbf{v}_{2}\int
\dd\widehat{\boldsymbol {\sigma}}\,\Theta (\widehat{\boldsymbol{\sigma}}\cdot
\mathbf{g}_{12})(\widehat{\boldsymbol {\sigma}}\cdot \mathbf{g}_{12})^{2}\widehat{\boldsymbol {\sigma}}\Big[4
(\widehat{\boldsymbol {\sigma}}\cdot {\bf G}_{ij})+(\mu_{ji}-\mu_{ij})(1-\al_{ij})(\widehat{\boldsymbol{\sigma}}\cdot
\mathbf{g}_{12})\Big]\Xi_{ij}\nonumber\\
& &-\sum_{i,j} \frac{m_i}{4} \sigma_{ij}^d \Delta_{ij}
\int \dd\mathbf{v}_{1}\int \mathrm{d}\mathbf{v}_{2}\int
\dd\widehat{\boldsymbol {\sigma}}\Theta (\widehat{\boldsymbol{\sigma}}\cdot
\mathbf{g}_{12})(\widehat{\boldsymbol {\sigma}}\cdot \mathbf{g}_{12})
\widehat{\boldsymbol {\sigma}}
\Big[4\mu_{ji}^2\Delta_{ij} +4\mu_{ji}^2\al_{ij} (\widehat{\boldsymbol {\sigma}}\cdot \mathbf{g}_{12})
-
4\mu_{ji} (\widehat{\boldsymbol {\sigma}}\cdot \mathbf{G}_{ij})\Big]
\Xi_{ij}\nonumber\\
&=&\mathbf{q}_c^{(\Delta=0)}+\mathbf{q}_c^{(\Delta\neq 0)}.
\eeqa
Here, $\mathbf{q}_c^{(\Delta=0)}$ denotes the contribution to ${\bf q}^\text{c}$ when $\Delta_{ij}=0$. This contribution has been determined in previous papers.\cite{GDH07}

We focus now on the term $\mathbf{q}_c^{(\Delta\neq 0)}$. Interchanging the labels $i$, $j$ and the variables $\mathbf v_1$,
$\mathbf v_2$ in the terms of $\Xi_{ij}$ proportional to
$f_j^{(0)}\boldsymbol{\sigma}_{ij}\cdot\nabla f_i^{(0)}$ and
$f_i^{(1)}f_j^{(0)}$ (which implies that $\widehat{\boldsymbol {\sigma}}\to -\widehat{\boldsymbol {\sigma}}$), the expression of $\mathbf{q}_c^{(\Delta\neq 0)}$ can be rewritten as
\beqa
\label{a4}
q_{c,\lambda}^{(\Delta\neq 0)}&=&-\sum_{i,j} \frac{m_i}{4} \sigma_{ij}^d \Delta_{ij}
\int \dd\mathbf{v}_{1}\int \dd\mathbf{v}_{2}\int
d\widehat{\boldsymbol {\sigma}}\Theta (\widehat{\boldsymbol{\sigma}}\cdot
\mathbf{g}_{12})(\widehat{\boldsymbol {\sigma}}\cdot \mathbf{g}_{12})\widehat{\sigma}_\lambda\Bigg\{\Big[4\mu_{ji}(\mu_{ji}-\mu_{ij})\Big(\Delta_{ij} +\al_{ij} (\widehat{\boldsymbol {\sigma}}\cdot \mathbf{g}_{12})\Big)\nonumber\\
& & -8\mu_{ji}(\widehat{\boldsymbol {\sigma}}\cdot \mathbf{G}_{ij})\Big] \chi_{ij}\Bigg[\frac{1}{2}f_{i}^{(0)}(\mathbf{v}_{1})\boldsymbol {\sigma
}_{ij}\cdot \nabla f_{j}^{(0)}(\mathbf{v}_{2})+f_{i}^{(0)}(\mathbf{v}_{1})f_{j}^{(1)}(\mathbf{v}_{2})\Bigg]\nonumber\\
& & +\frac{1}{2}\chi_{ij}\Big[4\mu_{ji}^2\Delta_{ij} +4\mu_{ji}^2\al_{ij} (\widehat{\boldsymbol {\sigma}}\cdot \mathbf{g}_{12})
-
4\mu_{ji} (\widehat{\boldsymbol {\sigma}}\cdot \mathbf{G}_{ij})\Big]f_i^{(0)}(\mathbf{v}_1)f_j^{(0)}(\mathbf{v}_2)\sum_{\ell=1}^s I_{ij\ell}\boldsymbol {\sigma}_{ij}\cdot \nabla \ln n_\ell\Bigg\},
\eeqa
where $\mathbf{G}_{ij}=\mu_{ij}\mathbf{V}_1+\mu_{ji}\mathbf{V}_2$.
In Eq.\ \eqref{a4}, according to the constitutive equation \eqref{2.17} for the heat flux, the only contributions to $\mathbf{q}^\text{c}$ are proportional to temperature and species density gradients:
\beq
\label{a5}
\nabla f_j^{(0)}=\sum_{\ell=1}^s n_\ell \frac{\partial f_j^{(0)}}{\partial n_\ell}\nabla \ln n_\ell+T \frac{\partial f_j^{(0)}}{\partial T}\nabla \ln T + (\text{irrelevant terms}).
\eeq
To perform the angular integrations in Eq.\ \eqref{a4}, we will use the results
\beq
\label{a6}
\int
\dd\widehat{\boldsymbol {\sigma}}\Theta (\widehat{\boldsymbol{\sigma}}\cdot
\mathbf{g}_{12})(\widehat{\boldsymbol {\sigma}}\cdot \mathbf{g}_{12})^k\widehat{\sigma}_\lambda =B_{k+1}g_{12}^{k-1}g_{12,\lambda},
\eeq
\beq
\label{a7}
\int
\dd\widehat{\boldsymbol {\sigma}}\Theta (\widehat{\boldsymbol{\sigma}}\cdot
\mathbf{g}_{12})(\widehat{\boldsymbol {\sigma}}\cdot \mathbf{g}_{12})^k\widehat{\sigma}_\lambda \widehat{\sigma}_\mu=\frac{B_k}{d+k}g_{12}^{k-2}\left(k g_{12,\lambda}g_{12,\mu}+g_{12}^2 \delta_{\lambda\mu}\right),
\eeq
\beq
\label{a8}
\int
\dd\widehat{\boldsymbol {\sigma}}\Theta (\widehat{\boldsymbol{\sigma}}\cdot
\mathbf{g}_{12})(\widehat{\boldsymbol {\sigma}}\cdot \mathbf{g}_{12})^k\widehat{\sigma}_\lambda \widehat{\sigma}_\mu\widehat{\sigma}_\beta=\frac{B_{k+1}}{d+k+1}g_{12}^{k-3}\left[(k-1)g_{12,\lambda}g_{12,\mu}g_{12,\beta}
+g_{12}^2\left(
g_{12,\lambda}\delta_{\mu\beta}+g_{12,\mu}
\delta_{\lambda\beta}+g_{12,\beta}\delta_{\lambda\mu}\right)\right],
\eeq
where the coefficients $B_k$ are\cite{NE98}
\beq
\label{a9}
B_k\equiv\int \mathrm{d} \widehat{\boldsymbol{\sigma}}\;\Theta (\widehat{{\boldsymbol {\sigma}}}\cdot {\bf g}_{12})(\widehat{{\boldsymbol {\sigma}}}\cdot \widehat{{\bf g}}_{12})^k=\pi^{\frac{d-1}{2}}\frac{\Gamma\left(\frac{k+1}{2}\right)}{\Gamma\left(\frac{k+d}{2}\right)},
\eeq
for positive integers $k$.

Let us consider first the term in $q_{c,\lambda}^{(\Delta\neq 0)}$ proportional to $\nabla f_j^{(0)}$ in Eq.\ \eqref{a4}. After integrating over $\widehat{{\boldsymbol {\sigma}}}$, one gets the result
\beqa
\label{a10}
& & -\sum_{i,j} \frac{m_i}{8} \sigma_{ij}^{d+1} \Delta_{ij}\chi_{ij}
\int \dd\mathbf{v}_{1}\int \dd\mathbf{v}_{2}\,f_{i}^{(0)}(\mathbf{v}_{1})\partial_\beta f_{j}^{(0)}(\mathbf{v}_{2})\Bigg\{
4\mu_{ji}(\mu_{ji}-\mu_{ij})\Bigg[\frac{B_1}{d+1}\Delta_{ij}g^{-1}(g_\lambda g_\beta+g^2\delta_{\lambda \beta})\nonumber\\
& & +\frac{B_2}{d+2}\al_{ij}\left(2g_\lambda g_\beta+g^2\delta_{\lambda \beta}\right)\Bigg]-\frac{8B_2}{d+2}\mu_{ji}G_{ij,\mu}
\left(
g_{\lambda}\delta_{\mu\beta}+g_{\mu}
\delta_{\lambda\beta}+g_{\beta}\delta_{\lambda\mu}\right)\Bigg\}\nonumber\\
&=& -\sum_{i,j} \frac{m_i}{8d} \sigma_{ij}^{d+1} \Delta_{ij}\chi_{ij}
\int \dd\mathbf{v}_{1}\int \dd\mathbf{v}_{2}\,f_{i}^{(0)}(\mathbf{v}_{1})\partial_\lambda f_{j}^{(0)}(\mathbf{v}_{2})
\Bigg[4\mu_{ji}(\mu_{ji}-\mu_{ij})\left(B_1 \Delta_{ij} g+B_2 \al_{ij}g^2\right)\nonumber\\
& & -8 B_2 \mu_{ji}(\mathbf{g}\cdot \mathbf{G}_{ij})\Bigg].
\eeqa
Here, we have adopted the notation $\mathbf{g}_{12}\equiv \mathbf{g}$ and in the last step in Eq.\ \eqref{a10} we have accounted for the isotropy in velocity space of the zeroth-order distribution functions.

We now evaluate the contribution to $q_{c,\lambda}^{(\Delta\neq 0)}$ coming from the first-order distribution $f_j^{(1)}$. By symmetry, the terms of $f_j^{(1)}$ contributing to $q_{c,\lambda}^{(\Delta\neq 0)}$ are $\boldsymbol{\mathcal{A}}_j \nabla \ln T$ and $\boldsymbol{\mathcal{B}}_{j\ell} \nabla \ln n_\ell$. Once the angular integration is carried out, this contribution to $q_{c,\lambda}^{(\Delta\neq 0)}$ is given by
\begin{align}
\label{a11}
    -\sum_{i,j} \frac{m_i}{4}\sigma_{ij}^d\Delta_{ij}\chi_{ij}\int \dd \mathbf{v}_1 \int \dd \mathbf{v}_2 \; f_i^{(0)}(\mathbf{v}_1)f_j^{(1)}(\mathbf{v}_2)\Bigg[4\mu_{ji}(\mu_{ji}-\mu_{ij})\left( B_2\Delta_{ij}g_{\lambda} + \alpha_{ij}B_3gg_{\lambda} \right) \nonumber\\
    - \frac{8B_1}{d+1}\mu_{ji}G_{ij,\mu}g^{-1}\left(g_\lambda g_\mu + g^2 \delta_{\lambda\mu}\right)\Bigg]
\end{align}
\end{widetext}
In the leading Sonine approximation, $\boldsymbol{\mathcal{A}}_j\propto \mathbf{V}_2 f_{j,\text{M}}(\mathbf{V}_2)$ and $\boldsymbol{\mathcal{B}}_{j\ell}\propto \mathbf{V}_2 f_{j,\text{M}}(\mathbf{V}_2)$, so that
\begin{equation}
\label{a12}
    f_j^{(1)}(\mathbf{V}_2)
    \to \frac{1}{n_j T_j^{(0)}}f_{j,\text{M}}(\mathbf{V}_2)
    V_{2,\beta}j_{j,\beta}^{(1)}.
\end{equation}
The contribution \eqref{a11} is explicitly determined below when one replaces $f_i^{(0)}(\mathbf{v}_1)$ by its Maxwellian approximation and $f_j^{(1)}(\mathbf{v}_2)$ by Eq.\ \eqref{a12}.

The last contribution to $\mathbf{q}^c$ proportional to $\delta \chi_{ij}$ is given by
\begin{widetext}
\beq
\label{a13}
-\sum_{i,j} \frac{m_i}{8d} \sigma_{ij}^{d+1} \Delta_{ij}\chi_{ij}
\int \dd\mathbf{v}_{1}\int \dd\mathbf{v}_{2}\,f_{i}^{(0)}(\mathbf{v}_{1})f_{j}^{(0)}(\mathbf{v}_{2})
\Big[4\mu_{ji}^2\left( B_1 \Delta_{ij} g+B_2 \al_{ij}g^2\right)-4B_2 \mu_{ji}(\mathbf{g}\cdot \mathbf{G}_{ij})\Big]
\sum_{\ell=1}^s I_{ij\ell}\nabla \ln n_\ell.
\eeq

Thus, according to Eqs.\ \eqref{a10}, \eqref{a11} and \eqref{a13}, $\mathbf{q}_{c}^{(\Delta\neq 0)}$ can be written as
\beq
\label{a14}
\mathbf{q}_{c}^{(\Delta\neq 0)}=-\sum_{i,j} \frac{m_i}{8d} \sigma_{ij}^{d+1} \chi_{ij} \Delta_{ij}\left(X_{ij}^T \nabla \ln T+\sum_{\ell=1}^s\; X_{ij\ell}^n \nabla \ln n_\ell \right),
\eeq
where
\begin{align}
\label{a15}
X_{ij}^T=\int \dd\mathbf{v}_{1}\int \dd\mathbf{v}_{2}\;f_{i}^{(0)}(\mathbf{v}_{1})T\partial_T f_{j}^{(0)}(\mathbf{v}_{2})\Bigg[4\mu_{ji}(\mu_{ji}-\mu_{ij})\left( B_1 \Delta_{ij} g+B_2 \al_{ij}g^2\right)-8 B_2\mu_{ji} (\mathbf{g}\cdot \mathbf{G}_{ij})\Bigg]\nonumber\\
- \frac{2}{\sigma_{ij}}\frac{\rho}{n_jT_j^{(0)}} D_j^T \int \dd \mathbf{v}_1 \int \dd \mathbf{v}_2 \; f_i^{(0)}(\mathbf{v}_1)f_{j,\text{M}}(\mathbf{v}_2)V_{2,\beta}\Bigg[4\mu_{ji}(\mu_{ji}-\mu_{ij})\left( B_2\Delta_{ij}g_{\beta} + \alpha_{ij}B_3gg_{\beta} \right) \nonumber\\
- \frac{8B_1}{d+1}\mu_{ji}G_{ij,\mu}g^{-1}\left(g_\beta g_\mu + g^2 \delta_{\beta\mu}\right)\Bigg],
\end{align}
\begin{align}
\label{a16}
X_{ij\ell}^n = \int \dd\mathbf{v}_{1}\int \dd\mathbf{v}_{2}\;f_{i}^{(0)}(\mathbf{v}_{1})\Bigg\{n_\ell \frac{\partial}{\partial n_\ell}
f_{j}^{(0)}(\mathbf{v}_{2})\Bigg[4\mu_{ji}(\mu_{ji}-\mu_{ij})\left(B_1 \Delta_{ij} g+B_2 \al_{ij}g^2\right)-8B_2\mu_{ji} (\mathbf{g}\cdot \mathbf{G}_{ij})\Bigg]\nonumber\\
+ \, I_{ij\ell}f_{j}^{(0)}(\mathbf{v}_{2})\Big[4\mu_{ji}^2\left( B_1 \Delta_{ij} g+B_2 \al_{ij}g^2\right)-4B_2 \mu_{ji}(\mathbf{g}\cdot \mathbf{G}_{ij})\Big]\Bigg\} \nonumber\\
-\frac{2}{\sigma_{ij}}
\frac{m_j\rho_\ell}{\rho n_jT_j^{(0)}}D_{j\ell}
\int \dd\mathbf{v}_1\int \dd\mathbf{v}_2\,
f_i^{(0)}(\mathbf{v}_1)f_{j,\mathrm{M}}(\mathbf{v}_2)
V_{2,\beta}\Bigg[4\mu_{ji}(\mu_{ji}-\mu_{ij})
\left(B_2\Delta_{ij}g_{\beta}+
\alpha_{ij}B_3gg_{\beta}
\right)
\nonumber\\
-\frac{8B_1}{d+1}\mu_{ji}
G_{ij,\mu}g^{-1}
\left(g_\beta g_\mu+g^2\delta_{\beta\mu}
\right)\Bigg]
.
\end{align}

According to Eq.\ \eqref{2.17}, the contributions to $D_{q,ij}^c$ and $\kappa^c$ coming from $\mathbf{q}_{c}^{(\Delta\neq 0)}$ are
\beq
\label{a18}
D_{q,ij}^{\text{c}(\Delta\neq 0)}=\frac{m_i}{8 d T^2}\sum_{\ell=1}^s\; \sigma_{i\ell}^{d+1} \chi_{i\ell} \Delta_{i\ell}X_{i\ell j}^n,
\eeq
\beq
\label{a17}
\kappa_\text{c}^{(\Delta\neq 0)}=\sum_{i=1}^s \kappa_{i,c}^{(\Delta\neq 0)}, \quad
\kappa_{i,c}^{(\Delta\neq 0)}=\sum_{j=1}^s\; \frac{m_i}{8 d T}\sigma_{ij}^{d+1} \chi_{ij} \Delta_{ij}X_{ij}^T.
\eeq

To evaluate $X_{ij}^T$ and $X_{ij\ell}^n$ one can use the identities
\beq
\label{a18.1}
T\partial_T f_j^{(0)}=-\frac{1}{2}\frac{\partial}{\partial \mathbf{V}_2}\cdot \left(\mathbf{V}_2 f_j^{(0)}\right)\Bigg(1-\frac{1}{2}\Delta^*\frac{\partial \ln \gamma_j}{\partial \Delta^*}\Bigg), \quad
n_\ell \frac{\partial}{\partial n_\ell}f_{j}^{(0)}=\delta_{j\ell}f_j^{(0)}-\frac{1}{2}\frac{\partial}{\partial \mathbf{V}_2}\cdot \left(\mathbf{V}_2 f_j^{(0)}\right)n_\ell \frac{\partial \ln{\gamma_j}}{\partial n_\ell}.
\eeq
Explicit expressions of $X_{ij}^T$ and $X_{ij\ell}^n$ can be obtained when one employs the relations \eqref{a18.1} and replaces the zeroth-order distributions by their corresponding Maxwellian forms \eqref{3.20}. After some algebra, the results are
\beqa
\label{a19}
X_{ij}^T&=&\frac{4\pi^{\frac{d-1}{2}}}{\Gamma\left(\frac{d}{2}\right)}\frac{n_i n_j T}{\overline{m}}\Bigg\{\mu_{ji}(\mu_{ji}-\mu_{ij})\Delta_{ij}^*
(\theta_i+\theta_j)^{-1/2}\theta_i^{1/2}\theta_j^{-1/2}\nonumber\\
& & +\sqrt{\pi}\frac{\overline{m}\gamma_j}{m_i}\mu_{ij}\left[\al_{ij}(\mu_{ji}-\mu_{ij})+
2\mu_{ji}\right]\Bigg\}\Bigg(1-\frac{1}{2}\Delta^*\frac{\partial \ln \gamma_j}{\partial \Delta^*}\Bigg) +\frac{8dn_i\mu_{ij}}{m_i\sigma_{ij}}\rho D_j^T\mathcal{Q}_{ij},
\eeqa
\beqa
\label{a20}
X_{ij\ell}^n&=&\frac{2\pi^{\frac{d-1}{2}}}{\Gamma\left(\frac{d}{2}\right)}\frac{n_in_j T}{\overline{m}}\Bigg\{
\left[4\mu_{ji}(\mu_{ji}-\mu_{ij})\delta_{j\ell}+4 \mu_{ji}^2 I_{ij\ell}\right]\Bigg[\left(\frac{\theta_i+\theta_j}{\theta_i\theta_j}\right)^{1/2}\Delta_{ij}^*+\frac{\sqrt{\pi}}{2}
\al_{ij}\left(\frac{\theta_i+\theta_j}{\theta_i\theta_j}\right)\Bigg]\nonumber\\
& & -\,4\sqrt{\pi} \frac{\overline{m}}{m_i+m_j}\mu_{ji}(\gamma_i-\gamma_j)
\left(\delta_{j\ell}+\frac{1}{2}I_{ij\ell}\right)
+\Bigg[2\mu_{ji}
(\mu_{ji}-\mu_{ij})\Delta_{ij}^*
(\theta_i+\theta_j)^{-1/2}\theta_i^{1/2}\theta_j^{-1/2}\nonumber\\
& & +\,2\sqrt{\pi}\frac{\overline{m}\gamma_j}
{m_i}\mu_{ij}\left[\al_{ij}(\mu_{ji}-\mu_{ij})+2\mu_{ji}\right]\Bigg]
n_\ell \frac{\partial \ln{\gamma_j}}{\partial n_\ell}\Bigg\}+ \frac{8dn_i\mu_{ij}}{m_i\sigma_{ij}} \frac{m_j\rho_\ell}{\rho}D_{j\ell}\mathcal{Q}_{ij},
\eeqa
\end{widetext}
where $\mathcal{Q}_{ij}$ is defined by Eq.\ \eqref{3.8.1}. To obtain Eqs.\ \eqref{a19} and \eqref{a20}, use has been made of the results
\beq
\label{a21}
\langle g \rangle =\frac{\Gamma\left(\frac{d+1}{2}\right)}
{\Gamma\left(\frac{d}{2}\right)}\left(\frac{\theta_i+\theta_j}{\theta_i\theta_j}\right)^{1/2}n_in_j \upsilon_\text{th},
\eeq
\beq
\label{a22}
\langle g^2 \rangle=\frac{d}
{2}\left(\frac{\theta_i+\theta_j}{\theta_i\theta_j}\right)n_in_j \upsilon_\text{th}^2,
\eeq
\beq
\label{a23}
\langle \mathbf{g}\cdot \mathbf{G}_{ij}\rangle =d \frac{n_i n_j T}{m_i+m_j}\left(\gamma_i-\gamma_j\right),
\eeq
where
\beq
\label{a23.1}
\langle F(\mathbf{V}_1,\mathbf{V}_2)\rangle= \int \dd\mathbf{v}_{1}\int \dd\mathbf{v}_{2}f_{i}^{(0)}(\mathbf{v}_{1})f_{j}^{(0)}(\mathbf{v}_{2})F(\mathbf{V}_1,\mathbf{V}_2).
\eeq

The expressions \eqref{3.8} and \eqref{3.11} for the collisional transport coefficients $D_{q,ij}^{\text{c}(\Delta\neq 0)}$ and $\kappa_\text{c}^{(\Delta\neq 0)}$ can be obtained by substitution of Eqs.\ \eqref{a19} and \eqref{a20} into Eqs.\ \eqref{a18} and \eqref{a17}.

\section{Expressions of $\omega_{ii}$, $\omega_{ij}$, and $\Lambda_i$}
\label{appB}

\begin{widetext}

Explicit expressions of $\omega_{ii}$ and $\omega_{ij}$ can be obtained when the zeroth-order distributions $f_i^{(0)}(\mathbf{V})$ are approximated by their Maxwellian forms \eqref{3.20} in their definitions \eqref{3.27} and \eqref{3.28}. After some algebra, one gets
\beqa
\label{b1}
\omega_{ii}&=&-\frac{\pi^{(d-1)/2}}{d\Gamma\left(\frac{d}{2}\right)}v_{\text{th}}\theta_i^{-1/2}\Bigg\{\frac{3}{\sqrt{2}}
n_i\sigma_i^{d-1}\chi_{ii}\left(1-\alpha_{ii}^2\right)-\sum_{j\neq i}^s n_j \sigma_{ij}^{d-1} \chi_{ij}\mu_{ji} \left(1+\alpha_{ij}\right)\left(\theta_i+\theta_j\right)^{-1/2}\theta_j^{-1/2}
\nonumber\\
& & \times \Big[3\mu_{ji}\left(1+\alpha_{ij}\right)
\left(\theta_i+\theta_j\right)-2\left(2\theta_i+3\theta_j\right)\Big]\Bigg\}
+\frac{2\pi^{(d-1)/2}}{d\Gamma\left(\frac{d}{2}\right)}\upsilon_{\text{th}}n_i\sigma_i^{d-1}\chi_{ii}\Delta_{ii}^*
\left(\sqrt{\pi}\alpha_{ii}+
\sqrt{\frac{\theta_i}{2}}\Delta_{ii}^*\right)\nonumber\\
& & +\frac{4\pi^{(d-1)/2}}{d\Gamma\left(\frac{d}{2}\right)}\upsilon_{\text{th}}\sum_{j\neq i}^s n_j \sigma_{ij}^{d-1}\chi_{ij} \mu_{ji}^2 \Delta_{ij}^*\Bigg[\sqrt{\pi}\left(1+\al_{ij}-\mu_{ji}^{-1}\right)+
\Delta_{ij}^*\left(\frac{\theta_i\theta_j}{\theta_i + \theta_j}\right)^{1/2}\Bigg],
\eeqa
\beqa
\label{b2}
\omega_{ij}&=&\frac{\pi^{(d-1)/2}}{d\Gamma\left(\frac{d}{2}\right)}\upsilon_{\text{th}} n_j \sigma_{ij}^{d-1}\chi_{ij}\mu_{ij}
\left(1+\alpha_{ij}\right)\left(\theta_i+\theta_j\right)^{-1/2}\theta_i^{-1/2}\theta_j^{-1/2}\Big[3\mu_{ji}
\left(1+\alpha_{ij}\right)
\left(\theta_i+\theta_j\right)-2\theta_j\Big]\nonumber\\
& & +\frac{4\pi^{(d-1)/2}}{d\Gamma\left(\frac{d}{2}\right)}v_{\text{th}}n_j \sigma_{ij}^{d-1}\chi_{ij} \mu_{ij} \mu_{ji} \Delta_{ij}^*\Bigg[\sqrt{\pi}\left(1+\al_{ij}\right)+\Delta_{ij}^* \left(\frac{\theta_i\theta_j}{\theta_i + \theta_j}\right)^{1/2}\Bigg].
\eeqa
In Eqs.\ \eqref{b1}--\eqref{b2}, it is understood that $i\neq j$. Note that we have found some typos involving the terms proportional to $\Delta_{ij}^{*2}$ in the expressions (C12) and (C13) of $\omega_{ii}$ and $\omega_{ij}$ reported previously in Ref.\ \onlinecite{GBS21}.  The forms provided in this paper correct these mistakes.

According to Eq.\ \eqref{3.29}, to determine $\Lambda_i$ one has to estimate the collision integral
\beq
\label{b3}
A_{ij}\equiv \int \dd\mathbf{v}\, m_i V^2 \mathcal{K}_{ij,\beta}\left[\frac{\partial f_j^{(0)}}{\partial V_\beta}\right]=A_{ij}^{(\Delta=0)}+A_{ij}^{(\Delta\neq 0)},
\eeq
where $A_{ij}^{(\Delta=0)}$ and $A_{ij}^{(\Delta\neq 0)}$ refer to the contributions to $A_{ij}$ when $\Delta_{ij}^*=0$ and $\Delta_{ij}^*\neq 0$, respectively. The contribution $A_{ij}^{(\Delta=0)}$ was previously obtained in Ref.\ \onlinecite{GGG19b} with the result
\beq
\label{b4}
A_{ij}^{(\Delta=0)}=-
\frac{\pi^{d/2}}{\Gamma\left(\frac{d}{2}\right)}\chi_{ij} n_i n_j \sigma_{ij}^d m_{ij}T(1+\al_{ij})\Bigg[3\mu_{ji}(1+\al_{ij})
\left(\frac{\gamma_i}{m_i}+\frac{\gamma_j}{m_j}\right)
-4\frac{\gamma_i}{m_i}\Bigg].
\eeq

The contribution to $A_{ij}$ coming from the terms proportional to $\Delta_{ij}^*\neq 0$ is
\beqa
\label{b5}
A_{ij}^{(\Delta\neq 0)}&=&-m_i \sigma_{ij}^d \chi_{ij}\Delta_{ij}\int \dd\mathbf{v}_{1}\int \dd\mathbf{v}_{2}\int
\dd\widehat{\boldsymbol {\sigma}}\,\Theta (\widehat{\boldsymbol{\sigma}}\cdot
\mathbf{g}_{12})(\widehat{\boldsymbol {\sigma}}\cdot \mathbf{g}_{12})\widehat{\sigma}_\beta f_i^{(0)}(\mathbf{V}_1)
\frac{\partial f_j^{(0)}(\mathbf{V}_2)}{\partial V_{2\beta}}\nonumber\\
& & \times \Big[4\mu_{ji}^2\Delta_{ij}-4\mu_{ji}(\widehat{\boldsymbol{\sigma}}\cdot
\mathbf{V}_{1})+4\mu_{ji}^2(1+\al_{ij})(\widehat{\boldsymbol{\sigma}}\cdot
\mathbf{g}_{12})\Big]\nonumber\\
&=&
-m_i \sigma_{ij}^d \chi_{ij}\Delta_{ij}\int \dd\mathbf{v}_{1}\int \dd\mathbf{v}_{2}\int
\dd\widehat{\boldsymbol {\sigma}}\,\Theta (\widehat{\boldsymbol{\sigma}}\cdot
\mathbf{g}_{12})f_i^{(0)}(\mathbf{V}_1)f_j^{(0)}(\mathbf{V}_2)\nonumber\\
& & \times \Big[4\mu_{ji}^2\Delta_{ij}
-4\mu_{ji}(\widehat{\boldsymbol{\sigma}}\cdot
\mathbf{V}_{1})+8\mu_{ji}^2(1+\al_{ij})(\widehat{\boldsymbol{\sigma}}\cdot
\mathbf{g}_{12})\Big].
\eeqa
\end{widetext}
As usual, to evaluate the last line of Eq.\ \eqref{b5} we make the replacements $f_i^{(0)}(\mathbf{V}_1)\to f_{i,\text{M}}(\mathbf{V}_1)$ and $f_j^{(0)}(\mathbf{V}_2)\to f_{j,\text{M}}(\mathbf{V}_2)$. After some algebra, one gets the result
\beqa
\label{b6}
A_{ij}^{(\Delta\neq 0)}&=&-\frac{2\pi^{\frac{d-1}{2}}}{\Gamma\left(\frac{d}{2}\right)}n_i n_j \sigma_{ij}^d \chi_{ij}\frac{m_i T}{\overline{m}}\Delta_{ij}^*\Bigg\{4
\sqrt{\pi}\mu_{ji}^2 \Delta_{ij}^*\nonumber\\
& & -\left(\frac{\theta_i+\theta_j}{\theta_i\theta_j}\right)^{1/2}\Big[4\mu_{ji}\theta_j(\theta_i+\theta_j)^{-1}-8\mu_{ji}^2\nonumber\\
& &
\times (1+\al_{ij})\Big]\Bigg\}.
\eeqa

\section{Some auxiliary quantities for hard disks}
\label{appC}

In this Appendix we provide the forms of some auxiliary quantities needed to illustrate the dependence of the transport coefficients associated with the mass and heat fluxes on the parameter space of the mixture. The case of hard disks ($d=2$) is considered.

For hard disks, a good approximation for the pair distribution function $\chi_{ij}$ is \cite{JM87}
\begin{equation}
\label{c1}
\chi_{ij}=\frac{1}{1-\phi}+\frac{9}{16}\frac{\phi}{(1-\phi)^2}\frac{\sigma_i\sigma_jM_1}{\sigma_{ij}M_2},
\end{equation}
where $\phi=\sum_i\; \pi n_i\sigma_i^2/4$ is the solid volume fraction for disks and
\begin{equation}
\label{c2}
M_n=\sum_{k=1}^2\; x_k \sigma_k^n.
\end{equation}
The expression of the chemical potential $\mu_i$ of the species $i$
consistent with the approximation \eqref{c1} is\cite{S16}
\begin{eqnarray}
\label{c3}
\frac{\mu_i}{T}&=&\ln (\lambda_{i}^2n_i)-\ln (1-\phi)+\frac{M_1}{4M_2}\left[\frac{9\phi}{1-\phi}+\ln (1-\phi)\right]\sigma_i
\nonumber\\
& &
-\frac{1}{8}\Big[\frac{M_1^2}{M_2^2}\frac{\phi(1-10\phi)}{(1-\phi)^2}-
\frac{8}{M_2}\frac{\phi}{1-\phi}+\frac{M_1^2}{M_2^2}\ln (1-\phi)\Big]\nonumber\\
& &\times \sigma_i^2,
\end{eqnarray}
where $\lambda_{i}(T)$ is the thermal de Broglie wavelength. \cite{RG73} 
As discussed in previous works,\cite{GGBS24,GMG26} the fact that granular fluids lack a thermodynamic description, the concept of chemical potential could be questionable. In fact, the expression \eqref{c3} has been obtained by employing thermodynamic relations that only apply for elastic systems. In this context, we expect that this approximation could be reliable for not too small values of the coefficients of restitution.

The presence of the chemical potentials $\mu_i$ in the theory is essentially due to the choice of the quantities $I_{i\ell j}$. These quantities (which are defined in terms of the functional derivative of local pair distribution function $\chi_{ij}$ with respect to the local partial densities $n_\ell$) are the origin of the primary difference between the standard and the revised Enskog kinetic theories for elastic collisions.\cite{BE73a,BE73b,BE73c,LCG99} For binary mixtures, the nonzero parameters $I_{i\ell j}$ are $I_{121}$ and $I_{122}$. They are given by\cite{GHD07,G19}
\beqa
\label{c4}
I_{121}&=&\frac{2}{\pi T n_2\sigma_{12}^2\chi_{12}}\left[n_1\left(\frac{\partial \mu_1}{\partial n_1}
    \right)_{T,n_{2}}-T\right]-2\frac{n_1\sigma_1^2\chi_{11}}
{n_2\sigma_{12}^2\chi_{12}}\nonumber\\
& & -
\frac{n_1^2\sigma_1^2}{n_2\sigma_{12}^2\chi_{12}}\frac{\partial\chi_{11}}{\partial n_1}
-\frac{n_1}{\chi_{12}}\frac{\partial\chi_{12}}{\partial n_1},
\eeqa
\beqa
\label{c5}
I_{122}&=&\frac{2}{\pi T\sigma_{12}^2\chi_{12}}\left(\frac{\partial \mu_1}{\partial n_2}
\right)_{T,n_{1}}-2-
\frac{\sigma_1^2 n_1}{\sigma_{12}^2\chi_{12}}\frac{\partial\chi_{11}}{\partial n_2}\nonumber\\
& &
-\frac{n_2}{\chi_{12}}\frac{\partial\chi_{12}}{\partial n_2}.
\eeqa
For mechanically equivalent particles ($m_1=m_2$, $\sigma_1=\sigma_2$), $I_{121}=I_{122}=0$, as expected since the standard and revised Enskog theories lead to the same Navier-Stokes transport coefficients for a monocomponent granular gas.\cite{GD99a}

Furthermore, the diffusion and heat flux transport coefficients are defined in terms of several derivatives. In particular, for the simple case $\Delta_{ij}=\Delta$, the derivatives $(\partial \gamma_1/\partial \Delta^*)$, $(\partial \gamma_1/\partial x_1)$ and $(\partial \gamma_1/\partial \phi)$ are given by\cite{GBS21}
\beq
\label{c6}
\left(\frac{\partial \gamma_1}{\partial \Delta^*}\right)=\frac{\sqrt{Y^2-4X Z}-Y}{2X},
\eeq
where $X=N \Delta^*$,
\beq
\label{c7}
Y=M \Delta^*-2 N \gamma_{1}+\gamma_{1}
\left(\frac{\partial \zeta_1^*}{\partial \gamma_1}\right), \quad Z=\gamma_{1}\left(\frac{\partial \zeta_1^*}{\partial \Delta^*}\right)-2 M \gamma_{1}.
\eeq
Here,
\beq
\label{c8}
M=\frac{1}{2}\Bigg[x_1 \gamma_1 \Big(\frac{\partial \zeta_1^*}{\partial \Delta^*}\Big)_{\gamma_1}
+x_2 \gamma_2 \Big(\frac{\partial \zeta_2^*}{\partial \Delta^*}\Big)_{\gamma_1}\Bigg],
\eeq
\beq
\label{c9}
N=\frac{1}{2}\Bigg(x_1 \gamma_1 \frac{\partial \zeta_1^*}{\partial \gamma_1}
+x_2 \gamma_2 \frac{\partial \zeta_2^*}{\partial \gamma_1}\Bigg).
\eeq
In addition, in the above equations
\beq
\label{c10}
\left(\frac{\partial \zeta_i^*}{\partial \Delta^*}\right)=\left(\frac{\partial \zeta_i^*}{\partial \Delta^*}\right)_{\gamma_1}+\left(\frac{\partial \zeta_i^*}{\partial \gamma_1}\right)\left(\frac{\partial \gamma_1}{\partial \Delta^*}\right).
\eeq
Similarly to Eq.\ \eqref{c10}, we have the relations
\beq
\label{c11}
\left(\frac{\partial p^*}{\partial \Delta^*}\right)=\left(\frac{\partial p^*}{\partial \Delta^*}\right)_{\gamma_1}+\left(\frac{\partial p^*}{\partial \gamma_1}\right)\left(\frac{\partial \gamma_1}{\partial \Delta^*}\right),
\eeq
\beq
\label{c12}
\left(\frac{\partial \zeta_0^*}{\partial \Delta^*}\right)=\left(\frac{\partial \zeta_0^*}{\partial \Delta^*}\right)_{\gamma_1}+\left(\frac{\partial \zeta_0^*}{\partial \gamma_1}\right)\left(\frac{\partial \gamma_1}{\partial \Delta^*}\right).
\eeq

The derivatives $\partial \gamma_1/\partial x_1$ and $(\partial \gamma_1/\partial \phi)$ can be written as
\begin{widetext}
\beq
\label{c13}
\frac{\partial \gamma_1}{\partial x_1}=-\frac{\gamma_{1}\frac{\partial \zeta_1^*}{\partial x_1}+\frac{1}{2}\left(x_1\gamma_{1}\frac{\partial \zeta_1^*}{\partial x_1}+x_2\gamma_{2}\frac{\partial \zeta_2^*}{\partial x_1}\right)\left[\Delta^* \left(\frac{\partial \gamma_1}{\partial \Delta^*}\right)-2\gamma_1\right]}
{\gamma_{1}\frac{\partial \zeta_1^*}{\partial \gamma_1}+\frac{1}{2}\left(x_1\gamma_{1}\frac{\partial \zeta_1^*}{\partial \gamma_1}+x_2\gamma_{2}\frac{\partial \zeta_2^*}{\partial \gamma_1}\right)\left[\Delta^* \left(\frac{\partial \gamma_1}{\partial \Delta^*}\right)-2\gamma_1\right]},
\eeq
\beq
\label{c14}
\frac{\partial \gamma_1}{\partial \phi}=-\frac{\gamma_{1}\frac{\partial \zeta_1^*}{\partial \phi}+\frac{1}{2}\left(x_1\gamma_{1}\frac{\partial \zeta_1^*}{\partial \phi}+x_2\gamma_{2}\frac{\partial \zeta_2^*}{\partial \phi}\right)\left[\Delta^* \left(\frac{\partial \gamma_1}{\partial \Delta^*}\right)-2\gamma_1\right]}
{\gamma_{1}\frac{\partial \zeta_1^*}{\partial \gamma_1}+\frac{1}{2}\left(x_1\gamma_{1}\frac{\partial \zeta_1^*}{\partial \gamma_1}+x_2\gamma_{2}\frac{\partial \zeta_2^*}{\partial \gamma_1}\right)\left[\Delta^* \left(\frac{\partial \gamma_1}{\partial \Delta^*}\right)-2\gamma_1\right]}.
\eeq
\end{widetext}



%
\end{document}